\documentclass[journal=cmatex,manuscript=article]{achemso}
\SectionNumbersOn

\usepackage[version=3]{mhchem} 
\usepackage{makecell}
\usepackage[sort&compress]{natbib}
\usepackage{tabularx}
\newcolumntype{L}{>{\raggedright\arraybackslash}X}
\usepackage{booktabs}
\usepackage{tabularray}
\usepackage{longtable}
\usepackage{multirow}
\usepackage{textgreek}
\usepackage[breaklinks]{hyperref}
\usepackage{breakcites}
\usepackage{xcolor}
\usepackage{soul}
\usepackage{microtype}
\usepackage{xcolor}
\usepackage{amssymb}



\author{Peter M. Piechulla}
\author{Mingliang Chen}
\author{Aristeidis Goulas}
\affiliation[TU Delft]
{Department of Chemical Engineering, Delft University of Technology, 2629 HZ, Delft, The Netherlands}
\author{Riikka L. Puurunen}
\affiliation[Aalto]
{School of Chemical Engineering, Department of Chemical and Metallurgical Engineering, Aalto University, FI-00076 Aalto, Espoo, Finland}
\author{J. Ruud van Ommen}
\affiliation[TU Delft]
{Department of Chemical Engineering, Delft University of Technology, 2629 HZ, Delft, The Netherlands}
\email{J.R.vanOmmen@tudelft.nl}

\title[pALD review]
  {Atomic layer deposition on particulate materials from 1988 through 2023:\\A quantitative review of technologies, materials and applications}

\keywords{atomic layer deposition, ALD, fluidized bed reactor, nanoparticles, particle coating, powder}

\begin{document}

\tableofcontents


\begin{abstract}
Atomic layer deposition (ALD) is widely studied for numerous applications and is commercially employed in the semiconductor industry, where planar substrates are the norm. However, the inherent ALD feature of coating virtually any surface geometry with atomistic thickness control is equally attractive for coating particulate materials (supports). In this review, we provide a comprehensive overview of the developments in this decades-old field of ALD on particulate materials, drawing on a bottom-up and quantitative analysis of 799 articles from this field. The obtained dataset is the basis for abstractions regarding reactor types (specifically for particles), coating materials, reactants, supports and processing conditions. Furthermore, the dataset enables direct access to specific processing conditions (for a given material, surface functionality, application etc.) and increases accessibility of the respective literature. We also review fundamental concepts of ALD on particles, and discuss the most common applications, i.e., catalysis (thermo-, electro-, photo-), batteries, luminescent phosphors and healthcare. Finally, we identify historical trends, and provide an outlook on prospective developments.
\end{abstract}

\section{Introduction}

Particles or particulate materials---pieces of solid material, sometimes referred to as powder---are widely used in various fields of science and engineering due to their unique properties, especially their large surface area. 
Modification of particle surfaces with a second material can greatly enhance their functionality. 
On the one hand, complete coverage with a film can protect the particulate materials from influences from outside. 
On the other hand, decorating the surface with clusters of a different material can introduce novel functionalities such as catalysis. 
Such coating processes have already been carried out for a long time using liquid-based techniques, typically with limited control over the structures (film thickness, cluster size). 
In contrast, gas-phase deposition processes, and in particular atomic layer deposition (ALD), enable much more control over the structure size, distribute coating materials more homogeneously, and allow the coating of smaller particles, down to nanoparticles.

The first article on ALD on particles that was assigned a digital object identifier (DOI) and that is therefore readily findable in an indexing database (e.g., Scopus), was published by Asakura et al.\ from Japan in 1988\cite{Asakura1988b}.
However, the branch of ALD technology that specializes on particulate materials started to develop already in the 1960s in the Soviet Union\cite{VPHAAleskovskii1974,Malygin2015}. 
The early technology development, also in Finland,\cite{Haukka1998, 2014Puurunen} was motivated by applications of that time, such as thermocatalysis, which is still one of the driving forces for the development of the field today.
However, the functionalities attainable through ALD on particles increasingly attract also newer applications of both high industrial relevance and research interest. 

ALD technology can be used to address multiple key challenges of our time.
Some of the upcoming applications, such as batteries or electrocatalysts, are essential elements in tackling decarbonization and electrification based on renewable energy sources.
Additionally, an increasing global challenge is the scarcity of raw materials, where ALD can be used to deposit these materials sparsely. 
An example is noble metals that are used as catalysts in various forms (thermo-, electro-, photocatalysts).
Here, metal atoms near the interface are taking part in the targeted reaction, and ALD seems suitable to deposit them exactly there. 
Hence, the consumption of these metals can be reduced, thereby preventing a bottleneck in scaling up the respective technology.
The bare scale of these recent particle-based applications, together with the unique features of ALD, could help ALD on particles to a similar breakthrough like wafer-based ALD has seen through semiconductor manufacturing.\cite{2023Kim,2009Ritala,2010George,2014Hwang,2014Puurunen,2021vanOmmen}
This calls for a thorough review of the field and its applications.

Despite the over a dozen review articles available today\cite{VPHAAleskovskii1974,1996Malygin,Lakomaa1994,Haukka1997a,Haukka1998,King2009e,King2012,Weimer2019,Liang2013c,2011Detavernier,ONeill2015a,Longrie2014b,Adhikari2018,2017Bui,vanOmmen2019,Cao2020a,Li2021b,
Hu2021}, no single review article has yet comprehensively covered the entire body of literature on ALD on particulate materials; 
some of them focus on a specific application field\cite{Lakomaa1994,Haukka1997b,Haukka1998,Lee2022i}, and only few cite more than 150 references\cite{Adhikari2018,Hu2021}. At the same time, as we will show, approximately 800 articles have been published on ALD on particles.

The goal of this review is to provide a broad overview through a data-driven analysis of the field of ALD on particulate materials.
We strive to be complete, by including all articles starting from the first publication that includes a DOI handle from 1988\cite{Asakura1988b} up to articles published until the end of 2023.
We will first give an overview of previous review works (Section~\ref{ch:stateoftheart}), before describing our systematic methodology (Section~\ref{ch:method}).
Based on our bottom-up approach, we will provide a historical overview of the field and its most active contributors, and the evolution of technology names over time (Section~\ref{ch:demographics}).
Before going into the specifics of the technology, we will discuss the principles of ALD in and its relation to particle technology (Section~\ref{ch:principles}).
The technology-oriented discussion will start by a review of the equipment used for ALD on particulate materials (Section~\ref{ch:equipment}). 
Then, we will review the various particulate substrate (support) materials that are used (Section~\ref{ch:supports}), as well as the processing parameters and typical characterization methods (Section~\ref{ch:processing}). Finally, we will discuss the applications of ALD on particulate materials (Section~\ref{ch:applications}), and give an outlook to future research directions~(Section~\ref{ch:outlook}).

\section{State of the art: previous review work}\label{ch:stateoftheart}

The review articles dedicated to ALD on particulate materials reflect the regional development of the field (Figure~\ref{fig:vstime}) towards a globally researched technology over the course of several decades.
The first review by Soviet researchers dates back to 1974 \cite{VPHAAleskovskii1974}. A book chapter on the topic from the same community and region (now Russia) was published in 1996\cite{1996Malygin}. 
Three early review articles from Finland from the 1990s \cite{Lakomaa1994, Haukka1997a,Haukka1998} focus on reaction mechanisms, with catalysis as the application. Of these three, Haukka et al.\cite{Haukka1998} is the most complete. 

It took more than a decade before the next review article was published \cite{King2009e}, i.e., after ALD on particles was established in the United States.
This article from the Weimer group has a very strong focus on the work of the own group; the same holds for two follow-up reviews\cite{King2012, Weimer2019}. From the same group, the review by Liang et al. \cite{Liang2013c}---a book chapter rather than a journal article---has a broader view. The main applications discussed are core--shell particles applied in batteries and tissue engineering. 

Around the same time, starting from the 2010s, two influential reviews by Detavernier et al.\cite{2011Detavernier}\ and O'Neill et al.\cite{ONeill2015a}\ with a focus on catalysts appeared, both of whom acknowledged that the community of ALD on particles is rooted more globally than just in the United States. 
Catalysts manufacturing is in most cases carried out on particulate materials, and we hence consider those reviews highly relevant for the field of ALD on particles.
In 2014, a thorough review was published by Longrie et al. \cite{Longrie2014b}, with attention to reactor design. Although not a real review article and not just focusing on particles, the bibliometric analysis by Alvaro et al. \cite{2018Alvaro} is interesting, showing a growing and interconnected ALD research community. 

From 2018, review articles on ALD on particulate materials were published at an increased pace. Adhikari et al. \cite{Adhikari2018} published a review on powder coating by ALD, but had to retract one section due to serious shortcomings in citing literature\cite{2017Bui}. Van Ommen et al. \cite{vanOmmen2019} published a brief review focusing on reactor design and operation. A review by Cao et al. \cite{Cao2020a} focused on surface modification. Li et al.\cite{Li2021b} reviewed the ALD literature with a focus on fluidization, especially of nanoparticles. Hu et al. \cite{Hu2021} reviewed surface modification of particulate materials by ALD for a range of applications. 

No single review article addresses the field as a whole, and typically no more than 160 articles are cited.
Most reviews focus on a specific application, set the scope to a limited part of the community, or anecdotally highlight a selection of notable articles.
While the latter may be justified, the selection criteria for which articles to include are often unclear. 

In addition to reviews focusing on ALD on particulate materials, of interest to the readers of this work may be general reviews on ALD chemistries. 
Such reviews have been published by Puurunen in 2005 \cite{2005Puurunen}, Miikkulainen et al. 2013 \cite{2013Miikkulainen}, and Popov et al.\ in 2025 \cite{2025Popov}. 
Unlike the earlier reviews, which presented an independent review of the scientific literature \cite{2005Puurunen, 2013Miikkulainen}, the review of 2025\cite{2025Popov} relied on the contents of an online ALD database\cite{AtomicLimitsALDdatabaseNote}; 
based on the work of Miikkulainen et al.\ from 2013\cite{2013Miikkulainen}, crowdsourcing has since been used to update the ALD process information in this database \cite{AtomicLimitsALDdatabaseNote}.
According to Popov et al. \cite{2025Popov}, the number of published thermal ALD processes found in the online ALD database\cite{AtomicLimitsALDdatabaseNote} is currently over 1700.
Merely a small fraction of those processes have been applied on particulate materials. 
Interestingly, according to our comparison \cite{AtomicLimitsCrossCheck}, less than 14\% of the 799 articles on ALD on particles identified in this work appear in the mentioned online ALD database\cite{AtomicLimitsALDdatabaseNote}.

\section{Methodology}\label{ch:method}

A four-step systematic approach was adopted in this quantitative review to populate the dataset and facilitate data-driven analyses. A schematic overview of the methodology described in the following section is illustrated in the Supporting Information file Figure~S1.

\subsection{Literature search method}

In the first step (keyword identification) a frequency analysis was conducted on keywords previously appearing in 15 selected relevant review articles (either in the title, abstract, or explicitly defined as keywords).\cite{Lakomaa1994, Haukka1997b, Haukka1998, King2009e, King2012, Longrie2014b, Liang2013c, Salameh2017, Adhikari2018, Weimer2019, vanOmmen2019, Cao2020a, Li2021b, Hu2021, Lee2022i}
A combinatorial approach was then used to generate search terms, pairing a technique description term (“atomic layer deposition”, “molecular layer deposition”, “atomic layer epitaxy”, “molecular layering”) with a term that served to narrow the focus to applications on particulate materials (particle(s), powder(s), porous, gel).

Two scholarly platforms, a subscription-based (Scopus) and an open-access (Google Scholar) were queried for the second step (scholarly platforms query).
Platform-specific search term customization was implemented.
Note, that a significant share of the results turned up exclusively on either one of the platforms.
Further, only indexed, electronically accessible articles that contained a digital object identifier (DOI) handle, were considered.
This search yielded approximately 4,000 articles as potential hits.
The final date of the systematic update of the search queries was 28-06-2024.
The surveyed span was limited to articles published up until the end of 2023. 

In the third step (literature selection refinement) the obtained article abstracts were manually screened by the co-authors, to assess their relevance to the review theme.
Here, it was opted to exclude coating studies involving: 1) immobilized particles (on flat substrates, wafers, sheets, sensors etc.), 2) particles that were contained in aerogels and foams, and 3) particles that were pressed or shaped (on disks, microscopy grids, fibers, electrodes etc.).
Only articles accessible through subscriptions of the affiliated institutes, or available as open-access, were considered.
The obtained articles were cross-checked against 465 articles reported in a previous literature review study.\cite{vanOmmen2019}
No exclusion was made based on the batch size of the coated particles.
A total of 799 indexed articles were available for the fourth (and last) step of the approach (dataset synthesis).

Additionally, a relevant literature list from the VPHA project\cite{2013VPHA} was considered and screened manually, to expand the scope of the obtained hits.
In that way, early ALD research activities (1965--1994) carried out mostly in the former Soviet Union could be addressed.
It should be noted that the VPHA list includes articles up until 1986.
From this list, 83 articles describing the use of particulate materials were identified, and are represented in the demographic data displayed in the history discussion (section~\ref{ch:demographics}).
However, these articles are excluded from the synthesis of the dataset described below, and therefore only discussed qualitatively.

\subsection{Dataset synthesis \& processing}\label{sec:datacollection}

After the screening and qualification of articles according to the criteria above, we condensed the bibliographic data, respective applications and most relevant ALD-specific information (identified through previous review work \cite{vanOmmen2019}) from the 800 indexed articles into a single dataset.
Accordingly, the data was grouped into seven main information categories: Demographics, applications, reactors, reactants, supports, processing and coating properties.
For each of these categories, we identified the most commonly reported qualities and quantities, for example, dosing time of reactants and process temperatures for process conditions, and collected the data from each of the considered articles into the single data table. 
Here, we limited our collection to data that could be directly read from the text or a graph of the publication and respective supplementary material.
Specifically, we did not attempt to infer data that was missing in a certain article according to the fields provided in our data table from equivalent data provided in the article, except for simple unit conversions.
For qualitative data, e.g. support materials, the data were subsequently classified.
The resulting data table is publicly available in a citable repository\cite{ALDpmZenodo} and contains approximately 74,000 numerical and categorical entries.

Based on this data table, we could then generate an overview of the field and answer specific questions by loading it into a scripting environment, i.e., Python with Pandas\cite{reback2020pandas}.
From there, the data was filtered to, e.g., a certain application, and create the respective overview graphs (timelines, support materials, etc.) that are presented throughout the present review. 
We provide a collection of Python scripts to generate some of the graphs from the manuscripts from the data table\cite{ALDpmZenodo}, available through a publicly accessible code repository\cite{2025GithubALDpm}.
The herein provided filtering methods can also be used to obtain specific parameter combinations and corresponding literature references, which is a valuable tool for further experimental research.

\section{History \& demographics}\label{ch:demographics}

The history of ALD on particulate materials can be traced back to the two independent inventions under the names Molecular Layering (ML) and Atomic Layer Epitaxy (ALE).\cite{2014Puurunen, Malygin2015} To clarify the early origins of ALD, the Virtual Project on the History of ALD (VPHA) was started in 2013 in an international collaboration, and it has resulted in several articles.\cite{2014Puurunen, Malygin2015, 2017Ahvenniemi, 2018Puurunen} As described in a VPHA-related essay on ML,\cite{Malygin2015} researchers from the USSR and Eastern Europe started a research program in the 1960s aiming to modify especially porous particles for use as sorbents, catalysts and rubber fillers. This program  resulted in tens of PhD-level theses.\cite{Malygin2015} As described in a VPHA-related essay on ALE,\cite{2014Puurunen} ALD was developed in Finland since 1974, initially aimed for electroluminescent display applications. Activities towards particulate materials in Finland started in 1987 with the founding of a new company, Microchemistry Ltd., with the initial goal to apply ALD for photovoltaics and catalysis. In addition to resulting in patents,\cite{1990PatentFI84562,1991PatentEP0525503B1} publications (e.g. references~\citenum{Lakomaa1992, Haukka1993a, Haukka1993b, 1993Haukka, Lakomaa1994, Haukka1994a, Haukka1994b, Haukka1994c, Lakomaa1996}) and PhD-level theses (e.g. references~\citenum{1993HaukkaPhD, 1994LindforsPhD, 1997KytokiviDr, 1999HakualiDr, 2002PuurunenDr, 2004LashdafDr}), the gained understanding of reaction mechanisms supported the adoption of ALD for semiconductor technology in the 2000s\cite{2013Lakomaa, HaukkaALDinnovationAward2016}. Interestingly, while during the past decades the majority of ALD publications have focused on films on planar substrates for semiconductor applications, it appears that in the early days of ALD, a large portion of the publications was, in fact, related to particulate materials.

\begin{figure}
  \includegraphics[width=\textwidth]{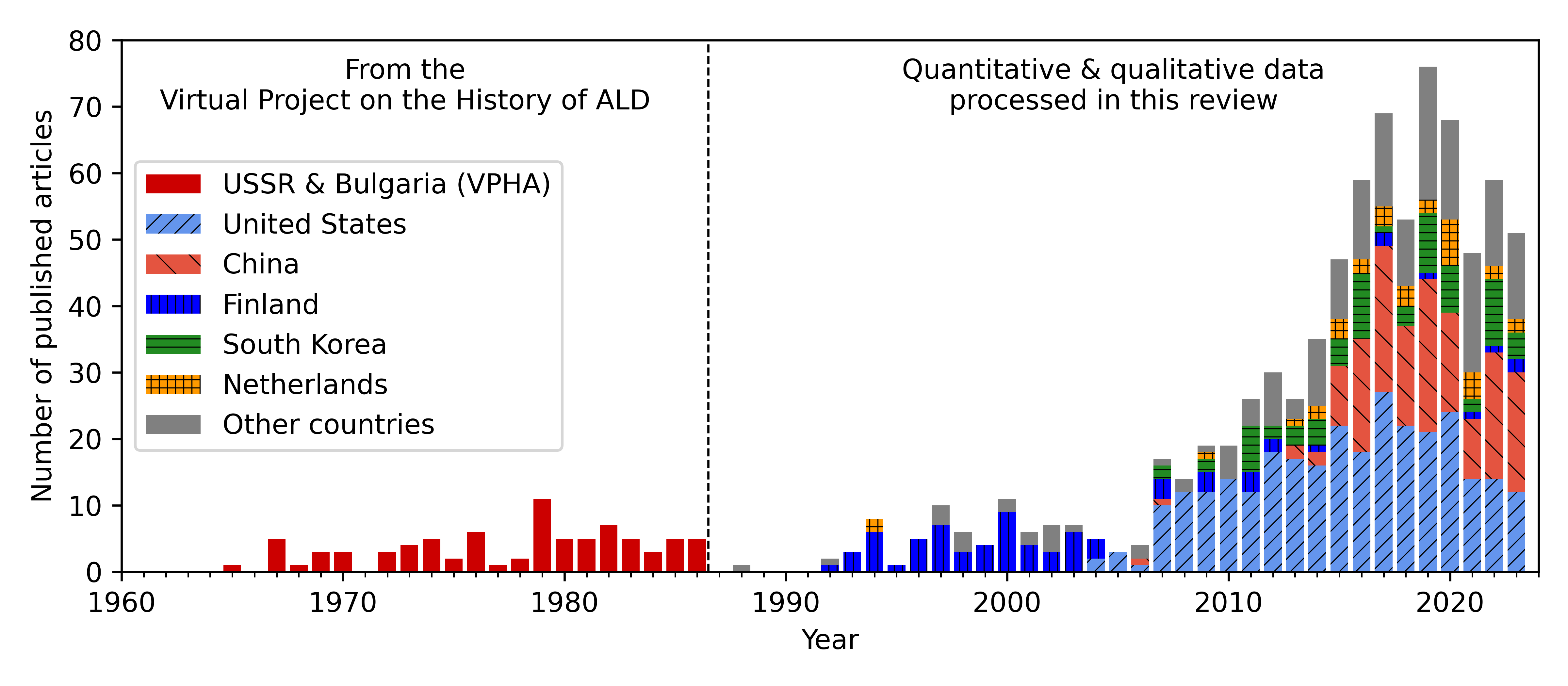}
  \caption{Number of published articles per year. The quantitative analysis in the present review starts from the earliest Japanese article from the year 1988\cite{Asakura1988b}, and is shown per country for the top five contributors, in order of number of contribution. Earlier contributions were identified in the Virtual Project on the History of ALD (VPHA), and came from the USSR and Bulgaria (investigation up to 1986, dashed line).}
  \label{fig:vstime}
\end{figure}

\begin{figure}
  \includegraphics[width=0.7\textwidth]{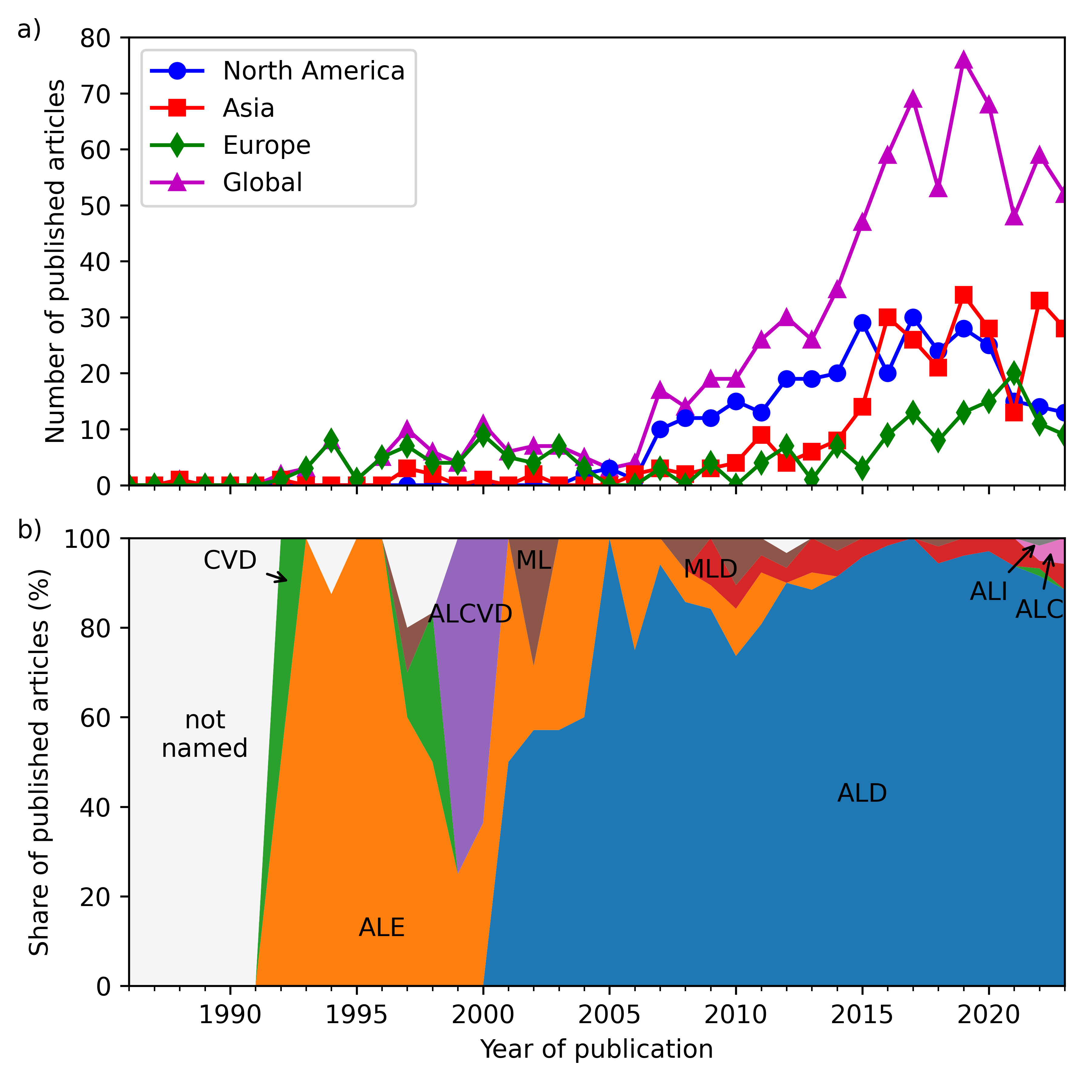}
  \caption{a) Number of published articles per year based on regions, and b) the ALD technology names used in the history for particulate materials.} 
  \label{fig:history}
\end{figure}

When examining global particle ALD research activities, they are primarily concentrated in Asia, Europe, and North America (Figure~\ref{fig:history}a) with the United States, China, Finland, South Korea, Canada, Germany, the Netherlands, Belgium and Russia ranking as the top 9 contributors (see Supporting Information Table~S3). As indicated by the VPHA project, active research on ALD on particulate materials was already being carried out by researchers from the USSR and Eastern Europe between the 1960s and 1980s (Figure~\ref{fig:vstime}). Afterwards (before 2005), particle ALD research was mainly conducted in Europe, particularly in Finland, Russia, and the Netherlands, with Finnish scientists leading efforts focused on catalyst synthesis. During this period, research activities in Asia (mainly Japan) and North America were minimal. Since 2006, research activities in Europe have become less concentrated, with the Netherlands, Germany, and Belgium emerging as the new leaders, taking over Finland's previous role. Meanwhile, North America has experienced a significant increase in research output, reaching over 10 articles per year, followed by a steady rise in subsequent years, ultimately surpassing Europe as the leader in global particle ALD research. Around 2010, Asia (i.e., South Korea and China) began a significant upward trend, surpassing both Europe and North America in the number of publications by 2015. Notably, China has emerged as the leader in Asia, with its research output far exceeding that of South Korea and approaching levels seen in North America (Figure~\ref{fig:vstime}). The peak publication year varied by region: Asia reached its highest point around 2022, North America peaked slightly earlier, and Europe peaked shortly afterward. Overall, the graph demonstrates a substantial rise in research output across all regions over the past decade, with Asia (in particular China) emerging as the leading contributor in recent years.

Throughout the historical development of ALD on particles, various technology names have been used in the ALD community (Figure~\ref{fig:history}b). 
In the pioneering Japanese work by Asakura et al.\cite{Asakura1988b} from 1988, the authors labeled their product 'one-atomic layer catalyst', while not directly assigning a name to the process.
In the following years, the term chemical vapor deposition (CVD) was employed by Japanese researchers in a few articles \cite{Asakura1992, Okumura1997b, Okumura1998a, Okumura1998b}.
Around the same time, the term ALE, coined by Tuomo Suntola, was widely used in Finnish publications\cite{Lakomaa1992,Haukka1993b,Lindblad1993}. In 1999 and 2000, researchers from Finland also started to use atomic layer chemical vapor deposition (ALCVD)\cite{Juvaste1999a,Juvaste1999b,Timonen1999,Puurunen2000a}, a term that was introduced as a trademark by the company ASM Microchemistry. 
Additionally, the term molecular layering (ML), named by Russian scientists, appeared in a few publications\cite{Malygin1997,Ermakova2002,Smirnov2008,Malkov2010}. In 2001, researchers from Finland and Belgium started to use the technology name “ALD” in their publications for the first time (based on our current dataset)\cite{Kanervo2001,Puurunen2001,Baltes2001}; however, we note that ALD has been used already in the 1990 in the broader field of ALD beyond particles. Over the following six years, the ALD community started to use “ALD” and “ALE” alternatively in the publications. Since 2007, the technology name “ALD” has been widely accepted by the ALD community, and is now predominantly used in numerous publications. Alongside the development of ALD technology, some new terms such as molecular layer deposition (MLD), atomic layer infiltration (ALI) and atomic layer coating (ALC), have also emerged in some publications\cite{Liang2009c, Liang2009d, Dong2019, Liang2010a, Barros2023, Swaminathan2023,Moseson2023a, Moseson2022, Duong2022, Qi2022}; those terms are either used as  synonyms for ALD, or refer to ALD-related techniques (see also Section~\ref{sec:otherALDtypes}).

\section{Principles}\label{ch:principles}
The following covers the basic principles of ALD processes, before moving on to the specific properties of and methods applied in ALD on particulate materials.
The aim of this tutorial-style section is to equip readers that are unfamiliar to ALD or particle technology, respectively, with the basic knowledge required to follow the rest of this review article.
For a more comprehensive introduction to the field of ALD, the reader is referred to the relevant literature\cite{2021vanOmmen,2014Leskela,2014Johnson,2010George,2002Ritala}.

\subsection{Basic characteristics of ALD processes}\label{sec:ALDbasic}

\begin{figure}
  \includegraphics[width=0.75\textwidth]{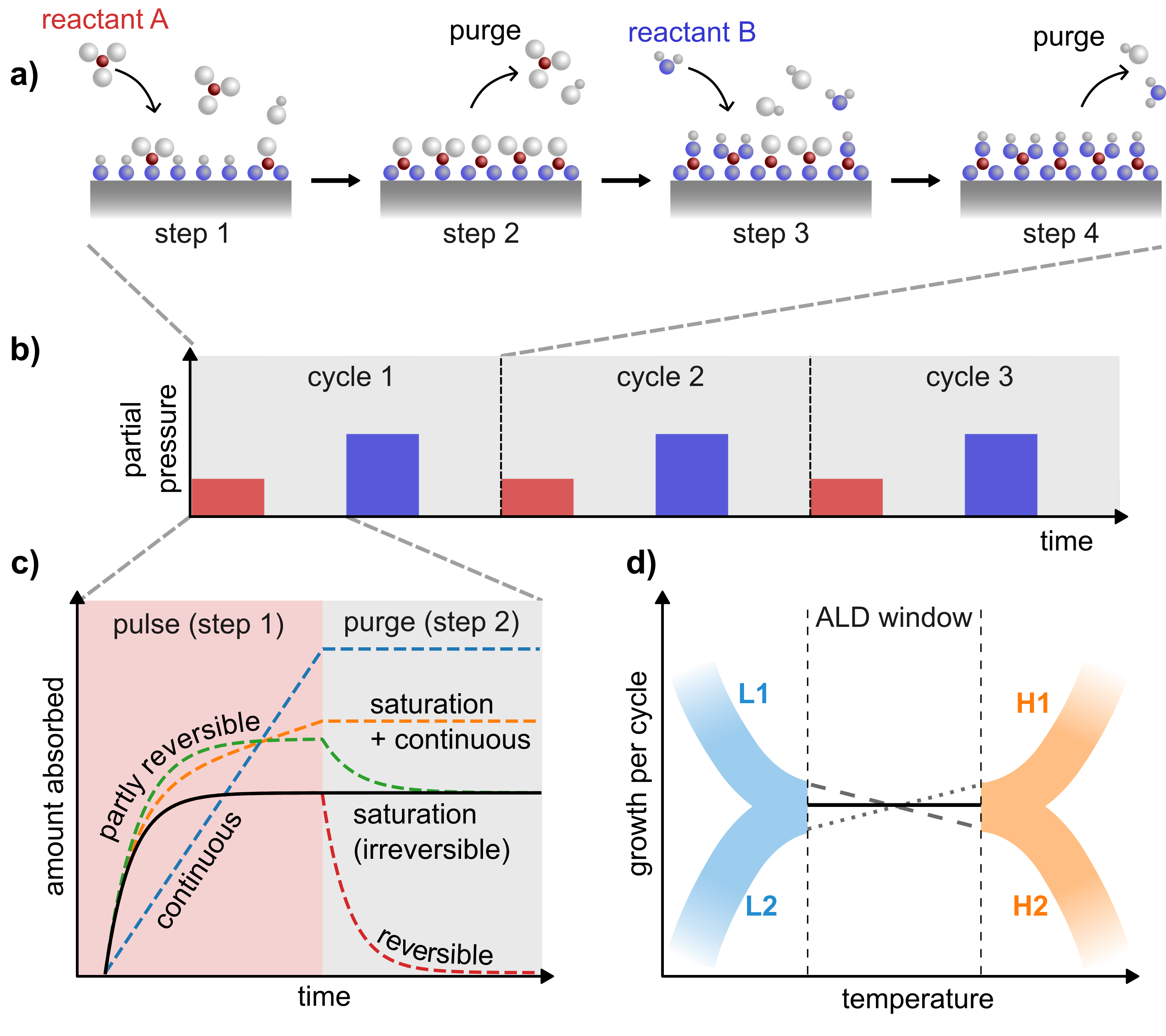}
  \caption{a) Schematic illustration of an ALD process during one cycle and b) timing diagram of the process (adapted from Ref. \citenum{GoulasALDbasics2020}), showing the partial pressures of the reactants A (red) and B (blue) in the reactor versus time. Reactant molecules A chemisorb to the substrate surface until it is saturated (step 1). A typically continuous purging gas flow removes excess reactant A and byproducts (step 2), before reactant B is introduced and reacts with chemisorbed A' (step 3), and purging again (step 4). Note, that A and B in the timing diagram (b) do not overlap, such that they do not react in the vapor phase. c) Amount of absorbed reactant A versus time (spanning over pulse and purge) for an ideal ALD process (saturation, solid black line), and deviations thereof (dashed lines). d) Schematic illustration of the ALD temperature window (adapted from Refs.\ \citenum{1989Suntola, OmmenALDwindow2020}). Within the ALD window the process meets the criteria of an ALD process, and the growth per cycle can be independent of temperature (solid black line), or increasing (dotted) or decreasing (dashed) with temperature. L1, L2, H1 and H2 indicate non-ALD growth behavior. Figure by the authors, distributed under a CC BY 4.0 license\cite{WikimediaALDprinciples}.
  }
  \label{fig:ALDscheme}
\end{figure}

The key components of any ALD process are a set of self-limiting gas-solid reactions.
The basic working principle is that pristine reactant molecules react exclusively at the substrate surface, in our case the particle surface (outer and pore surface, if present).
The most simple and most common case is an ALD process with reactants A and B.
In step 1 (Figure~\ref{fig:ALDscheme}a), the substrate is exposed to reactant A which chemically adsorbs, i.e. chemisorbs\cite{1990Calvert}, at the surface through a number of alternative reaction mechanisms\cite{2021vanOmmen}. 
The chemisorption process leaves partially reacted molecules A' bound to the surface (Figure~\ref{fig:ALDscheme}a) through the release of ligands which can form gaseous byproducts.
As the exposure time progresses (step 1, Figure~\ref{fig:ALDscheme}c), more of the possible binding sites on the substrate are occupied, or steric hindrance prevents the adsorption of further precursor molecules\cite{2021vanOmmen}. 
The chemisorption reaction is therefore self-terminating (saturating, solid line in Figure~\ref{fig:ALDscheme}c), since additionally the reactant does not react with itself (A-A or A-A' reactions do not occur).
Ideally, the chemisorption is also fully irreversible (solid line in Figure~\ref{fig:ALDscheme}c, gray shaded area).
Excess reactant and byproducts are subsequently removed by purging with an inert gas (step 2, Figure~\ref{fig:ALDscheme}a and c), leaving the chemisorbed layer of A' intact. 
This completes the first half-cycle.
In step~3 (Figure~\ref{fig:ALDscheme}a), the substrate is exposed to reactant B which reacts with A', ideally replacing all remaining precursor ligands; again, providing sufficient exposure to reactant B ensures that all possible reactions occur. 
Subsequent purging (step~4) with inert gases removes excess B and byproducts, completing the second half-cycle.

Reactants are usually transported in gaseous form by an inert carrier gas, such as nitrogen or argon.
The reactant partial pressure (colored boxes in Figure~\ref{fig:ALDscheme}b) is increased in the carrier gas for a defined amount of time, constituting a \textit{pulse}, whereas the total pressure is ideally kept constant (gray background Figure~\ref{fig:ALDscheme}b), though deviations from constant total pressure are possible\cite{2013Blomberg}. 
In many cases, the purging steps are carried out using the same uninterrupted flow of carrier gas in the absence of reactant pulses. 
Alternatively, purging can be achieved through a combination of interrupted gas flows and/or vacuum.

In ALD, one aims for saturating or near-saturating chemisorption through sufficient exposure times and partial pressures of the respective reactant, i.e.,\ sufficient exposures (product of partial pressure and exposure time, Pa$\cdot$s).
The dose of reactant (amount of substance or mass) is typically controlled by feeding it to the reactor in a pulse of defined duration.
The terms pulse duration, dosing time and exposure time are often used interchangeably; however, their meaning is not identical, as we will discuss in Section \ref{sec:dosing}.

Also, the amount of absorbed precursor may differ from the ideal saturation behavior (typical examples in Figure~\ref{fig:ALDscheme}c)\cite{2021vanOmmen}.
These non-idealities in some cases lead to a non-ALD process, e.g., continuous adsorption (like in regular CVD); in other cases, the process may in practice still be ALD-like, e.g., partly reversible adsorption.
Causes for the non-ideal adsorption behavior may be non-suitable process temperatures (see Section~\ref{sec:basictemperature}, below), or they may be related to the reactants, or insufficient purging time (step 2).

The described sequence of steps constitutes one ALD cycle (Figure~\ref{fig:ALDscheme}a), and the amount of material deposited per repetition of the sequence is called growth per cycle (GPC, see Section \ref{sec:GPC}, below).
By repeating the sequence for $n$ full cycles, the amount of deposited material can be increased. 
Note, that even though the term \textit{Atomic Layer Deposition} suggests otherwise, a single full cycle usually leads to less than a full monolayer of chemisorbed material due to steric hindrance and/or limited reactive sites \cite{2005Puurunen,2005Puurunen}.
Instead, the monolayers of chemisorbed reactants will typically lead to 10\,\% to 50\,\% of a monolayer of deposited material, or up to about five metal atoms per nm$^2$ in case of a metal or metal oxide coating (as we will show later, see Figure~\ref{fig:datadriven_thermocatalysis}b)\cite{2013Blomberg,2021vanOmmen}.

A classical example in both planar substrate and particle ALD\cite{Lakomaa1996,Uusitalo2000a,Puurunen2000a}, and the most reported combination for such A-B type reactions, is A = trimethylaluminum (TMA, \ce{(CH3)3Al}) and B = \ce{H2O}.
Here, in a simplified scene, a \ce{(CH3)3Al} reacts with an OH group at the substrate surface, forming an Al--O bond and releasing a methane molecule:
\begin{equation}
        \ce{
    Al-OH (s) + (CH3)3Al (g) -> Al-O-Al-(CH3)2 (s) + CH4 (g)
       }.
\end{equation}\label{chem:Alox1}
The remaining methyl groups react during the subsequent \ce{H2O} exposure to form \ce{CH4}, leaving behind an hydroxyl-terminated \ce{Al2O3} layer:
\begin{equation}
    \ce{
    2 Al-O-Al-(CH3)2 (s) + H2O (g) -> Al2O3-AlOH (s) + 2 CH4 (g)
       }.
\end{equation}\label{chem:Alox2}
Even though the reactions appear to be simple, the details of the chemistry are in reality more complex: multiple OH groups may react with TMA and surface metal-oxygen bonds can react, too.\cite{2005Puurunen}.

\subsection{Other ALD sequences \& molecular layer deposition}\label{sec:otherALDtypes}

For instructive purposes we limited the discussion above to the simple case of two reactants (A and B), but more complex sequences are possible, as for example A-B-C-B\cite{Cavanagh2009, Christensen2010, Gould2015b, Jin2019, Kaariainen2017, Kraytsberg2015, Lee2022b, Li2019b, Lichty2012, Liu2015a, Liu2021a, Malygin2002, Onn2017c, Qin2019, Scheffe2010, Scheffe2011, Shapira2018, Tian2023, Wiedmann2012, Yang2022a, Zhao2019a}. 
A representative of the latter are mixed oxide layers, where reactants A and C are precursors for (sources of) different metals, and B is an oxygen source \cite{Cavanagh2009,Jin2019, Kaariainen2017, Lee2022b, Li2019b, Lichty2012, Liu2015a, Malygin2002, Onn2017c, Qin2019, Scheffe2010, Scheffe2011, Shapira2018, Wiedmann2012, Yang2022a}.
If the pulse frequency for both metal precursors is not equal, i.e.\ in a supercycle (A-B)$_m$-(C-B), the stoichiometry can be controlled by adjusting $m$, and the overall amount of material is controlled by repeating the supercycle.
Note, however, that the stoichiometry (and growth per cycle)  does not necessarily scale linearly with the ratio (number) of (super-)cycles due, e.g., back-etching effects and different GPC on the substrate and on each of the film materials.

The intention behind ALD is usually to deposit inorganic materials such as metals or oxides.
Its organic counterpart \textit{molecular layer deposition} (MLD) allows growing organic layers, including polymers\cite{Myers2021,LaZara2020,Vasudevan2015,Qin2018,GilFont2020,Mahtabani2023}; the basic working principle with a set of self-limiting gas-solid reactions remains the same as in ALD (see Figure~\ref{fig:ALDscheme}), however, the details are beyond the scope of this work and are covered by previous review articles\cite{2017Bui,2014Sundberg}.
Additionally, hybrid MLD and ALD processes to produce metal-organic layers
are possible\cite{2021vanOmmen}.

On a side note, MLD is sometimes viewed as a subclass of ALD processes, and we also included it in our  literature dataset.
However, historically, MLD has been developed separately from ALD \cite{1991Yoshimura,2009George}.

\subsection{Effect of temperature}\label{sec:basictemperature}
All chemical reactions at the surface (chemisorption, A'-B reaction, see above) are temperature sensitive.
Additionally, the temperature ranges of potentially unwanted effects that lead to different growth behavior than expected from an ALD process have to be considered.

The temperature range where the process follows the basic ALD behavior outlined in Section~\ref{sec:ALDbasic} is referred to as the ALD window (Figure~\ref{fig:ALDscheme}d)\cite{2014Leskela,1989Suntola}.
If the process temperature is too low, multilayer adsorption (physisorption) of reactants can lead to multiple layers per ALD cycle. Thus, the process does not qualify as ALD anymore, and the apparent GPC is higher (indicated by L1 in Figure~\ref{fig:ALDscheme}d).
Alternatively, slow activated chemisorption of a reactant can lead to non-saturation and lower apparent GPC (L2).
At higher temperatures, thermal decomposition of reactants can occur and lead to more material being deposited and higher apparent GPC (H1), i.e., through fully continuous chemisorption or a continuous component (see also Figure~\ref{fig:ALDscheme}c).
Reversible adsorption, in contrast, can also lead to lower apparent GPC (H2) through desorption processes outside the ALD window.

Within the ALD window, the GPC is often depicted as approximately constant (solid line in Figure~\ref{fig:ALDscheme}d). 
However, even for an ideal ALD process there can be a temperature dependence of GPC (dotted, dashed line), as for example the number of surface groups (e.g., --OH) is often temperature dependent.

Note also, that the chemical reactions described above are typically strongly exothermic. 
While this remains unnoticed in wafer-based ALD, it can significantly increase the temperature of high-surface-area substrates, which underlines the need for particle agitation (see Section~\ref{sec:agitation}), and must be considered in the choice of reactor temperature (see Section~\ref{sec:reactortemperature}).

\subsection{Growth mode \& growth per cycle}\label{sec:GPC}

\begin{figure}[ht]
    \centering
    \includegraphics[width=0.6\linewidth]{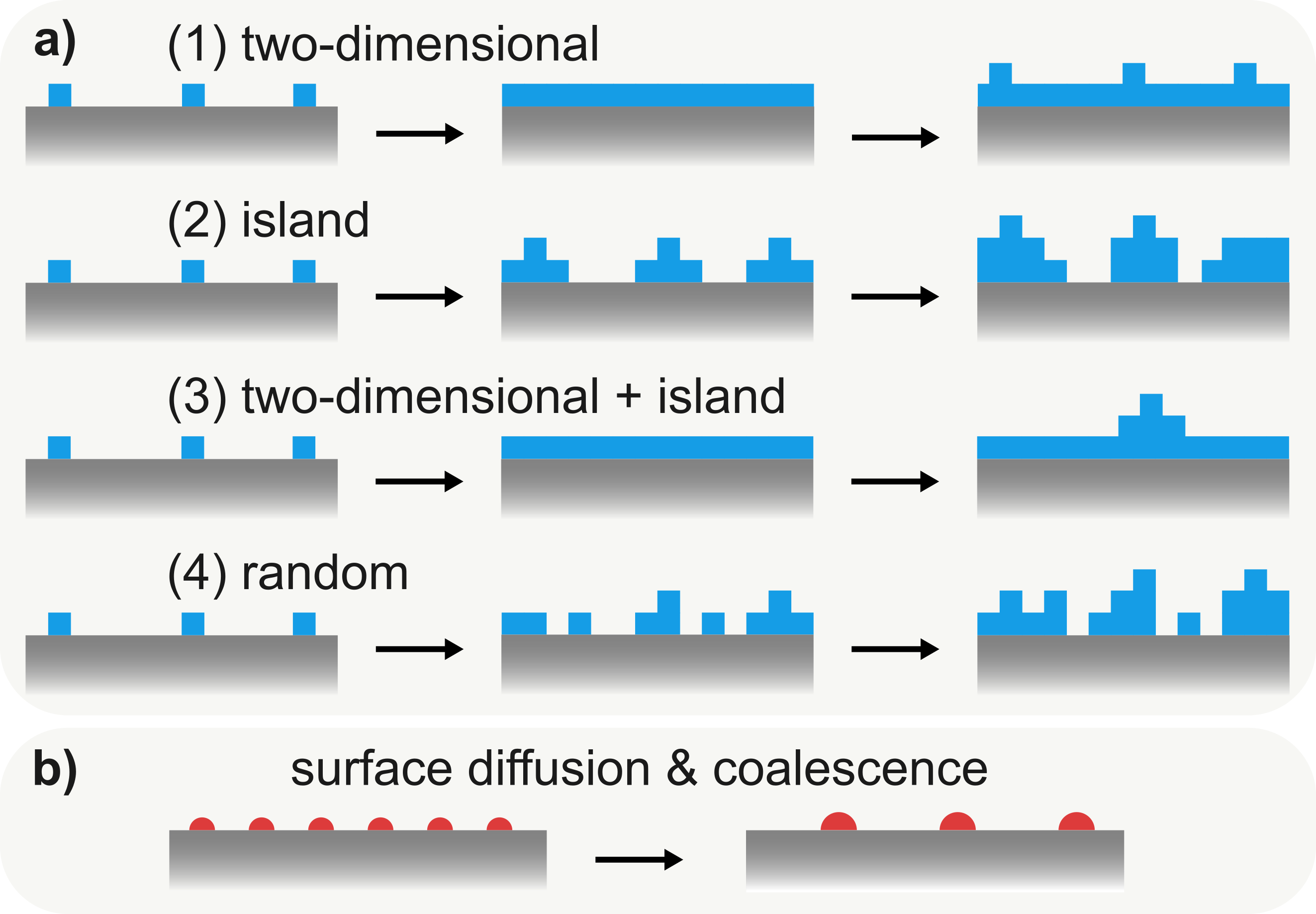}
    \caption{a) Potential growth modes in ALD in the absence of surface diffusion: two-dimensional growth, island growth, a combination of two-dimensional with island growth, and random growth. b) Effect of surface diffusion to increase particle size. Figure by the authors, distributed under a CC BY 4.0 license.\cite{WikimediaALDgrowth}}
    \label{fig:growthmodes}
\end{figure}

As described above, repeating the ALD cycle increases the amount of material deposited. 
However, the structure of the deposited material can depend on the growth mode\cite{2005Puurunen}, which will be discussed in the following alongside idealized representations thereof in Figure~\ref{fig:growthmodes}a.

For (1) two-dimensional growth (known as Frank–van der Merwe growth\cite{2010Lueth}), the deposited material forms a film, where one layer of material is filled before the next layer starts.
For (2) island growth (Volmer–Weber growth\cite{2010Lueth}), the deposited material starts growing at nucleation sites and new material is added preferably to ALD-grown material during the next ALD cycle.
The combination of two-dimensional + island growth (3) is also possible, where first a complete film grows (or multiple), but added material from the further ALD cycles tends to grow preferably at certain locations (Stranski--Krastanov growth \cite{2014Xie,2004Puurunen}).
This can occur due to a lattice mismatch between substrate and coating material\cite{2010Lueth}.
Initially, the lattice of the coating reacts to the mismatch with strain while still forming a film.
If the strain becomes too large, however, the growth mode transitions to island mode.
For random growth (4), the material is added at random positions during each ALD cycle \cite{2004Puurunen}.
This can be pictured as molecules dropped onto the substrates at random positions, with the exclusion that no two molecules can be dropped onto the same position within the same ALD cycle.

The growth mode depends on a number of different factors.
If only a limited density of binding sites is available on the substrate surface, the material will start growing in islands, as is often the case on, e.g., graphene surfaces.
For a high density of binding sites, as is typical for hydroxyl-terminated oxide surfaces (see Section \ref{sec:pretreatment}), two-dimensional growth can occur.
The material and condition of the substrate is therefore an important factor in determining the growth mode.
However, if chemisorption in the next ALD cycle preferentially occurs on the ALD grown material, island growth instead of two-dimensional growth can still occur\cite{2020Richey}.
Additionally, the growth mode can be affected by a number of other factors such as conditions during the ALD process itself, e.g., due to the deposition of carbonaceous species from ligands, or mobility of chemical species on the substrate.

Besides the growth mode, the structure of the coating can also be affected by surface diffusion and coalescence of deposited particles (Figure~\ref{fig:growthmodes}b)\cite{Grillo2018a}, either during the ALD process or in a subsequent process.
In both cases, thermal energy is required which increases the mobility of particles on the substrate surface.
Once mobile, the particles can coalesce and even sinter together to form larger particles. 

Regardless of the structure for the material, the amount deposited per cycle can be captured by the concept of growth per cycle (GPC), which is the equivalent to a growth rate in continuous gas-phase deposition methods.
For planar substrate ALD, where often two-dimensional growth is desired, the GPC is usually reported in terms of layer thickness increase per cycle (e.g. \AA\ or nm). 
For ALD on particles, the way the GPC is reported varies.
Often, weight loading of the coating material on the powder is used, or the number of atoms or molecules per unit area of substrate (i.e., areal number density). 
Sometimes the number of reference atomic layers is given, or added thickness per cycle, or amount of absorbed coating material in moles per weight of substrate.

\subsection{Agitation, mixing \& aggregation of particles}\label{sec:agitation}

When applying ALD to particulate materials, exposing the particles' surface to the reactants homogeneously is a challenge.
Here, we can make use of findings from a range of sectors outside the ALD field where particles or powders play an important role.
In fact, particle technology is a research field on its own, with conferences, journals, and textbooks. A good introductory text to this area is the book by Rhodes et al.\cite{2024Rhodes}.

When applying ALD to particulate materials, there are three important reasons to achieve effective mixing of the particles and/or the gas during the process. First of all, we want to give every particle the “same experience” in order to obtain a homogeneous product (i.e., a similar coating on each particle). 
In principle, the self-limiting nature of ALD makes it robust against differences. However, in reality, the ALD behavior may show deviations from ideality; mixing particles and gas, giving more homogeneous conditions throughout the reactor, will lead to a more homogeneous product. Insufficient mixing of the particles can lead to, e.g., variations in loading. \cite{Lindblad1994,vanOmmen2015}
Second, effective mixing will aid in using the reactants efficiently. Since particle ALD typically entails much larger surface areas than wafer coating---and thus much larger reactant consumption---it is crucial to have  precursor utilization efficiencies close to 100\% \cite{2015Grillo} rather than losing a major share of the precursor, as is typical for wafer reactors\cite{2023Weber}. Otherwise, the process will have a much larger environmental impact and will be harder to make cost-effective.
Third, mixing is important to distribute and dissipate heat released during the ALD reaction. ALD reactions are typically strongly exothermic, and the large reactant consumption in particle ALD means that extensive heat can be released\cite{Greenberg2020}. 
It is crucial to distribute this heat evenly over the reactor to prevent local hot spots, which could mean that parts of the reactor contents are outside the ALD temperature window leading to unwanted effects (see Section \ref{sec:basictemperature}), such as CVD.\cite{Hashemi2020}

The size and density of particles have an important influence on how easily mixing can be achieved. 
While the size can easily vary over orders of magnitude, the density can also vary significantly, since pores in the particles must be considered. 
Geldart proposed a regime map indicating flow behavior based on size and density\cite{1973Geldart,2024Rhodes}. While this map is primarily aimed at fluidization (suspending particles in an upward gas flow), it also gives an indication of how easily particles mix in general. For example, particles with a density around 3,000\,kg/m$^3$ and size below 30\,µm are strongly cohesive and thus are harder to mix; hence, proper agitation is crucial. 

Particle agglomeration and aggregation have a strong impact on the homogeneity of the coating. 
Cohesive particles can form large, non-moving lumps of material as well as smaller agglomerates. Very small particles---around 1\,µm or smaller---typically form very open agglomerates: more than 95\,vol\% open space between the particles is not uncommon in these structures\cite{2012vanOmmen}. In this case, the ALD reactants can quite easily diffuse into the agglomerate, and most of the surface can be coated\cite{2015Grillo}. 
Whether the touching points of neighboring particles are coated depends on the nature of the cohesive forces: if they are agglomerated with the particles held together by Van der Waals forces, the agglomerate's restructuring could occur during the processes, enabling coating of the full surface. If the applied (nano)particles are produced in a high-temperature process (like many ceramic oxide nanoparticles and carbon black), they will be sintered into aggregates, and the connection points will be permanent. \cite{2012vanOmmen} Consequently, not the full particle surface can be coated (i.e., not the touching points); to what degree this is problematic depends on the final application.

Mixing through agitation of the particles affects the transport of reactants and byproducts both directly (convection) and indirectly (diffusion).
Convective flow is the flow driven by pressure difference, often the flow between the particles, especially a higher pressure; diffusive transport is the transport driven by concentration gradients, typically inside a particle. Note, that for a non-agitated, packed bed of particles operated at vacuum, the dominant transport mechanism is diffusion rather than convection for the particle-to-particle length scale (as shown in Figure~\ref{fig:pressuretransport}).
Particle movement will lead to drag on gas, thereby directly influencing convective mass transport on a length scale of the reactor size down to the particle-to-particle length scale. An indirect effect of mixing on mass transport is that it helps to create a constant bulk concentration of reactants at smaller length scales, i.e., a similar concentration around the particles and/or aggregates, and thus at the entrance of the pores inside or between particles, respectively. 

On a side note, the powder nature of the material must be considered when assigning laboratory space to the respective experiments.
Powder handling equipment, safety considerations (see Section~\ref{sec:safety}), and possible influence of other experiments should be taken into account (e.g., a powder ALD will not be welcome in the clean room).

\section{Equipment}\label{ch:equipment}

The specific requirements related to ALD processing of particulate materials (see previous Section~\ref{sec:agitation}) has led to the development of specialized coating equipment.
In the following section, a historical overview of reactor development is introduced, followed by a brief discussion of the reactor operating principles for the most common batch reactor types.
In order to address the requirement of some of the key ALD applications for the processing of relatively large particle quantities, the potential for industrialization and upscaling is discussed, where continuous (spatial) ALD reactors are presented.
The section on equipment ends with a brief subsection describing online characterization tools available for monitoring ALD processes on particles. 

\subsection{Historical overview of reactor development}

\begin{figure}
\includegraphics[width=0.5\textwidth]
{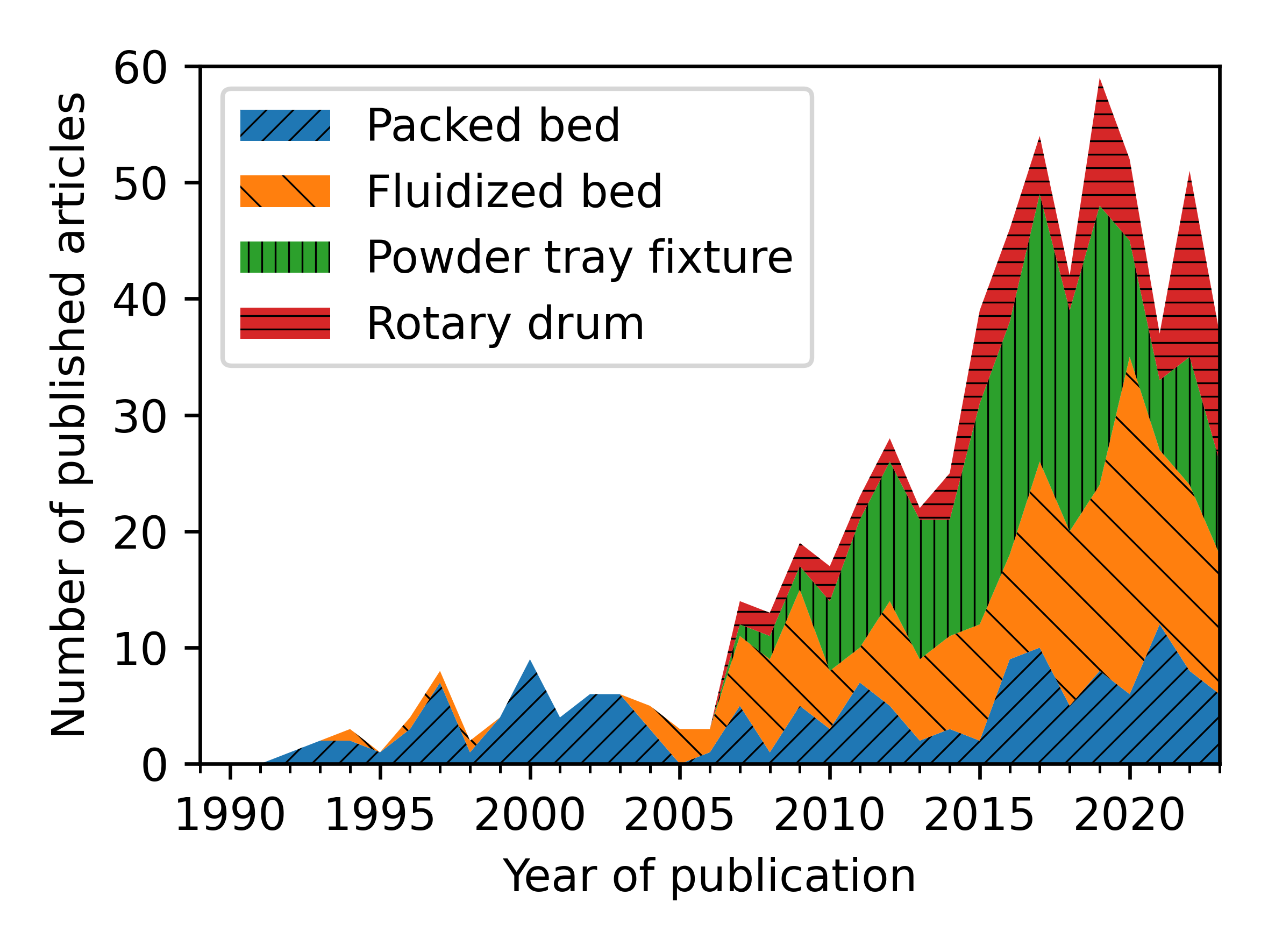}
\caption{Articles on ALD of particulate materials per year, grouped by reactor operating principle (packed bed, fluidized bed, powder tray fixture, rotary drum).}
\label{fig:reactors_years}
\end{figure} 

\begin{figure}
\includegraphics[width=\textwidth]{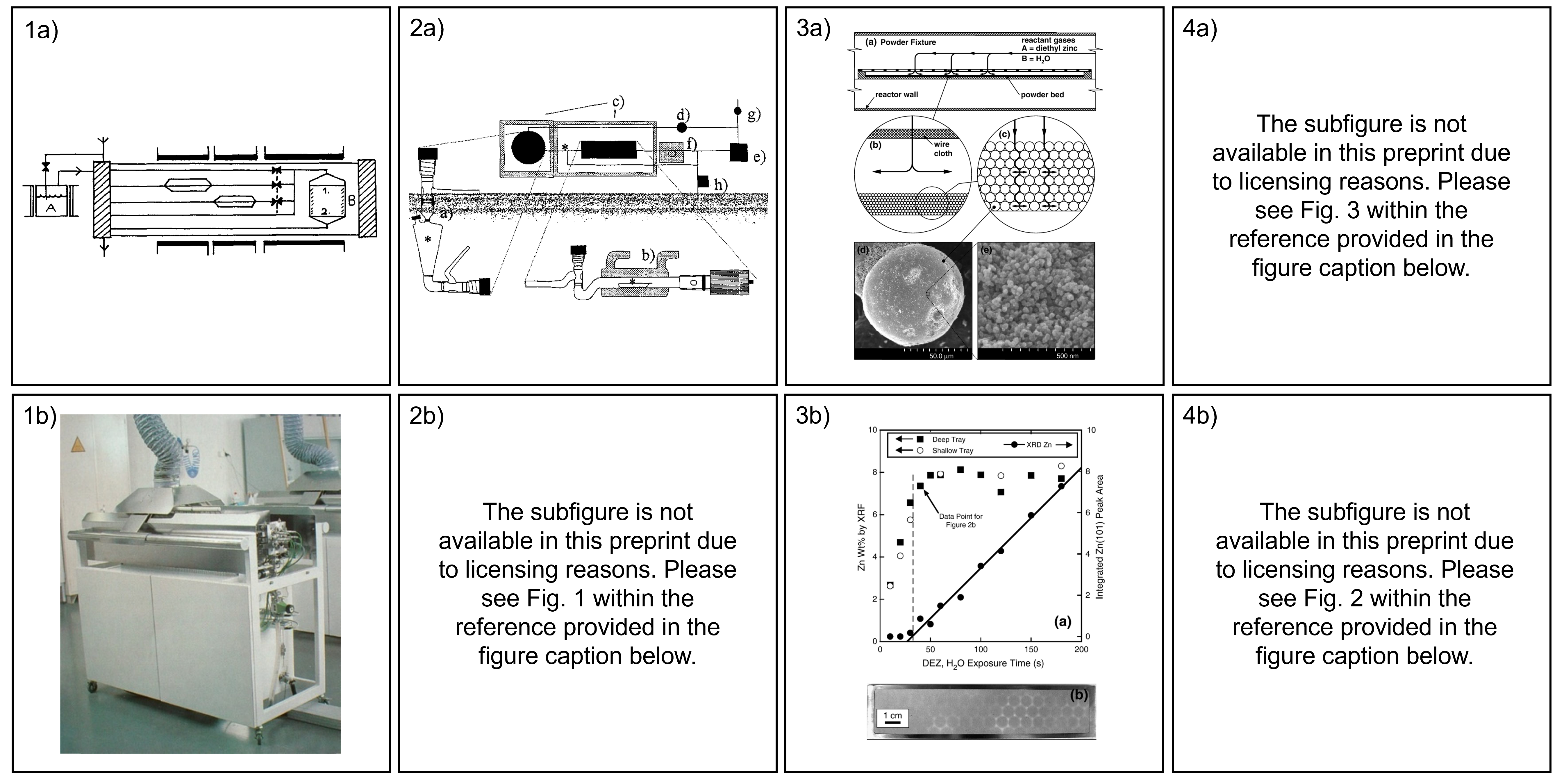}
\caption{Reactor schematics and indicative photographs of the four main reactor types used for research activities: packed bed (1a \& 1b) \cite{Haukka1993b, 2014Puurunen}, fluidized bed (2a \& 2b) \cite{Suvanto1997, Rauwel2011}, powder tray fixture (3a \& 3b) \cite {Libera2008} and rotary drum (4a \& 4b). \cite{McCormick2007a, Coile2020} Images reproduced with permission. \cite{Haukka1993b, Suvanto1997, McCormick2007a, Libera2008, Rauwel2011, 2014Puurunen,  Coile2020}}
\label{fig:reactors_schemes+photos}
\end{figure}

When examining the development of different types of ALD reactors for the processing of particulate materials over time (Figure~\ref{fig:reactors_years}), certain phases can be defined.
A first phase during the 1990s was marked by research activities in Finland particularly involving the use of the so-called flow-type reactor;\cite{Lakomaa1992} this type of reactor accounts for about 18\% of all reports.
By far, the most commonly reported reactor---still widely used today---is the F-120 model of Microchemistry (later ASM Microchemistry).
Approximately $5\%$ of all reported studies mention the use of an F-120 reactor (Figure~\ref{fig:reactors_schemes+photos}-1b) or one of its variants, such as the modified MC 120 reactor (Figure~\ref{fig:reactors_schemes+photos}-1a).
Operated as a packed bed (fixed bed), this reactor was characterized by the flow of the gaseous reactants through the powder bed (flow-through reactor).

A modification of packed bed reactors operated in fluidized bed mode \cite{Hirva1994} was developed in collaboration of academic institutions with Microchemistry, to ensure homogeneous mixing of the processed particles and prevent concentration gradients in the coated material.
The use of fluidized bed reactors for ALD expanded significantly in the 2000s, marking a second phase for the application of ALD on particulate materials.
As a result, the fluidized bed reactor became the most frequently reported reactor type, accounting for approximately $25\%$ of the total reported studies.
The reactor concepts developed by Microchemistry can be considered as modular extensions of their conventional flow-type reactors, initially built using simple reactor glassware (Figure~\ref{fig:reactors_schemes+photos}-2a) and later developing to customized designs such as powder cells (Figure~\ref{fig:reactors_schemes+photos}-2b).

A reactor implementing viscous flow of reactants over a powder tray fixture that contained a fixed bed of powder (Figure~\ref{fig:reactors_schemes+photos}-3a \& -3b) was introduced by scientists at Argonne National Laboratories\cite{Elam2007}, based on a custom reactor design.\cite{Elam2002} 
However, the powder tray fixture approach is also frequently implemented in commercial planar substrate ALD tools with minimal adjustments to the reactor; the most reported types in our dataset are the Veeco S-series (design by Ultratech/Cambridge Nanotech), the GEMStar-series of Arradiance and the R-series of Picosun.
Starting from about 2010, the powder tray fixture approach has gained popularity, with a total share of 24\% of all reports.
Presumably, the higher availability of planar substrate ALD tools in academic laboratories due to the increased interest in ALD in general (see Supporting Information Figure~S2) has made this reactor configuration readily accessible.

However, the challenge of effectively agitating particles during deposition continued to resurface, effectively leading to the introduction of the rotary drum (Figure~\ref{fig:reactors_schemes+photos}-4a \& -4b) reactor design.\cite{McCormick2007a}
Over the past decade, the rotary drum design has gained some momentum, as evidenced by the increasing number of studies reporting its use, now accounting for about 11\% of all reports.
Therefore, a third phase of research activities could be defined, characterized by the availability of all four major ALD reactor types, in some cases allowing researchers to select the most suitable reactor design based on the application requirements.
Notably, approximately $15\%$ of the analyzed articles in this review do not clearly specify the type of reactor used.

\subsection{Reactor operating principle}\label{sec:reactorprinciple}

\begin{figure}
\includegraphics[width=\textwidth]{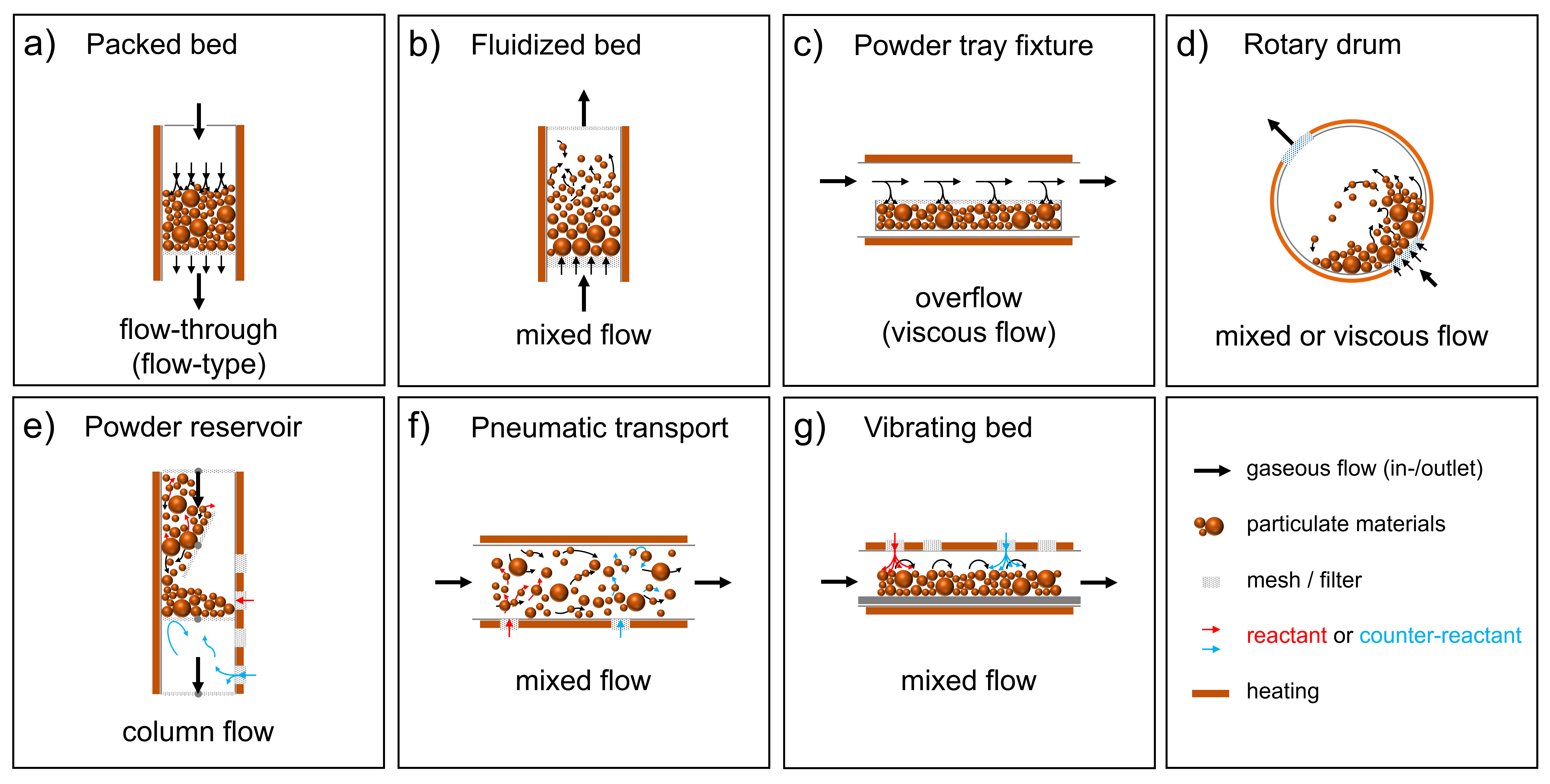}
\caption{Operation principle of the four main reactor types (top row): a) packed bed, b) fluidized bed, c) powder tray fixture, and d) rotary drum (d). Three spatial ALD concepts (bottom row): e) powder reservoir, f) pneumatic transport, and g) vibrating bed; all implemented for manufacturing scale-up. Figure by the authors, distributed under a CC BY 4.0 license.\cite{WikimediaALDreactors}}
\label{fig:reactors_schematics}
\end{figure}

The four batch reactor types shown in Figure~\ref{fig:reactors_schemes+photos} (top row) are all considered typical representatives for the equipment used in ALD on particles.
From the body of collected articles on ALD on particles, more than 160 schematics of reactors and coating setups were identified and screened for this review.
The vast majority of the reactors follow the same operating principles as the four types shown above, even if the exact implementation can differ from case to case.
Therefore, the working principle (schematically in Figure~\ref{fig:reactors_schematics}a-d), and variations of the four types are discussed here in more detail.

\subsubsection{Packed bed}

In the flow-through reactors (or flow-type, Figure~\ref{fig:reactors_schematics}a), the reactants are pumped or pulsed through a stationary bed of powder material.
Typically, no particle agitation takes place (packed-bed or fixed-bed mode of operation).
A stationary bed of the solid particulate material is being supported on a sintered metal filter plate, and the pulses of the vaporized reactants are directed through the bed for a sufficient time duration.\cite{Haukka1993b}
Early flow-through reactor designs accommodated the use of gas, liquid, or solid precursors.\cite{Lindblad1994}
Different heating zones were implemented for the reactants and the reaction chamber, while the possibility to operate at ambient pressure was also established.\cite{Haukka1995a}
Macroscopic deposition homogeneity (interparticle distribution) was verified by sampling different sections of the bed.\cite{Haukka1995a}
Cross-sectional SEM with elemental analysis was used to assess intraparticle distribution of the deposited species.\cite{Haukka1993a, Lindblad1994}
To improve gas--solid contact, a powder cell operated in (semi)fluidized bed mode was also introduced.\cite{Rauwel2011, Rauwel2012b}
In most cases, ALD requires the use of an inert carrier gas.
A flow-through reactor operated with dry air as the carrier gas was reported by researchers from Russia.\cite{Ermakova2002}
Additionally, a modification of the flow-through operation, with the gas flow operated in a discontinuous way inside a powder cartridge to agitate the particles, has also been described.\cite{Zhang2015c}
Moreover, a parallel reactor design for screening ALD reactor process conditions has also been described.\cite{Knemeyer2021a}
Based on the analyzed articles, flow-type reactors have been used for the coating of batches up to 100\,g.\cite{Nevalainen2009}

\subsubsection{Fluidized bed}

Fluidized bed reactors (Figure~\ref{fig:reactors_schematics}b) ensure particle agitation by means of introduction of a mixed flow of the carrier gas through the bottom of the particle bed.
The mixing characteristics of the fluidized bed enable favorable mass and heat transfer.
Particle elutriation can occur for high fluidization velocities, and in addition, the fluidization behavior of the particles can be influenced by the occurrence of certain species present in the fluidization gas (e.g., \ce{H2O} vapor).
The implementation of a fluidized bed for ALD was already reported by researchers in the former Soviet Union.\cite{VPHAYakovlev1979}
The use of a fluidized bed was also probed in Finland \cite{Hirva1994}, where the implementation of carrier gas circulation was evaluated.\cite{Suvanto1997}
A main contribution in the development of fluidized bed reactor technology came from researchers in the University of Colorado.
Several fluidization enhancing possibilities through particle agitation, including vibration\cite{Wank2004b}, stirring\cite{Hakim2007a} and gas pulsation\cite{King2007} have been described. 

Other contributions in the design and operation of the fluidized bed reactor include the use of a microjet for agitation\cite{Vasudevan2015}, reactor inclination for reduced particle elutriation \cite{Moghtaderi2006}, ultrasonic vibration assistance \cite{Li2022a} and the use of glass bead fillings to enhance the uniformity of the gas mass transfer. \cite{Lee2019c} Plasma-assisted ALD processing in a fluidized bed was also recently reported. \cite{Wang2022d, Tian2023} Up to 500\,g of particulate materials have been processed in a fluidized bed reactor \cite{OToole2019}. Moreover, it has been reported (without specifying further details) that scale-up in a fluidized bed at the Finnish company Neste involved the coating of 17\,kg of a silica-modified alumina catalyst. \cite{Parsons2013,2020Parsons}

\subsubsection{Powder tray fixture}\label{sec:powdertray}

An overflow ALD reactor (or viscous flow, Figure~\ref{fig:reactors_schematics}c) developed in the University of Colorado \cite{Elam2002} was modified to be equipped with a specifically designed powder tray fixture.\cite{Libera2008}
The top of the fixture contains a wire cloth that allows reactant access, while preventing disturbances of the powder layer by the introduced gas currents.
Thus, little to no powder agitation takes place.
Very thin powder layers, bounded by the depth of the tray (3.2\,mm), ensured rapid diffusion of the reactants.\cite{Libera2008}
The severe limitations of this setup with respect to scale-up have been clearly described in literature.\cite{Longrie2014a}
Static exposure of the reactants can be realized in these systems \cite{Onn2018} and a design of a layered configuration of several powder trays that can be rotated has been described.\cite{Sun2023}
Operation of overflow reactors in remote plasma-assisted mode has also been reported.\cite{Cao2018}
The largest batch processed in a powder tray fixture reactor is 10\,g\cite{Hood2023} although typically, the coating of less than 1\,g of particulate materials is being reported.

\subsubsection{Rotary drum}

The need to agitate the particulate materials during coating has also been addressed with the design of a rotary drum reactor (Figure~\ref{fig:reactors_schematics}d) proposed by researchers from University of Colorado.
\cite{McCormick2007a} In this way, the possibility to increase efficiency of reactant utilization, by implementing static reactant exposures, was combined with agitation of the powder bed.
The possibility of performing plasma-enhanced ALD in a rotary drum reactor has also been described.\cite{Longrie2012}
Using a double-layered cartridge for the containment of the particulate materials resulted in the introduction of a, so-called, fluidized coupled rotary reactor (or rotary fluidized bed).\cite{Duan2015}
In order to prevent agglomeration of the coated powders, powder stirring balls that could effectively de-agglomerate the powders without damaging them due to attrition have been used.\cite{Yoon2022b}
In addition, a dual-zone furnace, with the possibility of providing different temperatures at the two steps of the ALD reactions has been coupled with a rotary drum ALD reactor.\cite{Lee2022b}
Machine learning libraries (such as eXtreme Gradient Boosting, XGBoost) have been implemented for predicting the coating outcome response to changes in processing parameters of a rotary drum reactor.\cite{Yoon2023a}
  
\subsection{Industrialization \& upscaling}

Commercial equipment has been widely used for research activities on a laboratory scale, either by re-purposing planar substrate tools (as outlined in the historical overview above), or using dedicated ALD systems for particulate materials.
Examples of the latter are specialized versions of the F-120 reactor, more recently the Veeco S200 with powder option, tools by Forge Nano, or the Atomic-Shell model of CN1.
Most of the reported research activities involve coating of a small amount of particulate materials (e.g.\ $<$10\,g, see Supporting Information Figure~S3). 

However, the literature dataset also holds reports of the coating of more than 100\,g in a single batch (Table~\ref{tab:reactors}).\cite{OToole2019, Beetstra2009, Valdesueiro2017, Li2022a, Zhou2016, McNeary2021, Settle2019}
The ability to process very large amounts of total particulate surface area---as an extension of the batch mass descriptor---is an equally important factor to consider when studying industrialization activities.
A particle batch with a total surface area of up to $\sim$81,000\,m$^2$ (300\,g with 270\,m$^2$/g) was reportedly coated in Finland.\cite{Haukka1995a}
Another notable example involved the coating of 10\,g of activated carbon (1,275\,m$^2$/g) with TTIP-\ce{H2O2} for a total processable area of $\sim$13,000\,m$^2$\cite{Li2021}, and of 10\,g of high surface area \ce{SiO2} powder (506\,m$^2$) with TMA-\ce{H2O} for a total area of $\sim$5,000\,m$^2$.\cite{Strempel2018}

\begin{figure}
  \includegraphics[width=\textwidth]{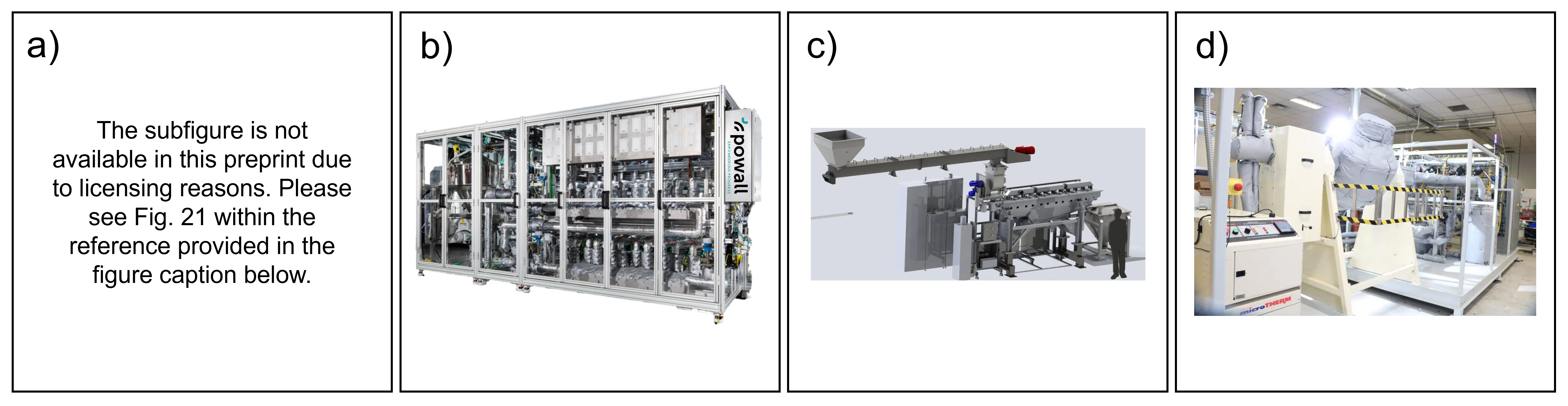}
  \caption{ALD equipment used for manufacturing at scale: a) powder reservoir (column flow unit, semi-continuous)\cite{Malygin2015}, b) pneumatic transport (continuous)\cite{2025PersonalPowall}, c) vibrating bed (continuous),\cite{Weimer2023} and d) rotary drum (batch).\cite{2024ForgeNano} Images obtained online or reproduced with permission \cite{Malygin2015, Weimer2023, 2024ForgeNano}.}
  \label{fig:reactors_photos}.
\end{figure} 

\begin{table}
\footnotesize
\begin{tabular}{lllllll}
Powder &
\makecell[l]{Batch\\{[}g{]}} &
\makecell[l]{SSA\\{[}m$^2$/g{]}} &
\makecell[l]{Operation\\principle}&
\makecell[l]{Reactants} &
\makecell[l]{Coating\\cycles} &
Reference
\\
\midrule
YSZ &
500 &
12.69 &
fluidized bed &
TMA/\ce{H2O} &
9 &
\citenum{OToole2019}
\\
\ce{SiO2} &
300 &
270 &
packed bed &
\ce{ZrCl4} &
6 &
\citenum{Haukka1995a}
\\
\ce{TiO2} &
200 &
- &
rotary drum &
\ce{SiCl4}/\ce{H2O} &
1 &\citenum{Yang2022b}
\\
\ce{Fe3O4} &
130 &
11.08 &
rotary drum &
TMA/\ce{H2O} &
50 &
\citenum{Duan2016b}
\\
\ce{LiMn2O4} &
120 &
1.9 &
fluidized bed &
TMA/\ce{H2O} &
28 &
\citenum{Beetstra2009}
\\
polyester resin & 
110 &
- &
fluidized bed &
TMA/\ce{H2O} &
9 &
\citenum{Valdesueiro2017}
\\
LiNi$_{0.8}$Co$_{0.1}$Mn$_{0.1}$O$_2$ &
100 &
- &
fluidized bed &
TMA/\ce{H2O} &
50 &
\citenum{Li2022a}
\\
\ce{ZrO2} &
100 &
15.2 &
rotary drum &
TMA/\ce{H2O} &
50 &
\citenum{McCormick2007a}
\\
polyamide &
100 &
- &
packed bed &
\ce{TiCl4}/\ce{H2O} &
50 &
\citenum{Nevalainen2009}
\\
YAG:Ce phosphor &
100 &
- &
fluidized bed &
TMA/\ce{H2O} &
5 &
\citenum{Zhou2016}
\\
ezetimibe/HPMCAS &
100 &
1.27 &
rotary drum &
TMA/\ce{H2O} &
- &
\citenum{Duong2022}
\\
Pd/\ce{Al2O3} &
100 &
115 &
fluidized bed &
TMA/\ce{H2O} &
10 &
\citenum{McNeary2021}
\\
Pd/\ce{TiO2} &
100 &
140 &
fluidized bed &
TMA/\ce{H2O} &
- &
\citenum{Settle2019}     
\end{tabular}
\caption{Reported cases of coating $>100$\,g batches of particulate materials by ALD.}
\label{tab:reactors}
\end{table}

As shown in Figure~\ref{fig:reactors_photos}, commercial up-scaling efforts aiming to address increasing manufacturing needs are implemented in either semi-continuous or fully continuous production schemes (Figure~\ref{fig:reactors_schematics}e, f, d). 
Already in the 1980s, researchers in the former Soviet Union implemented a multisectional reactor that was operated in a fluidized bed mode, with the powder being fed on the unit from its top and coated in successive sections of the column flow unit (Figure~\ref{fig:reactors_photos}a)\cite{Malygin2015, Sosnov2021}.
A vanadium oxide coating was deposited on silica gel, and the coated product (known as IVS-1) was used on an industrially-relevant scale as a high-sensitivity gas phase humidity indicator. 

This column flow unit (Figure~\ref{fig:reactors_photos}a and schematically in Figure~\ref{fig:reactors_schematics}e) constitutes an early example of a spatial ALD system.
In spatial ALD operation, the reactants are introduced in distinct sections of the equipment that can be separated by purge (inert gas) zones and/or purging ports.
Instead of alternating the reactants by alternating the gas flows in the temporal domain, the particles move through different reactant zones in the spatial domain. 

Other implementations of the spatial ALD principle followed.
A so-called a powder reservoir reactor\cite{ONeill2015a}, a high-throughput semi-continuous system, has been used by Forge Nano (formerly PneumatiCoat Technologies) for the coating of industrially-relevant amounts (450\,kg) of Li-ion battery cathode active materials.\cite{Mohanty2016}
Another continuous spatial ALD approach, relying on the pneumatic transport of particulate materials through distinct reaction sectors (schematically in Figure~\ref{fig:reactors_schematics}f) of a coiled tube was introduced by researchers at Delft University of Technology.\cite{vanOmmen2015}
Using the pneumatic transport approach, the production rate reported for the model photocatalyst produced (1\,g/min) has been increased by orders of magnitude in the commercialization efforts of Powall (formerly Delft IMP, Figure~\ref{fig:reactors_photos}b).\cite{2025PersonalPowall}
One additional continuous production scheme based on spatial ALD is pursued in the continuous vibrating reactor (CVR) of ForgeNano (Figure~\ref{fig:reactors_schematics}g and Figure~\ref{fig:reactors_photos}c).\cite{Hartig2021, Weimer2023} 

In certain cases, batch ALD systems relying on conventional, temporal ALD, operation have also been scaled.
That includes the early research in Finland on the atmospheric pressure packed bed reactors\cite{Haukka1995a}, and the fluidized bed reactors used for coating batches of up to 17 kg of catalysts.\cite{Parsons2013,2020Parsons}
Additionally, a double-cone rotary drum ALD reactor capable of processing up to 1 kg of pharmaceutical powders was also introduced recently by researchers from Applied Materials \cite{Swaminathan2023} and, in addition, by Forge Nano (Figure~\ref{fig:reactors_photos}d). 

Although several studies implemented different ALD reactor types\cite{Zhan2008, Tiznado2014, Longrie2014b, Mohanty2016, Settle2019, Zhang2019c, Bao2020, Lin2023, Swaminathan2023}, it is mostly the equipment availability, the required material scale, or other process implications that define the choice of the processing system.
Only a few cases have studied the influence of processing characteristics of the ALD reactors on the coating characteristics.
In one instance, static ALD was compared to dynamic, fluidized bed ALD\cite{Kim2018b} revealing significant differences in the growth mode of the deposited material.
The \ce{SnO2} coating layer was discontinuous for the case of static ALD in a flow-type reactor, while the dynamic ALD in a fluidized bed reactor resulted in a continuous film.\cite{Kim2018b}
The influence of agitation on coating uniformity was also studied in a packed bed (flow-type) reactor that could be operated with or without agitation.\cite{Dannehl2018}
Operating without agitation resulted in non-uniform surface coverage.\cite{Dannehl2018}
In addition, fluidized bed and rotary drum reactors have been compared \cite{Jung2022} revealing no differences in the coating characteristics in terms of thickness.
A shift towards larger sizes on the distribution of the particles coated in the rotary drum reactor was reported.

The possibility to handle large amounts of particulate materials, with total surface areas of many thousands of m$^2$ by ALD has already been proven.
Still, scalable approaches are expected to be limited mostly to ALD chemistries that demonstrate adequate precursor vaporization and stability (absence of thermal decomposition) characteristics, such as the case of metal-alkyl or halide precursors.
Packed bed reactors might have so far spearheaded small proof of concept research.
However, in order to ensure scalable technology  approaches, solutions that rely on agitation are expected to gain more momentum when aiming for industrially applicable coating schemes.
A better understanding of the influence of powder motion (e.g. agitation, fluidization) to the coating characteristics on the particle level still has to be established further to enable harnessing the potential of ALD to its full extent.

\subsection{Online characterization \& process monitoring}

By far the most common technique implemented for monitoring the ALD process is residual gas analysis  by means of quadrupole mass spectrometry. The technique has been routinely used for tracking ALD reaction completion in all four main reactor types, namely fluidized bed \cite{Kim2007, King2008a, King2008b, King2008c, King2008d, King2009a, King2009b, King2009c, King2009d, Scheffe2009, Li2010, Liang2010b, Kilbury2012, Lubers2017, Hoskins2018, Bull2021, Azizpour2017}, overflow \cite{CamachoBunquin2015, CamachoBunquin2017}, rotary drum \cite{Duan2015, Duan2016a, Duan2016b, Li2021, Longrie2012, Longrie2014a, Rampelberg2014, Coile2020, Duong2022, Swaminathan2023, Chazot2022} and packed bed reactors. \cite{Strempel2017, Laskar2017}

Packed bed reactors can also rely on gravimetric \cite{Strempel2017, Sosnov2017, Bodalyov2019, Ingale2020} and IR spectrometry based \cite{Sosnov2010, Najafabadi2016, Daresibi2022, Snyder2007} approaches for monitoring either the mass gain or the surface changes during ALD processing. The surface changes have also been monitored by means of Raman spectroscopy. \cite{Mittal2022}

\section{Particulate supports}\label{ch:supports}
If material is deposited onto particles instead of planar surfaces, the term \textit{support} instead of \textit{substrate} is commonly used. The term was coined during the early years of the ALD on particles field, when the particles were thought of as catalyst supports (see Section \ref{ch:thermocatalysis} on thermocatalysis). 
Besides being more accurate (referring to the etymological origin of the term \textit{substrate}), we here use \textit{support} to emphasize the marked differences of these particulate materials compared to planar surfaces.

These differences call for a dedicated discussion of the particulate support characteristics.
Besides obvious properties such as chemical composition and particle size, processing and application-relevant characteristics are due to the often porous nature of particles, leading to high specific surface areas (SSA).
Additionally, the pore diameter affects the mass transport and therefore the reactant adsorption behavior, which needs to be considered when choosing processing conditions. 

\begin{figure}
  \includegraphics[width=70mm]{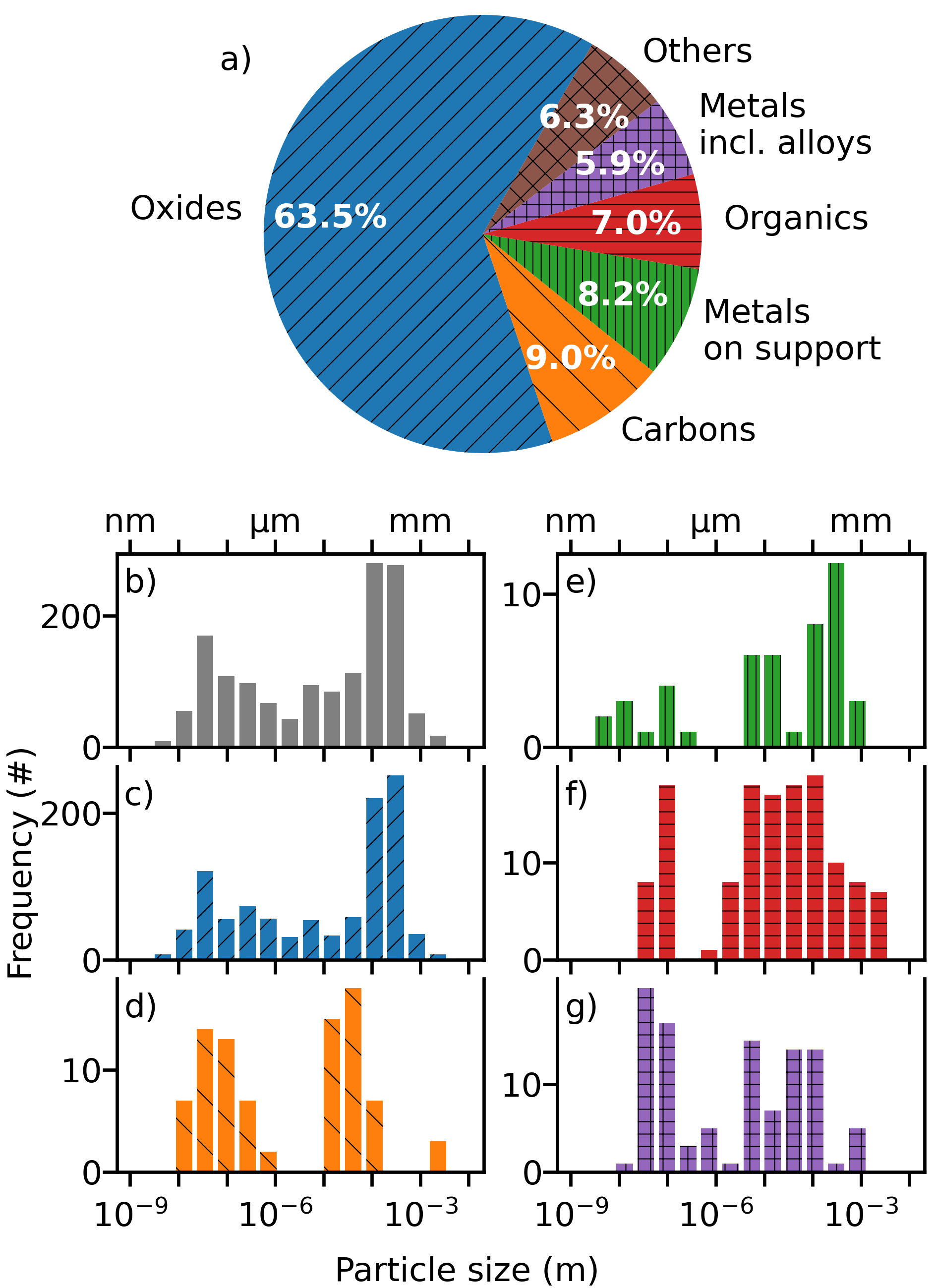}
  \caption{a) Overview of reported support materials: Oxides (metal, metalloid), carbons (carbon black, graphene, etc.), metals on support (e.g., a catalyst metal on a support), organics, metals incl.\ alloys and other types of materials. b) Histogram of all reported particle sizes. If a range was given, we counted the mean value. c)-g) Breakdown of particle size according to material (hatching and colors as in a).}
  \label{fig:supports}
\end{figure}

\subsection{Chemical composition}\label{sec:supportschem}

To provide an overview of support chemistries, we have grouped the reported materials into six categories (Figure~\ref{fig:supports}a).
Due to the relevance for ALD processing, we mostly discuss the surface chemistries in this particular section; 
however, we note that the chemistry of the particle core strongly affect the bulk properties of the powder (e.g., density, heat capacity, conductivity), with implications for some aspects of powder handling and applications.

The category \textit{oxides} is typical for both ALD on planar substrates and on particles, and accounts for $63.5\%$ reported values.
Within this category, \ce{SiO2}, \ce{Al2O3} and \ce{TiO2} are the dominant support materials since they are commonly used supports for thermocatalysts (see Section~\ref{ch:thermocatalysis}), and are therefore the focus of the following discussion about surface chemistry.
\ce{SiO2} surfaces are terminated by siloxanes (oxygen bridges), and isolated and H-bonded hydroxyl (--OH) groups \cite{Haukka1994a, 2000Zhuravlev}.
Depending on the respective precursor, all of these groups are available as binding sites (e.g.\ for TMA)\cite{Puurunen2000a}, or only hydroxyl groups (e.g.\ \ce{TiCl4})\cite{Haukka1993b, Lakomaa1992}.
Similarly for \ce{Al2O3} surfaces, relevant surface groups are hydroxyl groups, but also unsaturated Al or O ions (coordinatively unsaturated sites, c.u.s.)\cite{1978Knozinger,Lindblad1998}.
Like for \ce{SiO2} and \ce{Al2O3}, hydroxyl groups are usually assumed to be the primary binding sites for precursors at \ce{TiO2}\cite{Jackson2014}, though coordinatively unsaturated sites may also be possible (see also Section \ref{sec:pretreatment}).

\textit{Carbons}, including various types of carbon black, graphene platelets, but also particles in different shapes such as carbon nanotubes (CNTs), are common supports in ALD on particles (9.0\% in Figure~\ref{fig:supports}), in contrast to their rare use as substrates in planar substrate ALD\cite{2016Karasulu}.
Carbon materials often exhibit a relevant degree of graphitization, with implications for the binding of precursors molecules to the support surface: 
The graphene structure of sp$^2$-hybridized carbon is rather inert, and many reactants can only bind to defects (vacancies, edges, doped atoms, etc.) through chemisorption.
The limited density of binding sites leads to an island-like or particle growth on the carbon support particles for all reported noble metal precursors\cite{Cheng2015b, Gan2020, Gong2015, Grillo2017, Huang2017, Jiang2014, Lee2019c, Lee2020a, Lee2022d, Li2015, Li2022b, Liang2013b, Lubers2015, Lubers2016, Lubers2017, Luo2019, Rikkinen2011, Sairanen2012, Sairanen2014, Sun2013b, VanBui2017, Wang2016a, Wang2016f, Wang2017b, Wang2018, Xu2018a, Yan2015, Yan2017b, Yan2018a, Zhang2019c, Zhao2019a} and some metal oxide precursors, such as \ce{TiCl4} for TiO$_x$ growth\cite{Grillo2018b, Liu2020a, Sun2012, Wang2017g, Yang2017b}.
However, an alternative TiO$_x$ precursor, TTIP, has been reported to produce both particles and films \cite{Jang2019a, Jang2019b, Justh2018, Li2021, Liang2017, Liu2015a, Meng2011}.
TMA tends to facilitate AlO$_x$ two-dimensional (film) growth\cite{Liang2008a, Wang2017d, Yoon2022preprint, Azaceta2020, Zhan2008, Cavanagh2009, Lichty2013, Jaggernauth2016, Lu2013b, Sun2013a, Zhang2019c, urrehman2018, BorbonNunez2017, MunozMunoz2015, Park2014, Tiznado2014, Devine2011} on carbon, while DEZ often leads to ZnO$_x$ particles\cite{Dominguez2018, Jang2019a, Hilton2017, Lu2017, Luo2015, Sun2014}.

The third most common category is \textit{metals on support} (8.2\%), and a classical example is platinum on carbon (Pt/C) that is used as electrocatalyst. 
Here, ALD stands out as a tool to further modify the catalyst, e.g. by adding another metal to make bimetallic catalysts, or to modify both the support and catalyst to increase stability (see Section \ref{ch:electrocatalysis}).

The category \textit{organics} ($7.0\%$) includes organic materials for pharmaceutical applications (see Section \ref{ch:pharma}), as well as polymers for a variety of applications, such as fillers for ceramics and composite materials \cite{Miller2020,Liang2007a,Nevalainen2009,Liang2007b,Nevalainen2012}. The chemical nature of their surface is often not as well defined as for inorganic particles. Another complicating factor is that for several organic materials, certainly for polymers and pharmaceuticals, infiltration can occur: reactant molecules do not react with the surface, but diffuse into the materials and react with sub-surface molecules \cite{LaZara2021b, Moseson2022}.

Metal particles, though accounting for only a small share of reported support materials ($5.9\%$), are used in a range of over 20 different applications, the most reported of which are thermites and thermoelectric materials.
Especially for non-noble metals, the relevant surface for ALD is in fact more likely oxidic than metallic due to the native oxide growth. Hence, similar surfaces properties as for the oxides (see above) can be expected. 

The remaining support materials (\textit{others}, $6.3\%$, listed in order of occurrence) were metal compounds (nitrides, carbides, sulfides, hydrides, halides, excluding oxides), not chemically specified (e.g., proprietary materials), carbons coated with another material (e.g., metal oxides), metal organic frameworks (MOFs) or semiconductors.
MOFs take on a special role in the group of reported support materials. 
First, they have the highest reported SSA values of $>$\,2000\,m$^2$/g \cite{Kim2018a,Peters2015,Palmer2018}.
Second, they are typically not coated with a film or with particles dispersed over the surface, but rather ALD-like processes are used to deposit individual atoms at a specific site within the framework\cite{Kim2018a,Peters2015,Palmer2018}. In that sense, the use of the term atomic layer deposition is debatable, which led to the new term ALD in MOF (AIM)\cite{2023Zhou}.
Interestingly, metalloid semiconductors which are typical substrates for planar ALD, are of minor relevance for ALD on particles.
Note, however, that these materials often terminate with an oxide surface such that they effectively behave similarly to the oxides discussed above.

\subsection{Particle sizes}\label{sec:particlesize}
The literature covers a wide range of particle sizes (Figure~\ref{fig:supports}b), spanning more than six orders of magnitude, from nanometers to millimeters. The distribution is approximately bimodal, occurring either in the sub-micrometer range (typically 10\,nm to 1\,µm) or in the sub-millimeter range, while intermediate (µm-sized) particles appear underrepresented.
Note, that we grouped the particle sizes as reported in the original works, even if the meaning of the reported number is ambiguous in most cases (see Section \ref{sec:morphology}).

The overall distribution is dominated by oxide particles (compare Figure~\ref{fig:supports}b and c).
Oxides are popular supports for thermocatalysis applications, where often larger particles are used, therefore sub-millimeter particles is the most reported size (see also Figure~S5).
Oxides from the intermediate size range (around 10\,µm) are often used in battery materials.
Finally, smaller size oxide particles ($<1$\,µm) are used photocatalysts.

Though the other support material categories (d-g) show similarly wide size ranges, the distributions and most common sizes differ.
For carbons (d) the particles are typically either in the sub-micron (e.g. carbon black) or the 10--100\,µm range (e.g. carbon nanotubes\cite{Tiznado2014,Hilton2017}, activated carbon\cite{Li2021,Gong2015}, graphene nanoplatelets\cite{Grillo2017}). 
The category metals on support (e) shows no prevalent particle sizes, as these materials are used in both thermocatalysts and electrocatalysts which are applications with very different typical particle sizes.
Similarly for metal particles (g), where the application range is even wider (see Section \ref{ch:otherapps}).
The primary range of µm to mm of organic particles (f) is due to the contribution of pharmaceutical applications (see Section~\ref{ch:pharma} and Figure~S5).

Regarding processing equipment, each of the major temporal ALD reactor types reported (Figure~\ref{fig:reactors_schematics}, top row) has been used for nearly the entire range of particle sizes. However, the flow type reactor is more often used for larger particles (sub-millimeter range) than for smaller particles. The fluidized bed reactor in contrast was often chosen for nanometer-sized particles (see Supporting Information Figure~S4).

\subsection{Specific surface area \& pore diameter}\label{sec:surfacearea}
\begin{figure}
    \centering
    \includegraphics[width=0.6\linewidth]{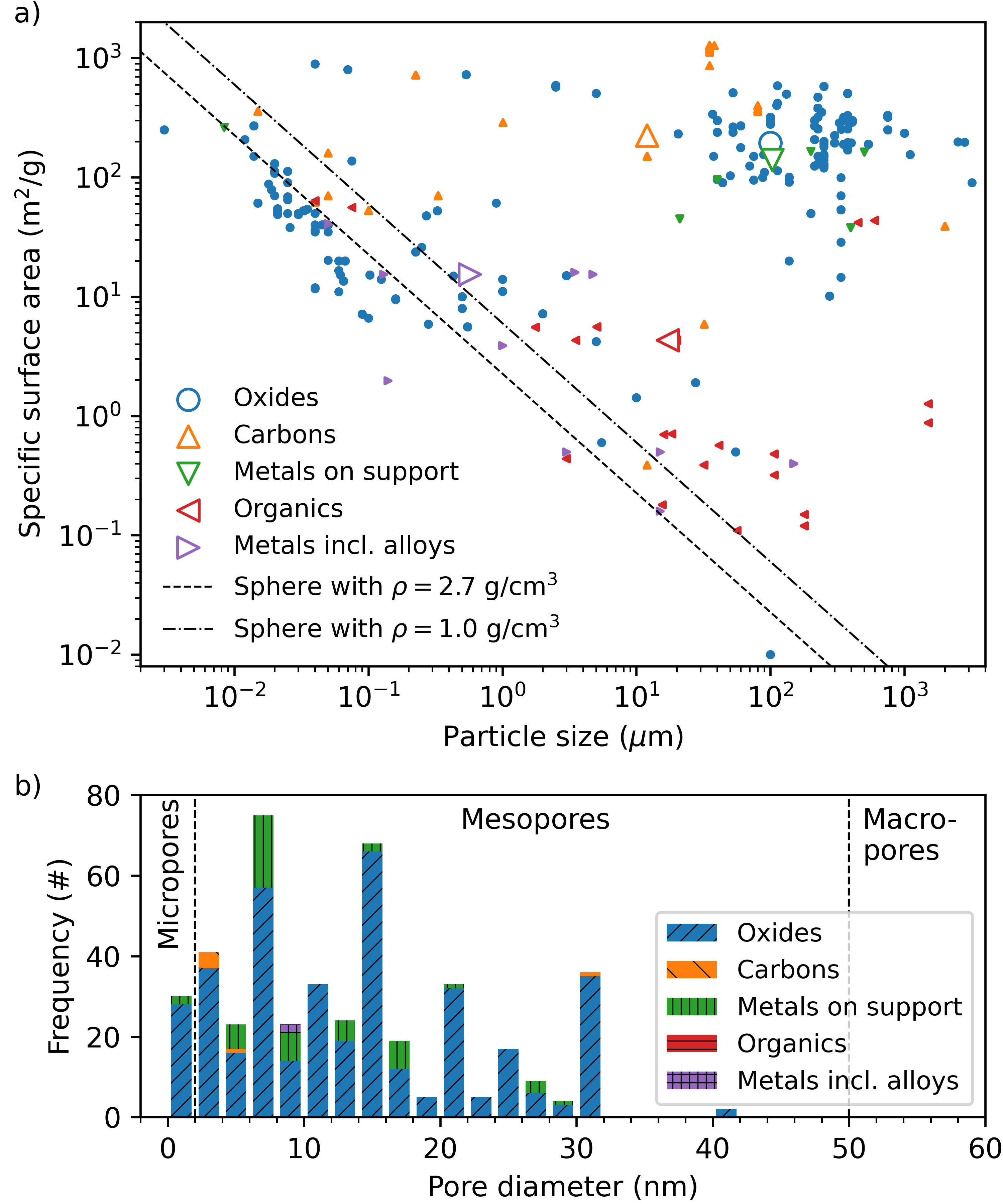}
    \caption{a) Specific surface area (SSA) versus particle size per type of support material (where both values were reported). The large open symbols (see legend) mark the median values of the smaller filled symbols of the same color and shape. Additionally, the geometrical surface areas of spheres of two different densities $\rho$ are shown. b) Pore diameter and IUPAC classification (micro-, meso-, macropores)\cite{2025IUPACMesopore}. Note, that the pore diameters are more often reported for oxides than for other support materials.}
    \label{fig:SSAvsSize}
\end{figure}

The specific surface area at a given particle size is shown in Figure~\ref{fig:SSAvsSize}a for the most common support materials. 
For comparison, we also plotted the geometric surface area of spheres of two different densities, where $\rho$=2.7\,g/cm$^3$ is typical for an oxide particle such as \ce{SiO2} (dashed line), and $\rho$=1.0\,g/cm$^3$ is typical for a lighter, organic material (dash-dotted line).
Non-porous, spherical particles of similar density as \ce{SiO2} will align with the dashed line.
Porous particles exhibit larger surface areas, i.e., are found above the dashed line.
Elongated particles, for which the length of the particle was reported, are found below the dashed line, since they have a smaller surface area compared to a sphere with diameter equal to the reported length.

The oxide particle supports ($\bigcirc$ in Figure~\ref{fig:SSAvsSize}) typically have SSA values in either the 10 to 100\,m$^2$/g range (close to the geometric surface area) for small particles $<1$\,µm, or in the low- to mid-100\,m$^2$/g range for larger particles $>10$\,µm.
Extreme examples of high surface area oxides are mesoporous silica \cite{Singh2016,Sree2012} and zeolite \cite{Gong2019} particles, with SSA greater than 1000\,m$^2$/g (not shown in Figure~\ref{fig:SSAvsSize} since no particle size was reported).
These two different SSA vs.\ particle size regimes originate from the different applications for oxide particles (see Supporting Information Figure~S6).
Larger, porous oxide particles are often used in thermocatalysis (see also Figure~\ref{fig:datadriven_thermocatalysis}g), whereas smaller oxide particles are used in other applications such as photocatalysis and batteries.

The reported carbon particle supports ($\triangle$) typically exhibit relatively high SSA values in the range of hundreds of m$^2$/g for carbon black, or even up to 1500\,m$^2$/g for activated carbons\cite{Zhang2017a,Gong2015,Li2021}.
An exception from this is diamonds, where the SSA is close to the geometrical surface area\cite{Zang2006,Liang2008b,Lu2011b}.

Metals on support particles ($\triangledown$) are often either based on oxides or on carbons. 
The SSA values therefore cover a similar range as these support materials combined.
Metal particles ($\triangleright$) typically show SSA values of $<50$\,m$^2$/g, that are close to the geometrical surface area, since these particles are usually not porous.
Organic support particles ($\triangleleft$) typically show larger SSA values than the geometrical surface area of a light sphere (dotted-dashed line), indicating that these support particles are often porous, though to a lower degree than, e.g., oxides.
The main application of organic particles is healthcare (see Supporting Information Figure~S6), a large part of which are pharmaceuticals (see section~\ref{ch:pharma}).

The porosity can be further characterized by the pore diameter (Figure~\ref{fig:SSAvsSize}b).
The majority of reported pore diameters put the particles in the mesoporous range ($>2$\,nm pore diameter $<50$\,nm), and few in the microporous range ($\leq2$\,nm). 
As already indicated by the high SSA values, porosity is of greater interest in oxide and metal on support particles, supposedly due to their use in thermocatalysis; hence, pore diameters are also reported more frequently for these materials.
However, the pore diameter, in combination with ALD process pressure, also affects the mass transport phenomena (see Section~\ref{sec:pressure}), and must therefore be considered in the choice of processing conditions.

\subsection{Pretreatment of supports}\label{sec:pretreatment}

In many cases, the support material is conditioned or pretreated prior to the ALD process for a number of different reasons.
A common objective is to ensure a homogeneous particle ensemble (powder) as a starting material;
to this end, the particles are often sieved, which can narrow the size distribution (in case of large primary particles), and reduce agglomeration. 
If the ALD process involves agitation, the powder can also be homogenized in the ALD reactor (in-situ) before the coating process. 

Another objective of pretreating the supports is to modify the particle surface.
For oxide supports, some researchers chose pretreatment in an oxidative environment, either by using reactive species such as ozone at 150 to 300\,°C\cite{Chen2016, Christensen2009, Cronauer2012, Elam2010, Enterkin2011b, Kennedy2018, Li2016b, Libera2008, Lobo2012, Lu2010, Lu2012b, Wang2016e, Wang2017c, Xie2013}, or at high temperatures in an \ce{O2} containing atmosphere\cite{Liu2023,Ermakova2002, Jacobs1994a}.
In many cases, the effect of these conditioning steps was not studied in detail, but it is assumed that sufficient temperatures and oxidative stress will remove physisorbed water\cite{1995Kytokivi,Suvanto1998,Backman2001} and hydrocarbons, respectively.
However, earlier and less application-oriented studies provide more details with respect to the effect of pretreatment processes. 
For \ce{SiO2} surfaces, the hydroxyl density is temperature sensitive and has been shown to decrease from more than 6.5 OH/nm$^2$ at 200\,°C to less than 1.1 OH/nm$^2$ at 820\,°C \cite{Haukka1994a,Haukka1994b}.
Similarly for $\gamma-$\ce{Al2O3}, Kytokivi et al. observed a decrease in OH surface groups from 8.2 OH/nm$^2$ at 200\,°C to 2.0\,OH/nm$^2$ at 600\,°C pretreatment temperature\cite{1995Kytokivi,Haukka1997a}, which is consistent with results from Puurunen et al.\cite{Puurunen2001}.
Given that hydroxyl groups are important binding sites for many precursors (see section~\ref{sec:supportschem}), the heat pretreatment will likely affect the growth behavior during ALD (see Section \ref{sec:GPC}).\cite{2005bPuurunen}

As mentioned above, the sp$^2$-hybridized carbon structure of many carbon supports renders them rather inert to reactions with precursors, and often leads to island growth starting from defects rather than two-dimensional (film) growth during ALD.
In order to tune the growth behavior on carbon, researchers apply pretreatment steps, such as a slight oxidation of the carbon before ALD by either \ce{O2} plasma \cite{Xu2018a} or \ce{O3} (ozone)\cite{Li2022b,Grillo2018b,VanBui2017,Grillo2017,Meng2015}.
Meng et al. have shown that pretreatment in \ce{O3} indeed increases the density of oxygen defects in multiwall carbon nanotubes (MWCNTs) and facilitates nucleation in ALD \cite{Meng2015}.
The effect of plasma pretreatment in particle ALD has so far not been studied in detail\cite{Xu2018a}.
However, for flat carbon surfaces, Karasulu et al.\ have shown both experimentally and numerically that oxygen plasma treatment prior to ALD effectively facilitates the growth of closed Pt films\cite{2016Karasulu}.
Alternatively to dry methods, wet chemical pretreatments were also employed, such as in \ce{HNO3}\cite{Gan2020, Gong2015, Lee2022d}, or citric acid\cite{Liu2015a}.
The latter led to an increase of oxygen functional groups on the carbon surface and subsequently a finer dispersion of Pt particles\cite{Liu2015a}.
Instead of an oxidative pretreatment, Cheng et al.\ used nitrogen-doped graphene and carbon nanotubes to increase the dispersion of deposited Pt particles, and their stability in electrocatalytic reactions\cite{Cheng2015b,Cheng2016}.

For metals, the surface is often oxidic due to native oxide growth.
To obtain metallic surfaces, particles can be pretreated in a reducing environment with hydrogen, at high temperatures, where the required temperature depends on the metal in question \cite{Cao2019,King2009a,Manandhar2016,Manandhar2017}. 
Clancey et al.\ used hydrogen plasma as an intermediate cleaning step to remove unreacted precursor molecules between ALD of different metals. Though they were not specifically aiming at reducing the metal, the treatment may have had such effect\cite{Clancey2015}.

On a side note, blocking potential binding sites on the particle surface with another chemical could also be considered pretreatment.
This approach has been pursued in the context of thermocatalysis (see Section~\ref{ch:thermocatalysis}), and we here consider it as part of the ALD sequence.

\subsection{Safety considerations}\label{sec:safety}

In addition to the safety considerations associated with ALD processes (see Section \ref{ch:processing}) that are also required when processing planar substrates, the powder form of the material requires additional precautions due to high specific surface area and the mobility of the fine particles.
These qualities increase both the risk of uncontrolled chemical reactions (e.g. fire or explosions) and the health risks of exposure.

The health risk of working with particles is not regularly addressed in the literature of ALD for particles, and a comprehensive review is beyond the scope of this work.
However, since the topic is often overlooked, we include some general remarks in this review, and strongly suggest considering the respective literature in the planning and safety assessment of any particle ALD experiment, see, e.g., Rhodes et al. for a broad view \cite{2024Rhodes}.

On the one hand, particles of all sizes (incl.\ nanoparticles) are a natural part of the environment and for a wide range of them, the human natural defense seems to be well-equipped to handle them.
On the other hand, in the context of ALD on particles, we are often dealing with materials and particles that do not occur in everyday life.
Assessing the risk of human particle exposure can be challenging, since the information on safe handling of exotic particles is typically not condensed in an official guideline, as issued for other materials by national work safety authorities\cite{2013NioshFlour}.
Even where, e.g., official exposure limits are available, they often cover an entire category of particles and should therefore be considered with caution.
This is further complicated by the fact that we are modifying the powders with our ALD processes.
Hence, it is our responsibility as researchers to ensure safe handling on a case-by-case basis, and based on general guidelines\cite{2016NioshNano}.

Specifically, the toxicology of nanoparticles (nanotoxicology) is an extensive field of its own\cite{2015Qiao,2016dePaula,2013Donaldson,2013Ferreira}.
Nanoparticles range from low levels of toxicity in moderate doses, to toxic and carcinogenic even in trace amounts\cite{2010EUComissionNanomaterials}.
Toxicity depends on several aspects such as size, shape, chemical composition and dose of the particles\cite{2015Qiao}. 
Sharing most properties of a relatively safe to handle material does not render another material equally safe. 
For example, exposure to (i.e., inhalation of) graphite nanoplatelets showed no increase of inflammation markers in rats, while exposure to MWCNTs showed an increase of the same markers by more than two orders of magnitude, despite the chemical similarities\cite{2013MaHock,2013Ferreira}.
Similar to asbestos fibers, there is a strong shape effect, where elongated particles beyond a certain length are particularly inflammatory and likely carcinogenic\cite{2010EUComissionNanomaterials}.
To further complicate the matter, a change in surface properties of carbon materials, e.g. by oxidation or doping (both potential effects of pretreatment or ALD) can influence toxicity \cite{2016dePaula}.

The high surface area of a bioactive compound is associated with high bioavailability when inhaled or ingested\cite{LaZara2021b}. 
While this is a positive feature of particles in pharmaceutical applications \cite{LaZara2021b}, it is generally an unwanted effect of these materials in a laboratory setting. 
For other materials that can potentially form highly toxic compounds, such as chromium \cite{Li2017a,Puurunen2003,Haukka1994b,Korhonen2007,Kytokivi1996a,Zhang2014a,Zhang2015b}, the high bioavailability is generally undesirable.

The fire hazard is a concern for metal particles and some pharmaceuticals, as these show potential for ignition due to electrostatic discharge\cite{Gupta2022}. 
ALD is used for both types of materials to reduce such risk by adding, e.g., oxide layers for charge management\cite{Gupta2022} and as oxidation 
barriers\cite{Hakim2007a,Qin2019,Cremers2018}, but care must be taken during the processing.
The same is true for catalysts on combustible supports.
While separately both the catalyst (e.g.\ Pt) and the support (e.g.\ carbon) may be stable even at elevated temperatures, the combination may be prone to combustion.

A safe option to handle these powders is to use a glove box with a low oxygen atmosphere around or connected to the ALD system, which enables handling of even radioactive materials as shown by Bhattacharya et al.\cite{Bhattacharya2019}. 
Besides risk mitigation, the low oxygen environment of a glove box also enables coating of metal surfaces instead of metal oxides that typically occur upon exposure to air\cite{Manandhar2017}. 
Another option to reduce both risk and oxidation is in situ generation of metal particles in the ALD reactor. 
An example of this is the thermal decomposition of iron oxalate to Fe nanopowder with particle sizes of 50\,nm to 80\,nm, as demonstrated by Hakim et al.\cite{Hakim2007a}.

In addition to the risk posed by the bare support particles upon exposure to an oxidant (incl.\ air), the combination of metal particles with ALD oxide layers bear the potential for violent exothermic reactions.
This is deliberately used in ALD-made thermites, where reactions can be extremely fast due to the nanoscale-mixing of the reducing metals with the oxides \cite{Ferguson2005,Qin2017}.
Ferguson et al.\ mitigated the risk during the fabrication of thermites by first optimizing their ALD process on inert \ce{ZrO2} particles before switching to reactive Al particles, and by keeping the batch size below 1\,g of powder \cite{Ferguson2005}.

\section{Processing \& characterization}\label{ch:processing}

In this section, we discuss the ALD-relevant reactants and processing conditions.
For each of the parameters that we deem relevant based on previous review work\cite{vanOmmen2019}, we first display a condensed overview of the values found in literature (see Section \ref{sec:datacollection}), followed by a brief discussion, respectively. 

\subsection{Reactants}

From the articles in our dataset, we identified 189 reactants that have been used for ALD on particles. 
We further categorized these reactants as metal or metalloid sources ('precursors'), 
reactants for ligand removal ('counter-reactants'), organic reactants (for MLD) or dopant sources. 
Some compounds may serve a dual function or be used for a different purpose depending on the ALD process; therefore, the classification and thus the terminology remain inherently ambiguous. 
Acknowledging this ambiguity, we still find the categories helpful in further discussing the reactants, where we focus on precursors and counter-reactants.

\subsubsection{Metal \& metalloid precursors}

\begin{figure}
  \includegraphics[width=0.85\textwidth]{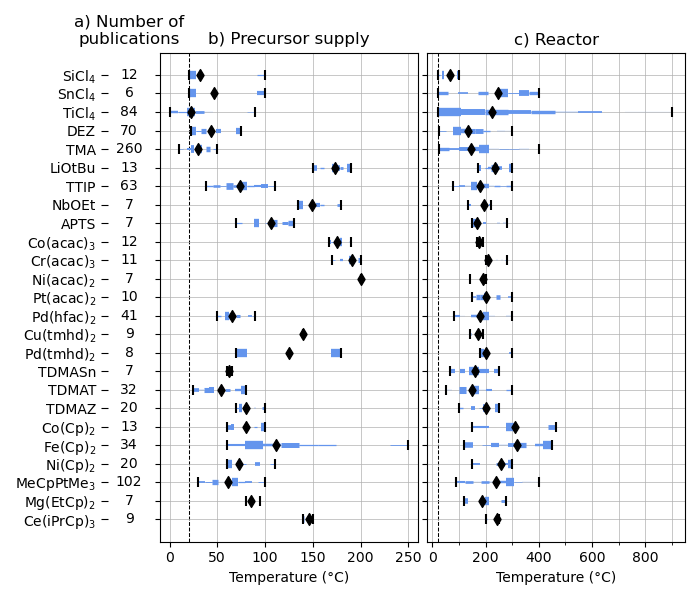}
  \caption{a) 25 most reported precursors (counted once per publication, grouped by similarity), constituting about 79\,\% of all reports involving metal and metalloid (compound) coatings. See Supporting Information Table~S4 for abbreviations. b) Precursor supply temperatures. c) Reactor temperatures. Black diamonds and bars mark the mean of all reported temperatures values and extremes, respectively. The thickness of the blue line indicates (per precursor) how often a specific temperature range was reported (multiple values per publication possible). The dashed lines mark room temperature (20\,°C).}
  \label{fig:Trange}
\end{figure}

Approximately 130 compounds were identified as sources of metal and metalloid that remained in the deposited layer, and were thus categorized as precursors.
The 25 most common are listed in Figure~\ref{fig:Trange}a, and include chlorides and n-alkyls such as \ce{SiCl4} and TMA, respectively, but also more complex compounds such as β-diketonates (e.g. acac, hfac, tmhd) or cyclopentadienyl (Cp) complexes.
The full names of these complexes and precursors abbreviations from Figure~\ref{fig:Trange} are given in the Supporting Information Table~S4.

An important aspect of precursor supply is the temperature at which the precursor is kept during the process.
Heating the precursor can increase the vapor pressure to ensure that sufficient precursor amounts (doses) reach the reactor.
Figure~\ref{fig:Trange}b provides an overview of precursor supply temperatures in reported ALD processes on particles. 
The simplest precursors in terms of structure (chlorides and alkyls) are often kept near room temperature (dashed line), or even cooled, as they often exhibit comparably high vapor pressures.
If heated, the temperatures are kept moderate, i.e., $\leq$100\,°C for chlorides \cite{Ermakova2002,Yang2022b,Lin2020b}, and lower for alkyls\cite{Wang2016b,Knemeyer2020,Ingale2021}.

The alkoxides (LiOtBu, TTIP, NbOEt) and γ-Aminopropyltriethoxysilane (APTS), in contrast, are typically heated to temperatures of up to 200\,°C.
Similarly, β-diketonates were heated in all reported cases. 
Temperatures of typically around 200\,°C were reported for acac-based, around 140\,°C for tmhd-based and $<$100\,°C for hfac-based precursors. 
Replacing the substituent methyl groups of the acac anion with other groups (such as tmhd or hfac) affects the intermolecular interactions, thereby lowers sublimation enthalpies and increases the vapor pressure at a given temperature\cite{2001Fahlman}.

Metal amides (TDMASn, TDMAT, TDMAZ) are typically supplied at temperatures $\leq$\,100\,°C, with TDMAT supply reported even at room temperature.

The cyclopentadienyl (Cp) complexes exhibit the widest range of all precursors in Figure~\ref{fig:Trange}b, ranging from 30\,°C for \ce{MeCpPtMe3} to 250\,°C for \ce{Fe(Cp)2} (ferrocene), however, these extremes are outliers (thin blue line indicates low number of reports).

ALD on particles often requires much higher precursor doses compared to planar substrates due to the higher surface areas and porosity of the particles (see Section \ref{sec:dosing}).
Therefore, precursors that are usually kept at room temperature for planar substrate ALD (e.g.\ DEZ) are sometimes heated to coat particles.
However, precursors can react to heating with decomposition, leading to the violent release of flammable gases and explosions, in some cases even at moderate temperatures (e.g., DEZ from 70\,°C)\cite{2023NouryonDEZ}.
The precursor supply temperature must therefore be chosen carefully, weighing in the required vapor pressure and stability.
Though the temperatures in Figure~\ref{fig:Trange}b indicate ranges from reported ALD processes, one should be careful to consider them safe.
Additionally, the technical implementation of precursor heating should be carefully chosen in order to exclude, beyond reasonable doubt, the possibility of runaway heating of, e.g., an electrical heater.

\subsubsection{Counter-reactants}
The role of the counter-reactants is to remove the ligands of the chemisorbed precursor, and in case of a binary compound, to deliver the other element.
In case of oxides, this is achieved by replacing ligands with oxygen atoms from sources such as \ce{H2O} vapor or \ce{O2}.
Except for some noble metal processes (e.g. Pt, Ir, Ru), the oxygen is incorporated into the layer, and the deposited material will hence be an oxide rather than reduced metal or metalloid.

In some cases, chemicals of higher reactivity such as \ce{O3}, \ce{H2O2}, or oxygen-containing plasma (in plasma enhanced ALD, PE-ALD) are used \cite{Longrie2012, Ramachandran2016, Rampelberg2014, Sun2019, Verstraete2019, Zhao2017a,Kikuchi2016}.
As an alternative, less oxidative reactants, such as \ce{CH2O} (formaldehyde) or \ce{NH3}, are sometimes used, or even \ce{H2} as a reducing reactant. 
Again, there are examples of PE-ALD processes, e.g., for \ce{NH3} \cite{Cao2018, Longrie2012, Longrie2014a, Rampelberg2014}, \ce{H2} \cite{Clancey2015, Deng2022, Tian2023, Wang2016a, Wang2022d} or \ce{N2} plasma \cite{Godoy2021}.
Like in the case of oxygen-based reactants, part of the counter-reactant may be incorporated into the layer, such as \ce{TiN}\cite{Cao2018, Godoy2021, Longrie2014a, Rampelberg2014}.

On a side note, we counted only 15 reports of PE-ALD on particles in total, in contrast to planar substrate processes where PE-ALD is regularly employed.
PE-ALD might be less common due to the challenges of combining particle- and plasma technology in a single reactor.
Also, many of the support materials are not temperature sensitive, especially since the largest application area are thermocatalysts, therefore thermal ALD can be employed.
However, applications with thermally sensitive materials (e.g. for luminescent phosphors, pharmaceuticals) may benefit from the availability of PE-ALD systems suited for particulate materials.

\subsubsection{Precursor requirements for coating particles}

A large share of the reactants (precursors and counter-reactants) used for ALD on particles are familiar from planar substrate processes. 
However, the differences in processing conditions and the different scope of applications can lead to a different set of requirements for a reactant.

The total surface area to be coated in one batch of particles can easily reach multiple hundreds of m$^2$ (see Figure~\ref{fig:datadriven_thermocatalysis}), therefore the dose of reactant to be delivered to the reactor is correspondingly large.
Hence, the vapor pressure of, e.g., metal precursor is required to be large in order to avoid exceedingly long pulse times. 
At the same time, the impact of the cost of a reactant on the overall cost of a process is higher than for low surface area processes.

The reactant exposure times (see following Section~\ref{sec:dosing}) are often longer than for planar substrates, such that self-decomposition of the reactant, even at moderate rates, can lead to a notable continuous growth component (see Figure~\ref{fig:ALDscheme}c).
This needs to be considered when assessing the stability of a reactant, especially for applications where self-limiting growth is essential.

Finally, the chemical compatibility of reactants and by-products must be taken into account with regard to the respective application. 
For example, carbon residues may be a concern for semiconductor applications, while this is often not problematic for thermocatalysts. 
At the same time, chlorine and fluorine residues can lead to poisoning of the catalyst.

\subsection{Reactor temperature}\label{sec:reactortemperature}

For the 25 most common precursors discussed above, we collected the reported reactor temperatures at which the support material is exposed to the precursor (Figure~\ref{fig:Trange}c).
The lowest temperatures, starting near room temperature, are used for the precursors with the simplest chemical structure (chlorides and alkyls).
The reported temperature range up to 900\,°C for \ce{TiCl4} is exceptional, and there are only few reports of temperatures $>$\,450\,°C \cite{Haukka1993a, Haukka1993b, Koshtyal2014, Lakomaa1992, Malkov2010, Malygin1997, Snyder2007, Sosnov2010, Sosnov2011}.

Many of the other precursor types, ranging from alkoxides through amides (Figure~\ref{fig:Trange}, list from TTIP through TDMAZ), show on average similar reactor temperatures (diamonds) of 150\,°C to 200\,°C, and maxima at 300\,°C.
The latter may well be due to the temperature limit of the equipment, since the number of different reactor implementations is fewer than the number of precursors, thereby imposing the same limit on many precursors.
For some of the Cp complexes (i.e., \ce{Fe(Cp)2} and \ce{Co(Cp)2}) higher reactor temperatures $>$\,400\,°C are relatively common. 
Also, the reactor temperatures for Cp complexes are higher on average.

Furthermore, the reactor temperatures in Figure~\ref{fig:Trange}c are values reported for the precursor exposure step only.
In some reports, the reactor temperature was indeed changed from one reactant to the next. 
For one-cycle ALD experiments, this is sometimes achieved by exposing the support material to reactant A (e.g., a metal precursor), and subsequently transferring the support material to a muffle furnace for calcination at much higher temperatures, i.e., using air as reactant B.
But there are also examples where the number of ALD cycles is $>$\,1 and the temperature between reactant A and B exposures is changed multiple times within the same reactor\cite{Backman1998, Backman2000, Backman2001, Clary2020, Ek2004, Ermakova2002, Hakuli2000, Haukka1997a, Huang2021, Knemeyer2021b, Kytokivi1996a, Kytokivi1997b, Lee2022b, Lee2022g, Lin2018, Lindblad1994, Lindblad1997, Lindblad1998, MakiArvela2003, Malkov2010, Milt2000, Milt2001, Milt2002, Molenbroek1998, Najafabadi2016, Okumura1998b, Onn2017b, Puurunen2000b, Puurunen2002a, Puurunen2002b, Puurunen2003, Rautiainen2002, Shen2023b, Shen2023c, Tsyganenko2000, Weng2018b}.
Likewise, in more complex ALD sequences than A-B, the reactor temperature is occasionally changed between reactant exposure steps.\cite{CamachoBunquin2018a, Cavanagh2009, Chen2010a, Haukka1997a, Heikkinen2021, Lee2022b, Lee2022c, Lei2012, Lin2018, Lin2020a, Lin2020b, Mao2020b, Shen2023a, Shen2023b, Xu2018b, Yang2022a, Zhang2017a, Zhang2019a}

The option to change the reactor temperature within the same ALD process arises from the long exposure times for particles, often on the order of minutes (see Section~\ref{sec:dosing}). 
For planar substrate ALD, in contrast, exposure times are commonly on the order of seconds and below, and therefore the reactor temperature is kept constant throughout the process.
In light of this finding, one has to reconsider the ALD temperature window concept (Section~\ref{sec:basictemperature}).
Usually, the same temperature window for the entire process is identified, i.e., the same window for all reactions.
However, one could also probe the temperature range for each of the reactions (and corresponding reactants), and thereby identify the ideal temperature window for each reaction.

Another relevant aspect in ALD on particulate materials is the effect of reaction enthalpy on reactor temperature. 
With large specific surfaces, especially with porous particles and nanoparticles, the exothermic chemical reactions at the surface can lead to a considerable release of heat\cite{Greenberg2020,2024OssamaMuhammad}.
At the same time, heat transport within these powders and to the reactor walls may be slower compared to planar substrates. 
A way to mitigate these problems is agitating the powder during the ALD process to increase the heat transfer (see Sections \ref{sec:agitation} and \ref{sec:reactorprinciple}), but even then a constant and uniform temperature may be difficult to ensure. 
Therefore, in practice, the apparent ALD temperature window may differ considerably between ALD processes on particles and planar substrates due to, e.g., parasitic decomposition at lower apparent temperatures.\cite{Knemeyer2021a}

It is important to note, that Figure~\ref{fig:Trange}c shows just the aggregated reported reactor temperatures of presumably successful ALD processes.
However, our data curation involved no assessment of reactor temperature in the context of the ALD temperature window (see Section~\ref{sec:basictemperature}).
For an explicit probing and discussion of the ALD temperature window, the reader is referred to the broader ALD literature including planar substrates, while taking into account the above discussed differences between particles and planar substrates\cite{Knemeyer2021a}.

\subsection{Process pressure}\label{sec:pressure}
\begin{figure}[ht]
    \centering
    \includegraphics[width=0.45\linewidth]{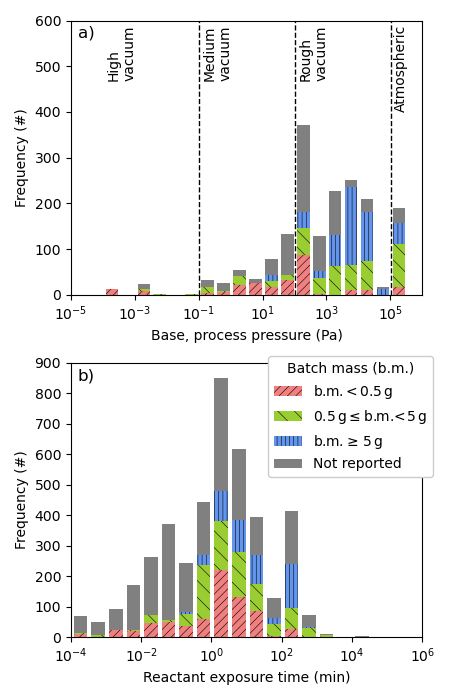}
    \caption{a) Histogram of reported values for 'base pressure' or 'process pressure', where the bar color indicates the batch size of the process (same legend for sub-figures a and b).
    If a range was given, we counted the minimum and the maximum separately. The different vacuum regimes\cite{2019IsoVacuum} are delimited by dashed lines. b) Reported values for reactant exposure times (irrespective of reactant type).}
    \label{fig:pressure}
\end{figure}

To provide an overview of the process pressures used in ALD reactors for particles, we collected reported pressure values (Figure~\ref{fig:pressure}a). 
In many cases, only approximate values were reported, or it was not clear if the reported value refers to the pressure during the ALD process or to the 'base' pressure to which the reactor is pumped before starting the ALD process.
However, a classification into different vacuum regimes is still possible.\cite{2019IsoVacuum} 

The majority of processes (Figure~\ref{fig:pressure}a) were carried out under rough vacuum (from $10^2$ to $10^5$\,Pa), and in a considerable number of processes even at atmospheric pressure  ($\sim10^5$\,Pa).
Medium vacuum ($10^{-1}$ to $10^2$\,Pa) processes, though still common, make up a minority of cases.
High vacuum processes ($<10^{-1}$\,Pa) finally, are rare in ALD on particulate materials. 

Although the process pressure is frequently reported, its effect is rarely discussed in the literature concerning ALD on particles.
An explicit experimental investigation of the process pressure is hindered by the fact that most reported reactors are designed for a limited pressure range, and examples of ALD systems that allow large pressure ranges are rare for both particles\cite{CamachoBunquin2015} and planar substrates\cite{2011Jur}.

The pressure can have an impact, for example, on the size distribution of Pt islands formed through ALD\cite{Grillo2018a}. The presumed mechanism is that high partial pressures of the \ce{O2} counter-reactant (which is only possible at a high process pressure) leads to faster nucleation, and formation \ce{PtO2} shells around the Pt islands (nanoparticles) that prevent aggregation\cite{Grillo2018a}.

\begin{figure}
    \centering
    \includegraphics[width=\linewidth]{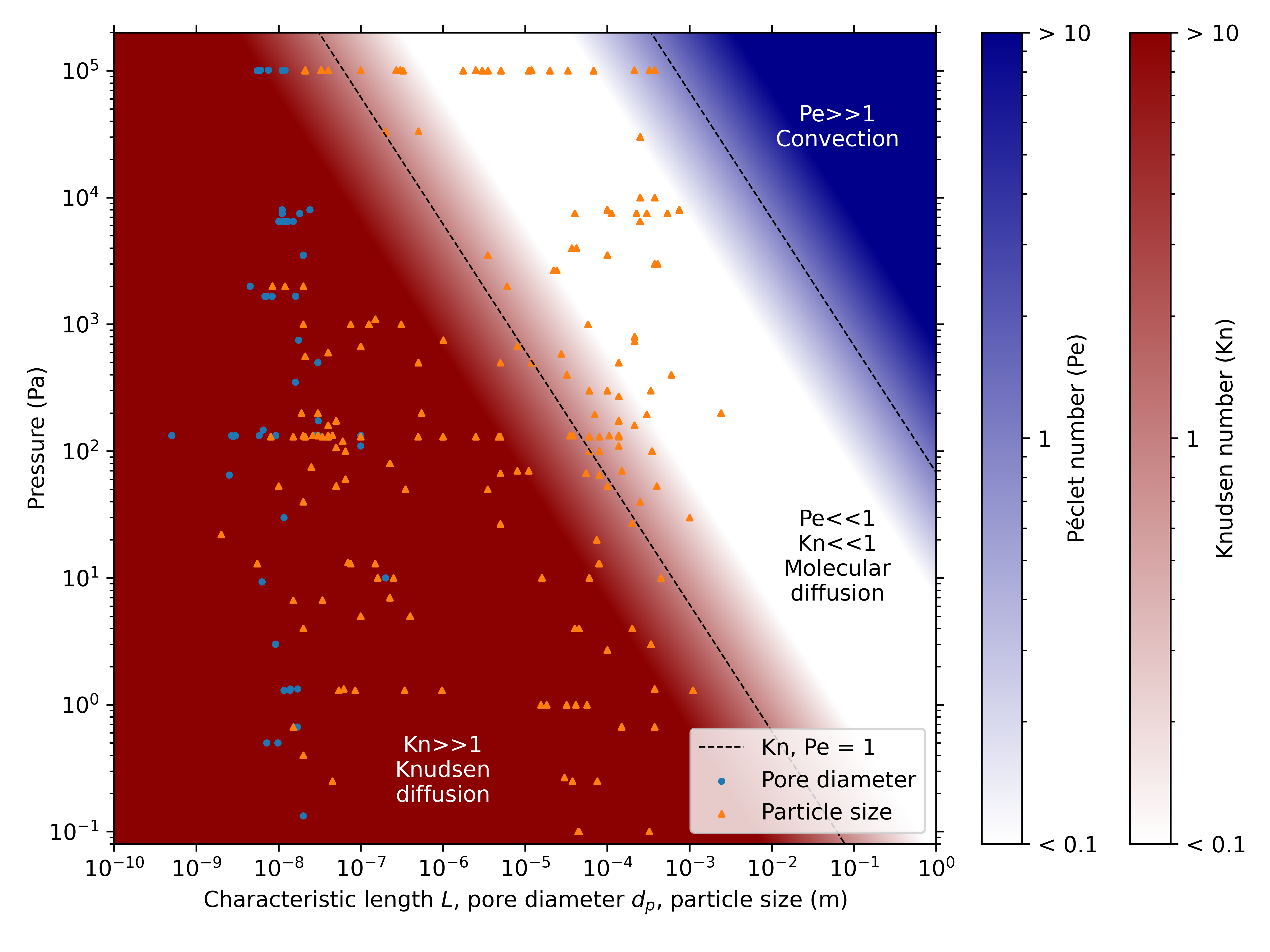}
    \caption{Mass transport regime classification using dimensionless numbers. The Péclet number indicates (molecular) diffusion-dominated or convection-dominated mass transport, depending on characteristic length $L$ and pressure; calculated assuming a flow of $u$=3.4\,cm/s, which corresponds to 1\,L/min process gas through a 2.5\,cm diameter tube (typical numbers from the dataset).
    The Knudsen number indicates a regime of molecular diffusion or confined diffusion inside a pore (Knudsen diffusion), depending on pore diameter and pressure; calculated using Equation~\ref{eq:eqKn}.
    Note, that for both calculations we assumed a pure TMA atmosphere (using literature values for $D_0$\cite{2016Poodt} and $\sigma_\text{TMA,TMA}$\cite{2018Cianci}), but the results are very similar for \ce{N2}. The scattered points (pore diameters and particle sizes) show experimental data from the literature dataset without further processing.}
    \label{fig:pressuretransport}
\end{figure}

However, the main focus of the discussion on the effect of pressure in the basic ALD literature is on mass transport of gaseous reactants and reaction products\cite{2010George}.
Here, it is important to identify which transport phenomenon is dominant, i.e, whether it is convection of species in a carrier gas stream, or diffusion. 
The dimensionless Péclet number compares the convective transport rate to diffusion\cite{2009Cussler}:
\begin{equation}
\text{Pe} = \frac{L u}{D}, 
 \label{eq:eqPe}
\end{equation}
where $L$ is the characteristic length, $u$ is gas flow velocity, and $D$ is the diffusion coefficient of a gaseous species.
Importantly, molecular diffusion increases with decreasing pressure\cite{2009Cussler}, i.e., $D\propto D_0/p$, where $D_0$ is the diffusion coefficient at a reference pressure.
This leads to convection being dominant (Pe$\gg1$) at high pressures and large spatial dimensions (Figure~\ref{fig:pressuretransport}, top right), and diffusion being dominant (Pe$\ll1$) at lower pressures and short length scales.

On the length scale of reactor dimensions (mm to m), convection is the dominant transport mechanism, in particular at near-atmospheric pressures and at rough vacuum.
On a length scale of particle sizes (values from literature dataset in Figure~\ref{fig:pressuretransport}) in contrast, mass transport is usually dominated by diffusion, which applies even to exceptionally large particles in the sub-mm size range.

Diffusion can further be dissected into different regimes. 
For a gas--solid mixture, diffusion can be dominated by collisions of gas molecules with solid walls, i.e., Knudsen diffusion, or with each other, i.e., molecular diffusion.
Whether Knudsen diffusion or molecular diffusion dominates, or whether the system is in the transition regime between them, can be analysed from the molecule's mean free path at the used pressure conditions. The mean free path of reactant ‘A’ in a binary system consisting of two gases---reactant ‘A’ and inert carrier gas ‘I’---can be calculated using the equation\cite{Cremers2019,1990Chapman,2018Ylilammi,2022Yim}
 \begin{equation}
 \lambda= \frac{k_\mathrm{B}T}{\sqrt{2}p_\mathrm{A}\sigma_\mathrm{A,A} + \sqrt{1 + \frac{m_\mathrm{A}}{m_\mathrm{I}}}p_\mathrm{I}\sigma_\mathrm{{A,I}}}\, .
 \label{eq:eq3}
\end{equation}
Here, $k_\mathrm{B}$ (JK\textsuperscript{-1}) is the Boltzmann constant, \textit{T} (K) is the temperature,  $m_\mathrm{A}$ and $m_\mathrm{I}$ (kg) are the masses of the molecules of reactant A and inert gas I, respectively; $p_\mathrm{A}$ (Pa) is the partial pressure of reactant A and $p_\mathrm{I}$ (Pa) is the inert gas partial pressure; and 
$\sigma_\mathrm{{A,A}}$ and $\sigma_\mathrm{{A,I}}$ are the collision cross-sections ($\mathrm{m^2}$) between the molecules A and I. The collision cross-section for components \textit{i} and \textit{j} is given by \cite{2022Yim} 
\begin{equation}
 \sigma_{i,j}= \pi {\left( \frac{d_i}{2} +\frac{d_j}{2} \right)}^2,
 \label{eq:eq4}
\end{equation}
where $d_{i}$ (m) and $d_{j}$ (m) are the hard-sphere diameters of the molecules \textit{i} and \textit{j}, respectively. Hard-sphere diameters can be estimated for example from the condensed phase volume \cite{2018Ylilammi}. If the mean free path $\lambda$ is much larger than the limiting dimension of the system $d_p$ (e.g. pore diameter, see Section~\ref{sec:surfacearea}), then Knudsen number \cite{2019Cremers,2009Cussler}
\begin{equation}
\text{Kn} = \frac{\lambda}{d_p}
 \label{eq:eqKn}
\end{equation}
is much larger than one (Kn$\gg 1$), and Knudsen diffusion dominates (Figure~\ref{fig:pressuretransport}, bottom left). 
If the mean free path is much lower than the limiting dimension (Kn$\ll 1$), then molecular diffusion dominates. 

When comparing the reported pore diameters to the diffusion regimes (Figure~\ref{fig:pressuretransport}), Knudsen diffusion is dominant within pores of a particle. 
We note, however, that the range of reported pore diameters may be limited by the typical characterization techniques used (e.g.\ BJH, see Section~\ref{sec:characterization}), and therefore pore diameters in the µm-range were potentially under-reported. 
Those would indeed lead to dominant molecular diffusion at atmospheric pressure, whereas Knudsen diffusion can be considered typical for vacuum-based ALD, even for such larger pores. 

Particles in a packed bed can be considered a porous solid with respect to mass transport.
For larger particles (sub-millimeter size), pores (voids) between the particles are typically of the same order of magnitude as the particle size.
Similarly, aggregations and agglomerates of nanoparticles, even if agitated (e.g., in a fluidized bed), can be considered porous particles.
To characterize the reactant mass transport in these cases, we plotted the particle sizes versus reactor pressure (where the combination was reported) in Figure~\ref{fig:pressuretransport}, thereby using the particle size as a measure for the pore diameter or length scale, respectively.
For all reported values, diffusion rather than convection is the dominant transport mechanism on a particle size scale. 
For particles in the size range below 100\,nm, Knudsen diffusion rather than molecular diffusion can be considered dominant for process pressures up to atmospheric ($10^5$\,Pa).
For larger particles up to about 10\,µm, Knudsen diffusion is dominant only for vacuum process in the $10^2$\,Pa range, and molecular diffusion is clearly dominant at atmospheric pressure.

We note, that the aggregation behavior of the specific particle type needs to be considered when identifying the particle size with pore diameter.
Especially nanoparticles tend to form aggregates with high void fractions and much larger pores than the particle size. 

Note, that the pressure dependence of the Péclet number Pe is a consequence of assuming constant, pressure-independent gas velocities $u$ for convection. 
If instead a constant mass flow is assumed, e.g., 1\,slm (standard liter per minute) instead of 1\,L/min, then both $D$ and $u$ scale $\propto 1/P$, and the pressure dependence of Pe cancels out (see Supporting Information Figure~S7). 
Importantly, this is due to an increase in both transport mechanisms, and therefore an overall more effective transport of reactants.
However, achieving low pressures at high mass flows of process gases (both reactants and carrier gas) comes with technical challenges.
Reducing the pressure leads to higher volume flows for the same mass flow (flow of substance) of gases, e.g., 1\,L/min at $10^5$\,Pa (atmospheric pressure, i.e., gas flow 1 slm) is equivalent to 1000\,L/min at $10^2$\,Pa; the flow rate is anti-proportional to the pressure, therefore requiring large pumps to achieve vacuum conditions. 

This circumstance also has to be considered when choosing a reactor for an envisioned experiment for which a certain process pressure is targeted. Some reactor types require comparably large gas flows for agitation and mixing of the particulate support material (see Section \ref{sec:agitation}). For example, agitation in a fluidized bed is typically achieved through carrier gas flows in the order of standard liters per minute, making lower pressures more difficult to achieve.

Regardless of the reactor type, the high surface area of particulate support material, especially for larger batches, will significantly slow down the reactor pump-down process, especially to low base pressures.  This may explain why batches $>5$\,g (blue in Figure~\ref{fig:pressure}) are more frequently processes at atmospheric pressure or rough vacuum. 

In summary, the choice of process pressure in ALD on particles is determined by both the conditions required for ALD, and those suitable for the processing of particles.
The latter may explain why typically higher process pressures are used for particles compared to planar substrate processes in the semiconductor domain.

\subsection{Reactant exposure times}\label{sec:dosing}

In Section~\ref{sec:ALDbasic} we outlined that ideal ALD processes incorporate a set of saturating chemisorption reactions through sufficient exposures (exposure time $\times$ partial pressure) of the substrate to the reactants.
The reactants are added to the carrier gas stream in a pulse of defined duration.
Especially for processes on non-porous, planar substrates in vacuum, pulse duration and exposure time are often near-identical, since the reactor is continuously evacuated while introducing the reactant. 
However, for ALD on particulate materials, it is also common to dose a certain amount of reactant in a pulse, followed by a soaking step in which the carrier gas flow is interrupted, and the substrate is exposed to the reactant\cite{Devine2011,Lee2022a,Yang2017a,Settle2019}.
In that case, the exposure time is obviously longer than the pulse duration, and is also the more relevant quantity to discuss.

In ALD on particulate materials, typical exposure times are on the order of seconds to minutes (Figure~\ref{fig:pressure}b); although sub-second exposure times have also been reported. 
Sometimes, exposure times up to hours are employed\cite{Asakura1992, Backman1998, Bahrani2023, Chae2018, Chen2006, Chen2008, Chen2009, Chen2010a, Chen2010b, Chen2011, Daresibi2023, Haukka1994b, Haukka1997a, He2018, Hussain2022, Iiskola1997a, Iiskola1997b, Juvaste1999a, Juvaste1999b, Keranen2003, Kim2018a, Koshtyal2014, KrogerLaukkanen2001, Lakomaa1992, Lin2013, Moulijn2023, Pallister2014, PlateroPrats2017, Rikkinen2011, Sairanen2012, Sairanen2014, Silvennoinen2007a, Silvennoinen2007b, Sosnov2017, Suvanto1998, Timonen1999, Uusitalo2000a, Valdesueiro2016, Voigt2019, Yim2023}.
However, in some cases only one full ALD cycle was performed, of which the first half-cycle was carried out in an ALD reactor, and the second half-cycle (ligands removal) was implemented through calcination, e.g., in air\cite{Backman1998, Backman2000, Chen2010a, Chen2010b, Keranen2002, Lin2013, Milt2002, Rautiainen2002, Yim2023}.

The exposure times for particulate materials are longer compared to planar substrate ALD.
The surface area to be coated is usually much larger, therefore requiring more reactant (see Section \ref{sec:surfacearea}).
This is particularly true for large batches of particulate material ($>5$\,g, blue in Figure~\ref{fig:pressure}), where reactants are dosed for minutes. 
In contrast, shorter reactant exposure times ($<$\,1\,min) are mostly chosen for smaller batches. 
In addition, the particulate materials are often porous in nature (see Figure~\ref{fig:SSAvsSize}) and longer exposure times are a means for reactants to reach into the pores through diffusion (see Section \ref{sec:pressure}).
Finally, longer exposure times can also be used to compensate for poor (or complete lack of) agitation of the support material during the ALD process (see for example Figure~\ref{fig:reactors_schematics}a and c). 
In this case, obviously much more reactant is dosed than would be needed theoretically to saturate the surface, and the excess reactant is lost.

Note that, in contrast, thorough mixing can lead to near-100\% precursor utilization and, especially if the support's surface area is much larger than the reactor's inner surface, negligible precursor losses \cite{2025YanguasGil,2015Grillo}.
This is particularly relevant for applications where ALD is used for its ability to save material through thin layers, e.g., for noble metals in (electro)catalysis.

The longer timescales compared to planar substrate ALD will also have to be considered when (re-)assessing some of the unwanted effects, such as the decomposition of precursor.

While exposure times (pulse durations) are reported and investigated regularly, i.e., regarding saturation, purge durations are often overlooked.
Albeit, the same mass transport limitations of reactant delivery (discussed above) in ALD on particles also apply to purging, leading to long purge durations.
Insufficient purging can lead to CVD-like behavior of the process, potentially giving inhomogeneous coatings \cite{2021vanOmmen}, for example through blocking of pores\cite{Weng2018a}.

\subsection{Characterization of supports \& coatings}\label{sec:characterization}

From the reviewed literature of ALD on particulate materials, we identified the most 12 commonly employed characterization techniques (Supporting Information Table~S1), and here provide a brief overview along with the respective quantities (Table~\ref{tab:characterization}).
Since we searched the literature for characterization techniques for only certain qualities and quantities, i.e., the imaging of coatings and the amount of deposited material, some prominent techniques from powder technology are not found in this ranking or appear in the lower positions, as indicated in footnotes.
Hence, we complemented Table~\ref{tab:characterization} accordingly with techniques for particle size and surface properties.

Some characterization techniques are more localized and aim at investigating individual particles (Table~\ref{tab:characterization}a). 
These allow forming a picture of the particles (often quite literally) and ALD coatings.
In contrast, powder and ensemble analysis techniques (Table~\ref{tab:characterization}b) are used to investigate the bulk of the powder or at least a representative powder sample.
We here make this distinction, since consistent coating results can be difficult to achieve in ALD on particulate materials, and proof thereof is facilitated through powder analysis techniques \cite{Puurunen2000a}. 

In the following, we will outline the practical benefits of the respective characterization methods for ALD on particles, regarding morphology (Section~\ref{sec:morphology}) and chemical composition (Section~\ref{sec:chemicalcomposition}).
A detailed examination of the mechanism of action of the respective technique is beyond the scope of this review.
A graphical overview covering some of the methods discussed here can be found in the Materials Characterization chart from EAG\cite{2024EAGbubblechart}, though not specific to particulate materials.
A particle-specific evaluation of a limited set of techniques in the context of battery research is found in Moryson et al.\cite{Moryson2021}.

\begin{center}
\footnotesize
{\setstretch{1.0}
\begin{longtblr}[
caption = {Commonly employed characterization methods in ALD on particulate materials and corresponding quantities. The typical ranges given here are relevant ranges for the scope of this review, and the attainable range of a given tool may be larger. DL: detection limit, SR: spatial resolution.},
label = {tab:characterization},
note{a} = {Top 12 technique identified from systematic evaluation of literature.},
note{b} = {Additional technique, and/or lower position.}
]
{
width=\textwidth, 
colspec={X[c,m]|X[c,m]|X[c,m]}}
 \SetCell[c=3]{c} \textbf{a) Individual particles \& microanalysis}\\\hline
\textbf{Method} & \textbf{Quantity} & \textbf{Typical range}
 \\\hline
 Scanning Electron Microscopy (SEM)\TblrNote{a}& Particle morphology&SR $\approx$ 10\,nm to 10\,µm\\
 Optical Microscopy\TblrNote{b}& Particle morphology&SR $\approx$ 0.5 µm to 100\,µm\\
 Transmission Electron Microscopy (TEM)\TblrNote{a}& Particle morphology, coating morphology&SR $\approx$ 1 nm to 1 µm\\
 High-resolution TEM (HRTEM)\TblrNote{a}& Coating morphology &SR at atomic level\cite{2009Williams}\\ 
 Scanning TEM (STEM)\TblrNote{a}& Coating morphology, material contrast, elemental mapping in combination with EDX&SR at atomic level\cite{2009Williams}\\ 
     Energy-dispersive X-ray spectroscopy (EDX, EDS)\TblrNote{a}& Elemental composition&DL $\approx$ 0.01\,wt\% to 1\,wt\%\cite{2009Williams,2018Goldstein}, SR\,$\approx$ 1\,nm to 1\,µm\cite{2009Williams}
 & &\\\hline
 & &\\
 \SetCell[c=3]{c} \textbf{b) Powder \& ensemble analysis}\nopagebreak\\\hline\nopagebreak
 \textbf{Method}& \textbf{Quantity}&\textbf{Typical range}\\\hline
 Sieving\TblrNote{b}& Particle size distribution&44 µm to 1 mm (325~to~16~mesh)\\

 Laser diffraction\TblrNote{b}& Particle size distribution&10\,nm to 1 mm\\
 Brunauer, Emmett, Teller (BET)\TblrNote{a}& Specific surface area&0.01 m$^2$/g to \>2,000\,m$^2$/g\cite{2014Rouquerol,2025HoribaBET}\\
 Thermogravimetric analysis (TGA)\TblrNote{a} & Composition & DL$\approx10$\,µg\\ 
 Combustion analysis (LECO)\TblrNote{a} & Elemental composition & DL\,$>0.6$\,µg\cite{2024LecoPeru}\\
Instrumental Neutron Activation Analysis (INAA)\TblrNote{a} & Elemental composition & DL = 0.1 to 1000\,ppm\cite{2016INAANist} (depending on element)\\
Nuclear magnetic resonance spectroscopy (NMR)\TblrNote{b} & Chemical structure & Quantitative structural information related to specific isotopes ($^{1}$H, $^{13}$C, $^{27}$Al, $^{29}$Si, …)\cite{2021Reif}\\
 Fourier Transform Infrared Spectroscopy (FTIR)\TblrNote{b} & Functional groups & DL varies depending on factors including bond polarity\cite{Weng2019}\\
Inductively coupled plasma mass spectroscopy (ICP--MS)\TblrNote{a}& Elemental composition & DL in ppb range\cite{2024EAGbubblechart}\\
 Atomic absorption spectroscopy (AAS)\TblrNote{a} & Elemental composition & DL in ppb range \\
 Inductively coupled plasma optical emission spectroscopy (ICP-OES, ICP-AES)\TblrNote{a}& Elemental composition &DL = 0.2 to 0.5\,wt\%\cite{2019Neikov} \\ 
 X-ray fluorescence spectroscopy (XRF)\TblrNote{a} & Elemental composition & DL in ppm range\cite{2001Rousseau}\\
X-ray photoelectron spectroscopy (XPS)\TblrNote{a} & Elemental composition (at surface), chemical state& DL = 0.1\,at\% to 1.0\,at\% \cite{2014Shard} $<10$\,nm~depth, SR\,$\approx100$\,µm (lateral)\\
Low Energy Ion Scattering (LEIS)\TblrNote{b} & Surface elemental composition & DL in ppm to at\%, depth $<10$\,nm\cite{2025Tascon}\\
Time-of-Flight Secondary Ion Mass Spectrometry (TOF--SIMS)\TblrNote{b} & Elemental composition & DL in ppm-ppb range, lateral     SR $<50$\, nm; depth SR $<1$\, nm \cite{2025Iontof}\\
 \hline
 \end{longtblr}}
\end{center}

\subsubsection{Morphology}\label{sec:morphology}
One obvious property is particle size, but on closer inspection it becomes clear that a simple definition of this property is only given for certain particles, e.g., the diameter of a spherical particle. 
Albeit, it is common practice to give the particle size in terms of a diameter, even for non-spherical particles. 
Although the diameter can be unambiguously defined as, e.g., the diameter of the smallest circumscribed perimeter or a volume-based equivalent diameter\cite{2019Neikov,2015Bagheri}, the definition is rarely reported in the ALD on particles literature.
In the field of powder technology, more attention is being paid to this problem, as discussed for example in Rhodes et al. \cite{2024Rhodes}.
Therefore, care must be taken when using the reported values for quantitative calculations of the powder properties.

Individual particles are readily observed and characterized using an optical or electron microscope (SEM or TEM) covering the spatial resolution (SR in Table~\ref{tab:characterization}a) appropriate for the given size range.
Note, that the SR ranges given here are rough estimates, and in reality depend on many factors \cite{2009Williams}.
To consider the often substantial variations in particle size in a powder, the size distribution can be derived from the microscopic images, provided they are representative, but powder analysis is often simpler and more time-efficient. Examples are laser diffraction or even sieving\cite{2019Neikov} (Table~\ref{tab:characterization}b), if a sufficient amount of powder is available.

For some support material-coating combinations with high material contrast (e.g. platinum on carbon), regular TEM imaging can provide valuable insights. 
However, for high resolutions or material combinations with lower contrast, more sophisticated derivatives of TEM are required, such as HRTEM for atomic resolution and STEM techniques (e.g. high-angle annular dark-field, HAADF) for material contrast. 
The latter can also be combined with EDX (see below) to obtain a spatial map of elements in the 2D image.

Closely related to the size and morphology of particles is the SSA per unit weight of powder which can be inferred from gas adsorption measurements as, e.g., in the Brunauer, Emmett and Teller method (BET)\cite{2014Rouquerol}.
For non-porous particles of known shape (requiring imaging techniques), the average particle size can be determined from the SSA\cite{2005Weibel}.
Gas absorption methods can also be used to obtain pore volumes, pore surface areas and pore diameters (e.g. Barrett, Joyner and Halenda method, BJH), however, requiring non-trivial calculations and reference measurements, depending on the specific method\cite{2014Rouquerol}.

\subsubsection{Chemical composition}\label{sec:chemicalcomposition}
The chemical composition can be obtained for individual particles, or for the particle ensemble average.
Starting again from the perspective of individual particles (Table~\ref{tab:characterization}a), a common technique to determine the elemental composition of a sample is energy-dispersive X-ray spectroscopy (EDX or EDS)\cite{2019Neikov}.
The sample is exposed to focused electron beams, which is why EDX is usually integrated with an electron microscope (SEM, or variations of STEM), and element-characteristic emission of X-rays is analyzed.
In principle, EDX provides spatially resolved information about the elemental composition of a sample, which entails the option to analyze individual particles and layer compositions.
However, due to the high energy of the electrons required to excite X-ray photon emission, the penetration depth of the beam is on the order of micrometers.
Therefore, the number of excited atoms is rather large in case of SEM-EDX, with a negative effect on spatial resolution, as unintended excitation of X-rays from the surrounding material adds to the background signal\cite{2018Goldstein}.
If the specimen is thin, as is the case typically in STEM-EDX, spatial resolutions as indicated in Table~\ref{tab:characterization}a can be reached.
A powder analysis technique (Table~\ref{tab:characterization}b) that similar to EDX relies on X-ray emission from the sample is X-ray fluorescence spectroscopy (XRF). 
In contrast to EDX, the sample is exposed to X-rays or gamma-rays to excite fluorescence, and the higher penetration depth allows detecting ensemble-averaged information of the sample while not providing spatial information. 

A more surface sensitive technique is X-ray photoelectron spectroscopy (XPS, Table~\ref{tab:characterization}b), with penetration depth below 10\,nm and detection limits of around 0.1 to 1.0\,at\%\cite{2014Shard}. Beyond that, XPS reveals information on the chemical state of surface-near atoms. 
Hence, besides deposited material, also the condition of the support material before the ALD process can be analyzed, such as the presence and density of functional groups which facilitate precursor adsorption.
However, due to the low lateral resolution, XPS provides ensemble averaged information on the powder surface and does not provide any information about individual particles.
In addition, the low penetration depth means that mainly atoms close to the surface are detected. Therefore, the XPS data on the atomic composition can only be used to calculate the mass fractions in special cases, e.g. for fine grades of carbon black. Similar characterization techniques such as low energy ion scattering (LEIS) and time-of-flight secondary ion mass spectrometry (TOF-SIMS) can also be used to determine the chemical composition of the sample with even higher surface sensitivity. 

Reliable methods to determine the bulk weight loading of an element (from coatings and other constituents) in a powder are inductively coupled plasma-optical emission spectroscopy (ICP-OES or interchangeably ICP-AES\cite{2021Nolte}) and -mass spectroscopy (ICP-MS).
Both methods involve dissolving (digesting) a small amount of powder (typically several 10\,mg) in a liquid and subsequently ionizing the sample to perform optical emission or mass spectroscopy. 
ICP-MS offers higher sensitivity and isotope analysis, while ICP-OES is typically cheaper.
A related method is atomic absorption spectroscopy (AAS).
Similar to the ICP methods, the sample is first dissolved in, e.g., aqua regina, and the element-specific absorption of light is measured by introducing the analyte into a flame\cite{Backman1998, Backman2000, Backman2009, Haukka1995a, Hirva1994, Hsu2012, Hsu2015, Nieminen1999, Puurunen2002a, Puurunen2003, Rautiainen2002, Suvanto1998}.

An alternative method to determine the elemental composition is instrumental neutron activation analysis (INAA), in which some stable isotopes of a given sample are converted to radioactive isotopes through irradiation with neutrons from a nuclear reactor. 
The sample can remain in powder form, hence INAA is essentially non-destructive.
However, not all elements form radioactive nucleotides under neutron irradiation and such elements cannot be detected. 

Combustion analysis, frequently referred to by the brand name LECO (for \textit{Laboratory Equipment Company}), is another option to determine the elemental composition for elements forming volatile compounds, e.g. carbon and hydrogen \cite{Ek2003a, Ek2003b, Kytokivi1997a, OToole2021a, Puurunen2001, Zhou2010a}.
The effluent combustion gases are analyzed via, e.g., infrared spectroscopy.
A related technique is thermogravimetric analysis (TGA), where the weight of a sample is continuously measured while increasing the temperature under inert or oxidative conditions\cite{2019Saadatkhah,Duong2022, Han2020, Jaggernauth2016, Lee2019c, Lee2020a, Lee2020b, Lee2022d, Lichty2013, Liu2020a, LopezdeDicastillo2019, Lu2017, Meng2015, Moseson2022, MunozMunoz2015, Shi2019, Sosnov2017, Sun2014, Swaminathan2023, VanBui2017, Wang2016a, Wang2017g, Xie2015a, Xie2015b, Yang2017b, Zhao2019c}.
If one constituent of the coated material (e.g., platinum on carbon, Pt/C) remains non-volatile (Pt) while the other forms a volatile compound under oxidative conditions (\ce{CO2}), the weight loading of the powder can be determined.
Note that the simpler method of weighing the powder before and after ALD (gravimetric method) only works for non-cohesive powders, where handling losses can be neglected.
In addition, TGA often allows some estimate of the chemical state of the powder constituents. 
For example, hydrocarbons will oxidize at lower temperatures than graphitized carbon, which can be observed in the characteristic sample weight vs. temperature curve of the sample.

Functional groups on high-surface-area materials before and after chemisorption reactions are commonly identified with Fourier transform infrared spectroscopy (FTIR) either in transmission or diffuse reflectance mode, and sometimes with solid-state nuclear magnetic resonance spectroscopy (NMR). FTIR sensitively detects functional groups with polar bonds such as hydroxyls, methyls and amines (e.g.\ Refs. \citenum{1993Haukka, Puurunen2000a}). FTIR can also be applied in a difference mode to detect changes from cycle to cycle (e.g., Ref. \citenum{2000Ferguson}). Also, probe molecules can be used in FTIR to reveal the characteristics of adsorption sites, such as the type of acidity via pyridine adsorption (e.g., Ref. \citenum{Weng2018b}). NMR tells about the chemical environment of specific isotopes ($^1$H, $^{13}$C, $^{27}$Al, $^{29}$Si, etc.). For example, sophisticated $^{29}$Si NMR  can  detect and quantify how many hydroxyl or methyl groups are attached to silicon before and after chemisorption  \cite{1993Haukka, Puurunen2000a, Pallister2014}. 

\section{Applications}\label{ch:applications}

\begin{figure}
  \includegraphics[width=0.7\textwidth]{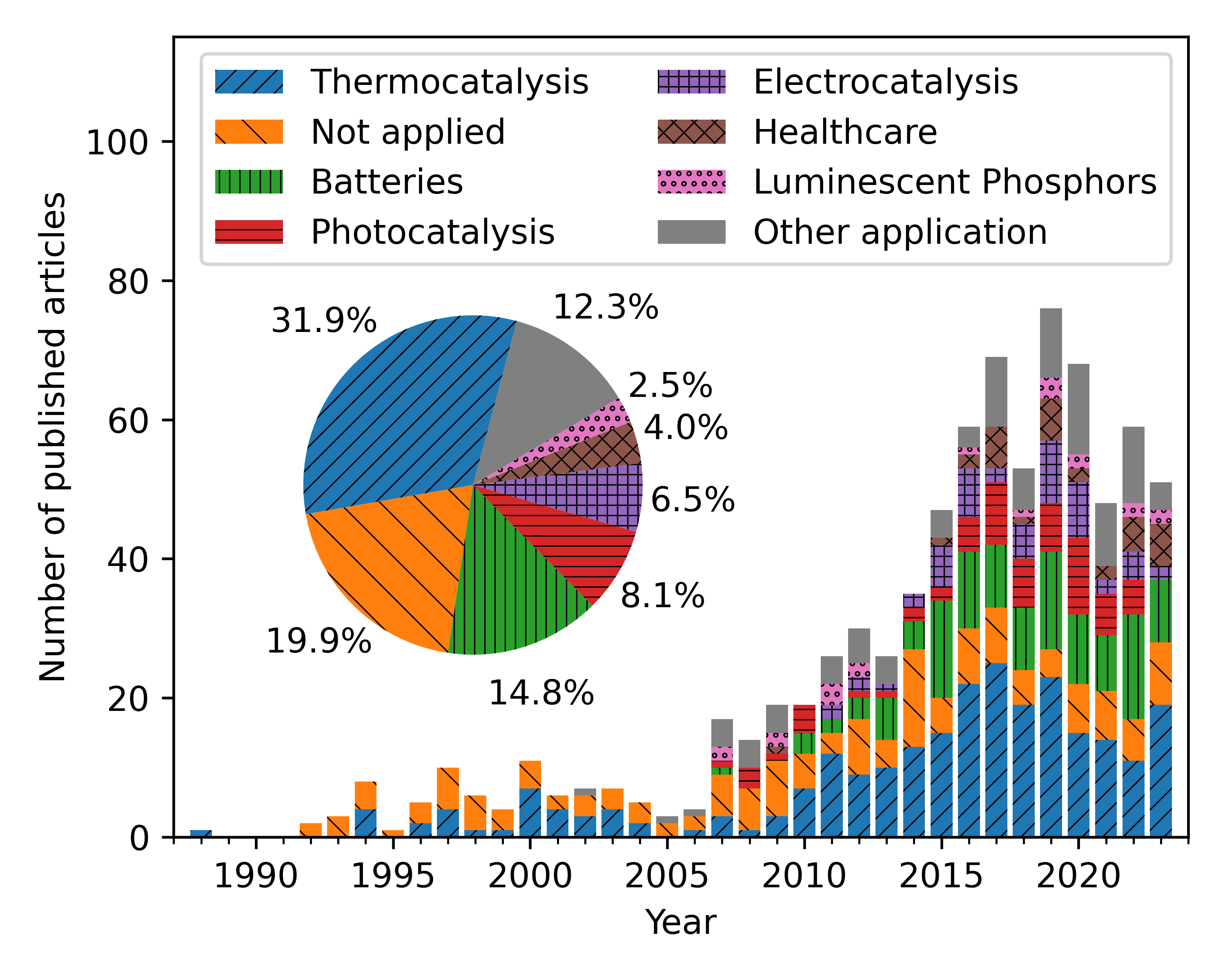}
  \caption{Number of articles published per year for ALD on particles in popular application fields. Inset: Cumulative share of applications for the period 1988 through 2023.}
  \label{fig:application}
\end{figure}

About 80\% of the articles within the scope of our data-driven review aim at a specific application, and we categorized the articles accordingly in Figure~\ref{fig:application} (pie chart, inset).
Historically, the field of ALD on particles was driven by thermocatalysis which remains the dominant application today (Figure~\ref{fig:application}).
However, starting from the late 2000s, also batteries, photocatalysis, medical, and luminescent phosphors appeared as applications; from 2011, electrocatalysis came up as an application for ALD on particulate materials.

Note that we required articles to show at least some application-relevant experimental data; in particular, the mere intention of using an ALD process for a given application was not sufficient to count the article as application-oriented. According to this requirement, about 20\% were counted as not applied.

In the following, we will summarize the respective developments of the main application fields.
Thermocatalysis is the dominant application for ALD on particles, has the longest history, and many of the basic concepts were developed in the context of thermocatalysis; we place emphasis on this topic, since many other applications draw from these developments.
This is particularly true for closely related application areas, i.e., photocatalysis and electrocatalysis.
ALD for batteries, including using particulate materials, has recently been reviewed extensively\cite{Lee2022i}, and here we limit our review to a short section.
Some less popular applications have yet to be reviewed thoroughly, and here we aim to provide a complete picture.
For these reasons, the subsections vary in length and set a different focus according to the respective application. 
A brief account of those applications not explicitly mentioned (others in Figure~\ref{fig:application}) will be given in the end of this section.

\begin{figure}
  \includegraphics[width=\textwidth]{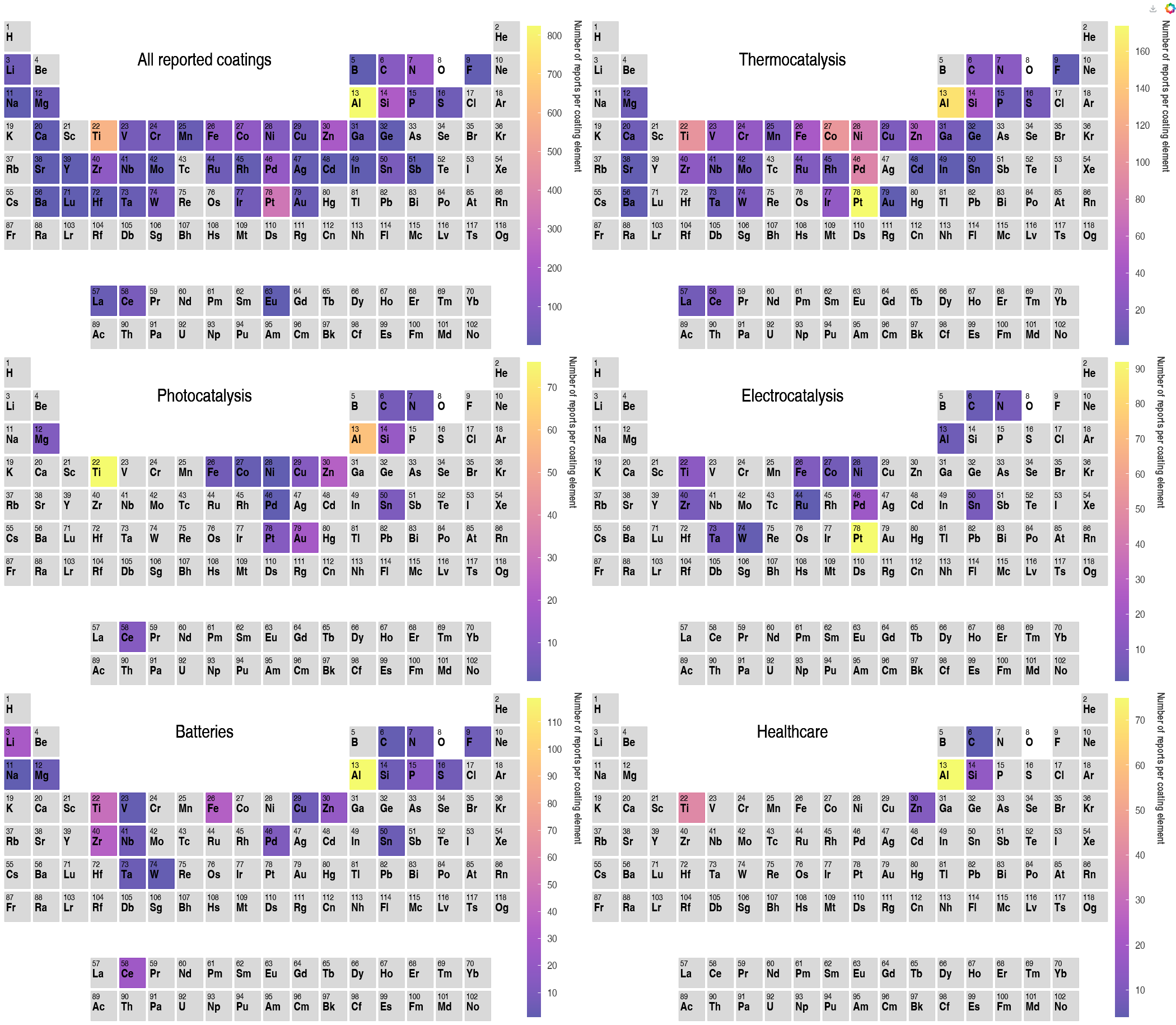}
  \caption{Chemical elements found in coatings from all ALD on particles articles covered in this review, and for specific applications. Note, that one article may include several coatings or the same coatings but different parameters, which are all counted separately. Oxygen is not quantified in this graph, as it would outshine all other elements.}
  \label{fig:periodic}
\end{figure}

\newpage\subsection{Thermocatalysis}\label{ch:thermocatalysis}

Catalysts increase the rate of chemical reactions, without affecting the chemical equilibrium and without being consumed in the process. Catalysts can be either in the same phase as the reactants, called \textit{homogeneous catalysts}, or in a different phase, called \textit{heterogeneous catalysts}. \textit{Biocatalysts} form yet their own class of catalysts, not discussed further here. \textit{Thermocatalysis} refers to catalytic processes where in addition to the catalyst only reaction temperature is used to control the activity (e.g., no electric voltage as in \textit{electrocatalysis}, or light as in \textit{photocatalysis}). Thermocatalysts are often simply referred to as "catalysts". Typically, industrial processes employ solid heterogeneous catalysts, which are easy to separate from the product mixture that usually consists of liquid phase, gas phase, or both. Solid heterogeneous catalysts characteristically consist of three types of components: \textit{active component}, \textit{promoter(s)}, and \textit{support}; not all catalysts have all three components. ALD has been used to make or modify all three types of catalyst components. In addition, overcoats (also called overlayers) have been applied by ALD to increase the stability of catalysts. 

According to the number of publications, thermocatalysis is the largest application area of ALD for particulate materials (Figure~\ref{fig:application}). In addition to the 255 primary publications, over forty review articles have already been published that, in one way or another, deal with catalysts made by ALD \cite{Lakomaa1994, 1995Haukka, 1996Malygin, Haukka1997b, Haukka1998, 2011Detavernier, 2012Stair, 2013Lakomaa, 2013Lu, 2013Zaera, 2014Dendooven, ONeill2015a, 2015Munnik, Malygin2015,  2015Sobel, 2016Lu, 2017Bui, 2017Ramachandran, 2017Singh, 2018Onn, 2018Cao, 2018Zhang, 2018Wang, 2019Mackus, 2019Chen, 2020DeCoster, 2021Otroshchenko, 2021Lu, 2021Fonseca, 2021Plutnar, 2021Lin, 2021Huo, 2021Xu, 2021Zaera, 2021Sarnello, 2022Zaera, 2022Li, 2022Hu, 2022bLu, 2023Lu, 2023Liu, 2023Zhou, 2023Dai, 2024Lausecker, 2024Olowoyo, 2024Zheng, 2024Abdelrahman, 2025Jung}. The main interest in using ALD for catalysis comes from the ability to synthesize catalysts with an exceptionally high level of control at the atomic scale. This control offers the possibility to improve catalytic activity, selectivity, and longevity, and paves the way to elucidating catalytic structure--activity relationships. The ALD synthesis is furthermore highly reproducible and scalable, does not use solvents, and is in principle applicable to a large part of the periodic table of elements (Figure~\ref{fig:periodic}; references \citenum{2005Puurunen, 2013Miikkulainen, 2025Popov}). 

\begin{figure}
    \centering
    \includegraphics[width=0.6\linewidth]{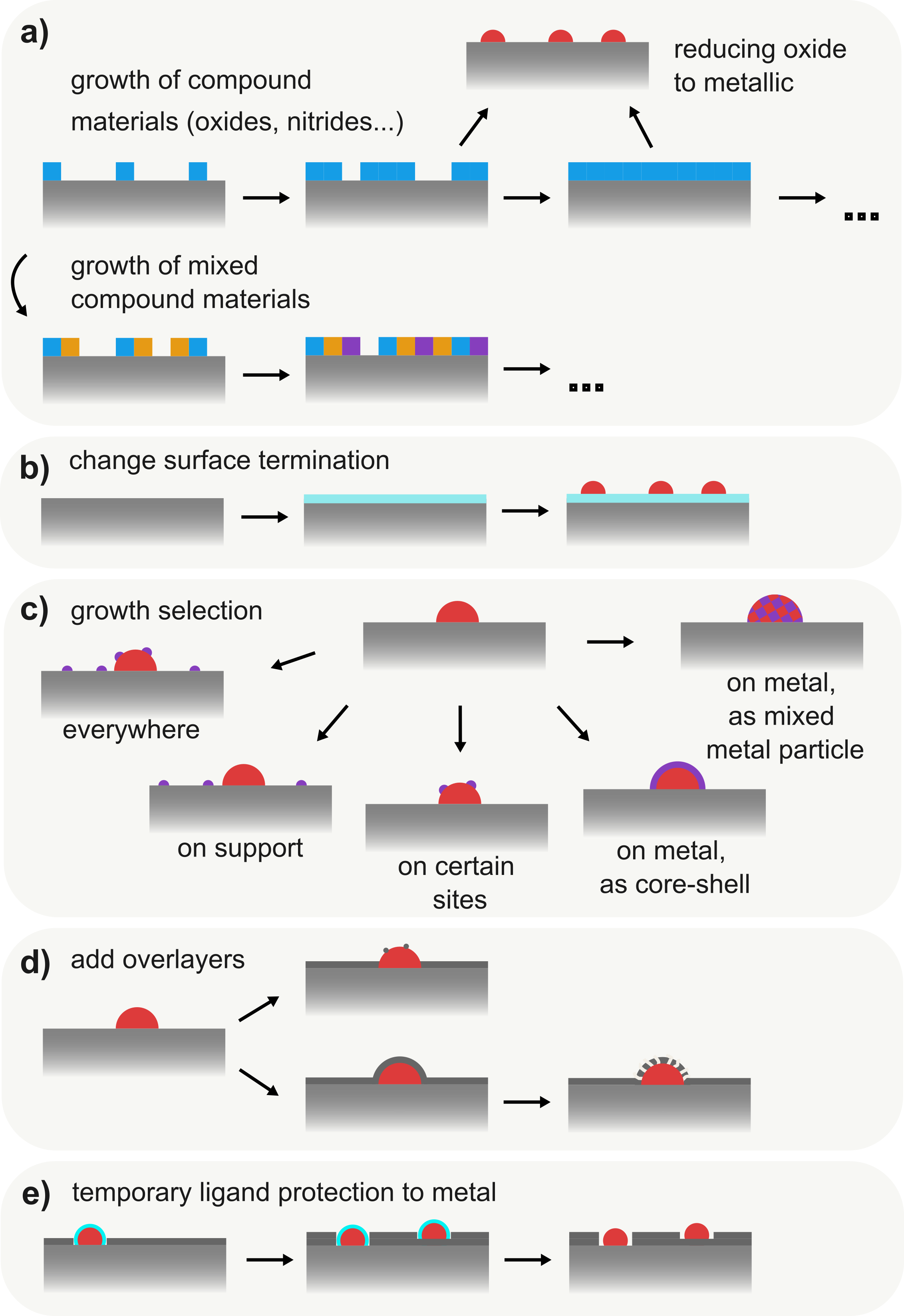}
    \caption{Various schemes of catalyst preparation by ALD, as reported in the literature. a) Growth of a compound layer in ALD cycles, either of the same material or mixed material. Optional reduction afterwards to make a metallic film. b) Change of the surface of a high-surface-area material to another, to be used as a support, with addition of active component on it afterwards. c) Variations of selective growth, with growth everywhere, or only on the support (typically oxide on oxide), growth on specific sites  of a metal nanoparticle (typically an oxide on undercoordinated sites such as edges and corners), and growth of metal on metal to create either a core--shell nanoparticle or a mixed alloy. d) Use of an overcoat (also called overlayer) to prevent catalyst deactivation, covering the support and the active component partly or fully. Heat treatment may be needed to expose the active component for catalysis through creation of porosity in the overcoat. e) ABC-type of sequence, where after the reaction of the metal compound of interest (A), ligand removal step is not made, but another compound layer is grown (BC). The sequence ABC can be repeated. Ligand removal step is performed as final step. Figure by the authors, distributed under a CC BY 4.0 license.\cite{WikimediaALDcatalyst}} \label{fig:ALD_catalyst_preparation}
\end{figure}

\begin{figure}
   \centering
  \includegraphics[width=0.75\linewidth]{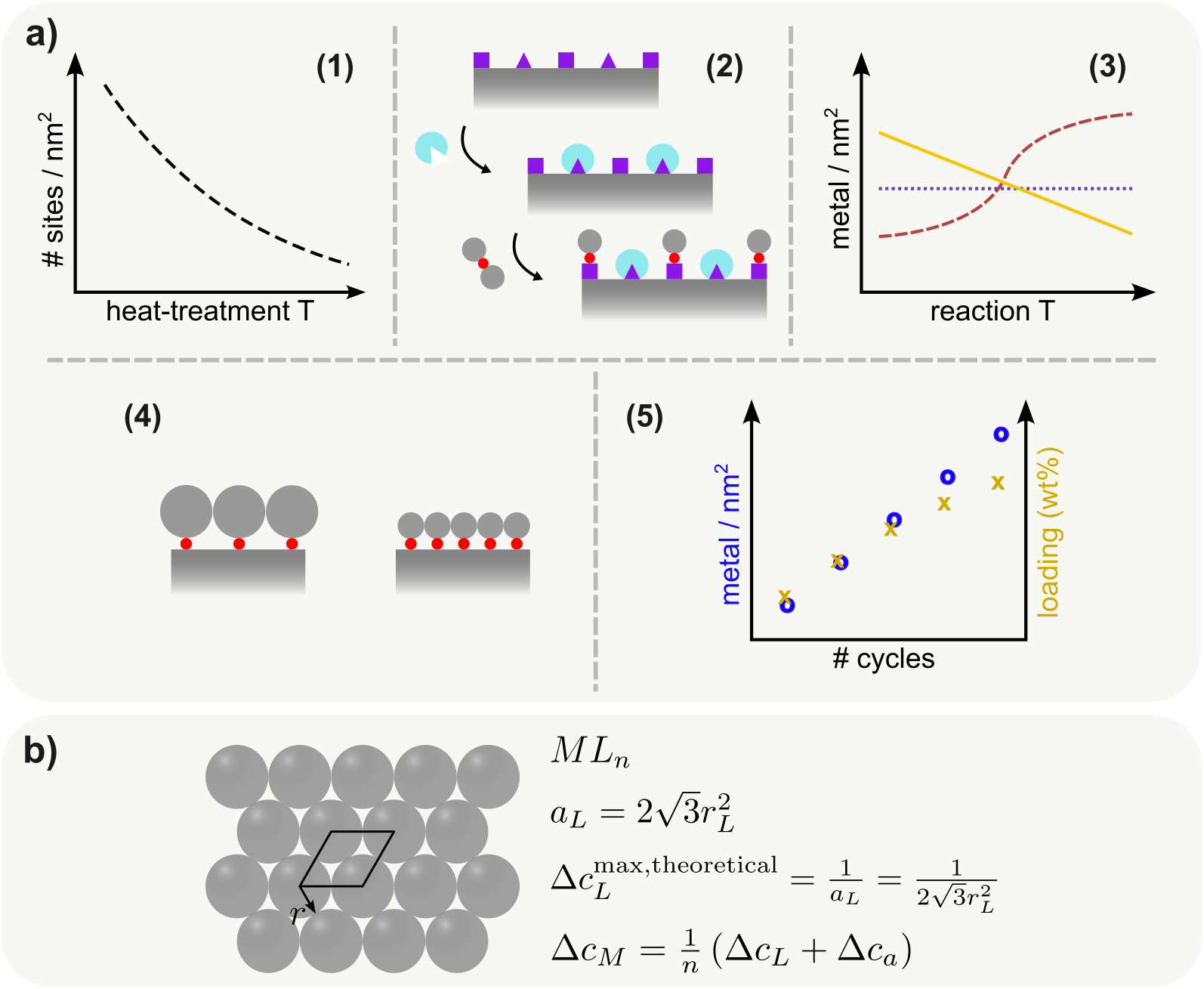}
 \caption{General ALD concepts developed in early catalysis-related studies, from Lakomaa \cite{Lakomaa1994} and Puurunen \cite{2003Puurunen,2003aPuurunen}. a) Five ways to control the metal loading obtained on particulate materials: (1) controlling the number of bonding sites, most typically hydroxyl groups on the support by heat treatment; (2) use other reactant molecules (also called inhibitors) to block selected bonding sites before binding the active metal species; (3) sometimes the reaction temperature can be used; (4) use the size of the molecule or its chemical character, and (5) grow compound layers (i.e., use repeated ALD cycles) to increase the metal concentration \cite{Lakomaa1994}. b) Model relating the size of ligand,  the number of ligands $n$ in the metal reactant molecule $ML_n$, and the reaction mechanism to the maximum theoretical resulting GPC  \cite{2003Puurunen,2003aPuurunen}. If the number of reactive sites on the surface is not limiting growth, steric hindrance of the ligands is limiting  \cite{2003Puurunen,2003aPuurunen}. Figure by the authors, distributed under a CC BY 4.0 license.\cite{WikimediaALDcatalysislearnings}} \label{fig:ALE_GPClearnings}
\end{figure}

In the next subsections, we first briefly review ALD for heterogeneous thermocatalysis from a historical perspective---sharing insights on the history never presented in a review article before---and thereafter provide a data-informed overview of the developments in the field. Because the field of ALD for thermocatalysis is already large, and making an exhaustive review of individual applications was not the purpose of this review of ALD on particles, individual articles are cited in the following subsections as mere examples. Through the section, we will discuss and give examples of design strategies that have been developed in the context of ALD for thermocatalysis, which are of general relevance to the field of ALD on particulate materials (Figures~\ref{fig:ALD_catalyst_preparation} and \ref{fig:ALE_GPClearnings}). 

\subsubsection{Brief history}

\begin{figure}
    \centering
    \includegraphics[width=0.75\linewidth]{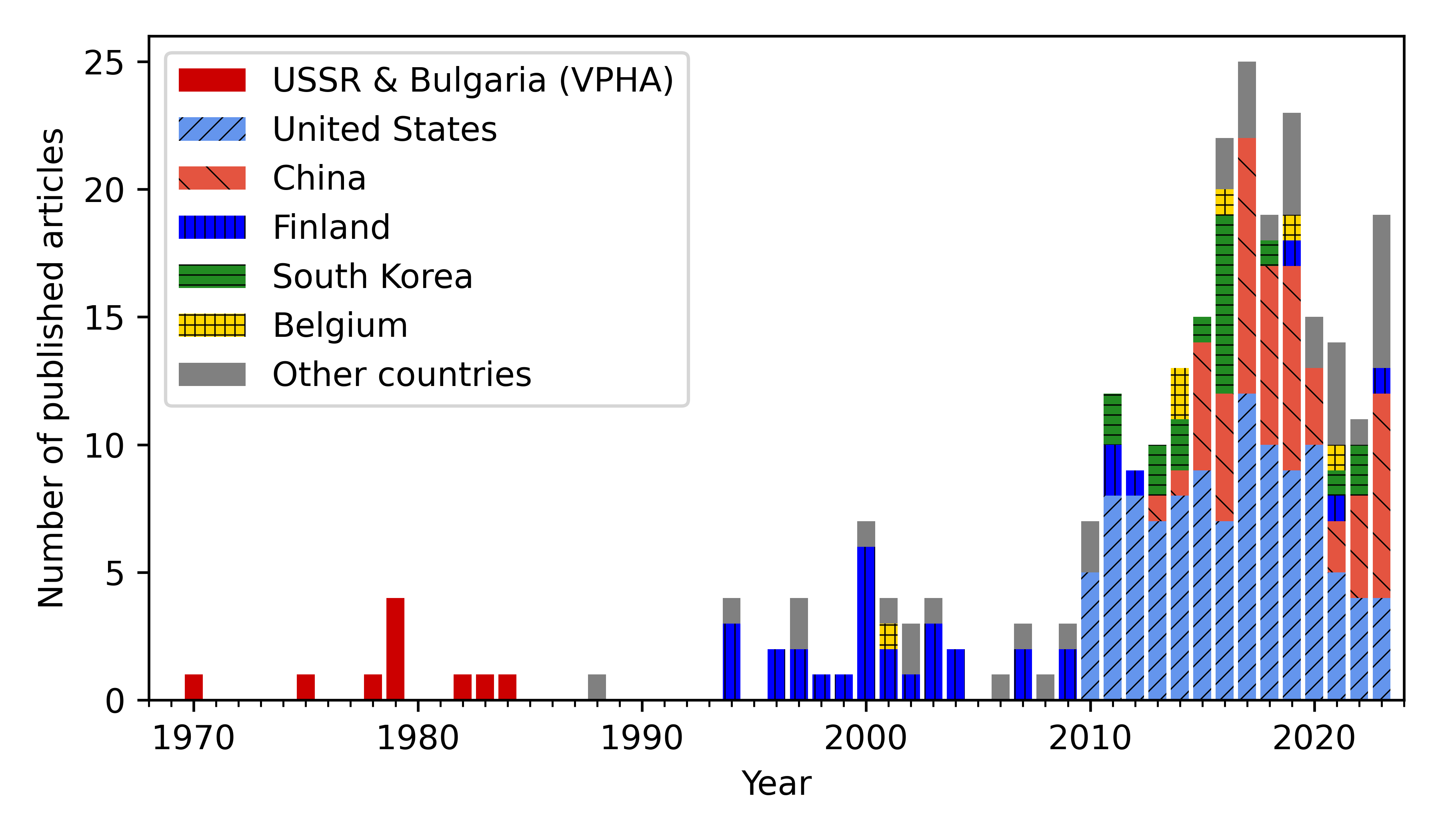}
    \caption{Number of thermocatalysis ALD publications in the dataset, per year and per country (five highest numbers of publications). The articles from USSR \& Bulgaria did not show up in the regular literature search forming the dataset, and have been added manually added on the basis of information collected in the VPHA.\cite{2013VPHA}}
    \label{fig:tc_timeline}
\end{figure}

Thermocatalysis is the most historical application area of ALD for particulate materials, although this appears not to be generally known in the field of ALD (or catalysis). Figure~\ref{fig:tc_timeline} presents the number of thermocatalysis-related publications per year per country from the current dataset, topped up with information collected from the VPHA until 1986.\cite{2013VPHA} The first catalyst ALD studies date back to the 1970s in the USSR, where ML--ALD research activities continued through the 1980s. The first publication from outside of the ML community (located in the USSR and Eastern Europe) came from Japan in 1988.\cite{Asakura1988b} More publications per year started appearing in the 1990s, especially from Finland. The first decade of 2000s continued with a rather similar publication volume per year, Finland still dominating the numbers but other countries appearing, too, e.g. Belgium. Starting from 2010, the number of publications per year suddenly increased, first from the USA and thereafter from China. Another country with notable activity in the past decades in ALD for thermocatalysis has been Belgium, and one with many publications in the catalysis ALD field since the 2010s is South Korea.

In both historical branches of ALD---atomic layer epitaxy (ALE) and molecular layering (ML)---catalysis was among the early applications \cite{2014Puurunen,Malygin2015}. Unawareness of the history of catalysis ALD is aligned with more general developments in the field of ALD, where the early historical developments, in particular of the ML branch, have been overlooked for a long time \cite{2005Puurunen, 2017Ahvenniemi, 2018Puurunen, Parsons2013NOTE, 2024ALDnoteML, CatALDhistoryNOTE}. In an attempt to bridge the fundamental knowledge gap regarding the early history of ALD for catalysis, we provide a brief but detailed summary of both the historical development branches of catalysis ALD, before we describe ALD for thermocatalysis more generally. 

First, about the older and less-known historical branch of ALD, molecular layering. The ML--ALD developments started on particulate materials in the 1960s,\cite{2005Puurunen, Malygin2015} and the first known catalyst ML--ALD studies date back to year 1970.\cite{VPHAKoltsov1970h} At the university "Lensovet Leningrad Technological Institute" in St.\ Petersburg, USSR, Koltsov et al.\cite{VPHAKoltsov1970h} studied the decomposition of CCl$_4$ through catalytic hydrolysis to carbon dioxide and water, using silica modified with phosphorus oxide and titanium dioxide by ML--ALD as a catalyst. Importantly, already this first publication\cite{VPHAKoltsov1970h} is available as an English translation.\cite{MLalsoInEnglishNOTE} The modifications were carried out at 180$^\circ$C, using PCl$_3$, TiCl$_4$ and water as reactants; modification with phosphorus oxide was repeated up to six cycles and modification with titanium dioxide up to ten cycles.\cite{VPHAKoltsov1970h} This is an early example of the growth of a mixed oxide material by ALD, as in Figure \ref{fig:ALD_catalyst_preparation}a. Interestingly, for both modifications (phosphorous and titanium oxide), highest rate constants for CCl$_4$ decomposition were found for one layer (ML--ALD cycle) of modification.\cite{VPHAKoltsov1970h} Follow-up studies were published (to our knowledge only in Russian) on CCl$_4$ hydrolysis:  mixed oxides were made and it was found that the order of adding the materials affects the catalytic activity.\cite{VPHAKoltsov1973c, VPHASmirnov1974a} Several other catalytic reactions were later studied, where ML--ALD modifications have typically been made from the inorganic metal chloride reactants and water: vanadium oxide and phosphorus-modified oxidation catalysts e.g. for benzene to maleic anhydride,\cite{VPHAMalygin1974, VPHABienert1974, VPHAKhalif1977, VPHAHanke1978a, VPHAPostnova1978} chromia/silica and Ti- or V-modified silica for ethylene polymerization.\cite{VPHADamyanov1975, VPHAMehandjiev1979, VPHADamyanov1976a, VPHADamyanov1979a, VPHADamyanov1979b} A fluidized bed was used, and a related mathematical modelling was made.\cite{VPHAYakovlev1979} Research was not only carried out the USSR, but also in the DDR (e.g. Bienert et al.\cite{VPHABienert1974} and Hanke et al.\cite{VPHAHanke1978a}) and Bulgaria (e.g. Damyanov et al.\cite{VPHADamyanov1975} and Mehandhiev et al.\cite{VPHAMehandjiev1979}). Several academic theses were completed, including a habilitation thesis;\cite{1988DamyanovHabil} for a comprehensive list of theses, the reader is referred to an essay on ML-ALD  by Malygin et al. from 2015.\cite{Malygin2015} 

Second, about the other, in general, better-known historical development branch of ALD, atomic layer epitaxy. After the commercialization of the first targeted application of ALE--ALD thin-film based electroluminescent flat panel displays (work started in 1974 and entered commercial production in 1984), particle-based catalysis became a new targeted application area (along with thin-film photovoltaics). \cite{2014Puurunen} This work began as an industrial endeavor with the founding of Microchemistry Ltd. in 1987 as a jointly owned company by Neste (Finnish oil company) and Lohja (owner of EL flat panel production) with Tuomo Suntola (the inventor of ALE) leading it.\cite{2014Puurunen} ALD catalysts prepared at Microchemistry were tested at Neste.\cite{2013Lakomaa, 2024Krause} In the first study, elemental zinc was added to ZSM-5 zeolite (in the exhaust tube of a F-120 reactor, as no dedicated reactors were available yet for particulate materials);  the catalytic activity was tested in butane aromatization, with promising results.\cite{2013Lakomaa, 2024Krause}  Catalyst development was continued, and a modification was made to the F-120 to allow ALD on porous particulate materials in a fixed bed with gases flowing through it.\cite{2013Lakomaa, 2024Krause} A patent application on ALD for catalysis was filed by Neste company on January 16, 1990,\cite{1990PatentFI84562} describing Zn/ZSM-5 for butane aromatization; Re/Al$_2$O$_3$; and Cr/SiO$_2$ for ethylene polymerization. In the examples mentioned in the patent, zinc was added on the zeolite from Zn or ZnCl$_2$; rhenium on alumina from Re$_2$O$_7$ and magnesium additive from Mg(thd)$_2$; and chromium on silica from CrO$_2$Cl$_2$ and titanium and aluminum additives from TiCl$_4$ and AlCl$_3$.\cite{1990PatentFI84562} All these are examples of catalyst preparation according to Figure \ref{fig:ALD_catalyst_preparation}a. 

A period of active research followed, and more institutes in Finland and elsewhere became involved. A master's thesis at Helsinki University of Technology (HUT) in 1992 considered the scale-up of catalyst synthesis by ALD up to bench scale (150 g of support),\cite{Lujala1992Msc}  followed by a general particle ALD article by Lakomaa in 1994.\cite{Lakomaa1994} The first case study of ALE--ALD catalysts with scientific publications is for alumina-supported nickel catalysts, tested for toluene hydrogenation.\cite{1993Lindfors, Lindfors1994, Lindblad1994} Nickel was added by repeating cycles of Ni(acac)$_2$ and synthetic air, up to six cycles. For activation, nickel was reduced with hydrogen, representing an early example of the growth of an oxide reduced to metal (Figure \ref{fig:ALD_catalyst_preparation}a). The size of the metal precursor was found to affect the areal density of nickel chemisorbed at saturation, the smaller acac ligand (acac = acetylacetonate) giving a higher loading than the larger thd (thd = 2,2,6,6-tetramethyl-3,5-heptanedionate) \cite{Lakomaa1994} (Figure{\ref{fig:ALE_GPClearnings}, panels (a4) and (b)). Other examples of catalysts made are cobalt for toluene hydrogenation, with cobalt added from Co(acac)$_2$ and Co(acac)$_3$ on silica and alumina (reduced for activation as catalyst);\cite{Backman1998, Backman2000, Backman2001, Rautiainen2002, Backman2009}  chromia for dehydrogenation, with chromium on alumina from Cr(acac)$_3$.\cite{Kytokivi1996a,Hakuli2000} Related to noble metals, the very first attempt is for palladium in  bimetallic Cu/Pd binary alloy catalysts  from 1998.\cite{Molenbroek1998} Thermal properties of the volatile $\beta$-diketonate complexes of ruthenium, palladium and platinum were characterized,\cite{2001Lashdaf} Pd and Ru catalysts made from $\beta$-diketonates on alumina and silica and Pt from (trimethyl)methylcyclopentadienylplatinum(IV) on $\beta$ zeolite and tested for activity in cinnamaldehyde hydrogenation.\cite{Lashdaf2003b,Lashdaf2003a, Lashdaf2004a, Lashdaf2004b} Also Pt(acac)$_2$ on carbon nanofibers was tested,\cite{2008Plomp} and iridium on various supports\cite{Silvennoinen2007a, Silvennoinen2007b, Vuori2009, Vuori2011} with catalytic testing for toluene hydrogenation \cite{Silvennoinen2007b} and decalin ring-opening.\cite{Vuori2009} To reduce the obtained iridium loading, acetylacetone (Hacac) was used, proving effective (90\% reduction of Ir content) on alumina, less effective on silica--alumina, and having almost no effect on  silica \cite{Silvennoinen2007a} (Figure{\ref{fig:ALE_GPClearnings}a2). Experiments were also made with tungsten from WOCl$_4$;\cite{Lindblad1993} for silica on alumina (typical acid catalyst) from hexamethyldisiloxane (HMDS) and air;\cite{Lindblad1998} for modifying silica and alumina with an AlN and adding cobalt and chromia on the AlN-modified surface by ALD. \cite{Puurunen2000a, Puurunen2000b, Puurunen2001, Puurunen2002b, Puurunen2003} The AlN modification represents an early example of the growth of changing the surface termination by ALD (Figure \ref{fig:ALD_catalyst_preparation}b).  For the AlN modification, trimethylaluminium and ammonia were used, with a significantly higher reaction temperature for ammonia (550$^\circ$C) than for TMA (150$^\circ$C).\cite{Puurunen2000b, Puurunen2001, Puurunen2002b}  Several PhD-level theses came from HUT\cite{1999HakualiDr, 2002PuurunenDr, 2004LashdafDr, 2009BackmanDr, KrauseHUTnote} and one at Åbo Akademi.\cite{1994LindforsPhD} Haukka et al. overviewed the early ALE--ALD catalyst preparation related learnings in 1995,\cite{1995Haukka} 1997,\cite{Haukka1997b} and 1999;\cite{Haukka1998} Krause gave an invited lecture on ALD for catalyst preparation at the EuropaCat 2005 conference,\cite{2005KrauseEuropaCat} and Lakomaa et al. published an overview in a book chapter in 2013.\cite{2013Lakomaa} 

In addition to the pioneering works in the USSR and Finland, there have been noteworthy early works on ALD related to thermocatalysis at least in Japan and the Netherlands. Japanese researchers from the University of Tokyo worked on ALD catalysts already in late 1980s, making zirconium oxide through cycles of zirconium ethoxide and calcination, to modify ZSM-5 zeolite with zirconia (the first article in the current dataset).\cite{Asakura1988b} The catalyst was found to produce isopentane selectively from methanol.\cite{Asakura1988b} In another work, titania was grown on silica from titanium isopropoxide and calcination.\cite{Asakura1992} It was found that the formed anatase could be transformed to rutile by interaction with Pd particles.\cite{Asakura1992, CatALDcitationJAPANnote} Researchers from Shell, The Netherlands, made titanium-on-silica catalysts by an ALD-like process, which showed high activity in propylene epoxidation \cite{2004Buijink}. 

\subsubsection{Design strategies from thermocatalysis}\label{sec:thermocatalysisdesgin}

The research on catalyst preparation by ALD resulted in publications with concepts of general significance not only to the field of ALD for particulate materials, but in general ALD. 
This concerns fundamental studies on controlling material growth, but also material design strategies for thermocatalysts that have been transferred to other applications.

Lakomaa in 1994\cite{Lakomaa1994} presented five ways to control the areal number density (then called surface density) of metal compounds obtained on silica and alumina particles. 
These five ways are still valid and apply to other support materials as well, and are illustrated in Figure~\ref{fig:ALE_GPClearnings}a: 
(1) controlling the number of bonding sites, most typically hydroxyl groups on the support by a heat treatment (Figure~\ref{fig:ALE_GPClearnings}a1); 
(2) use other reactant molecules (also called inhibitors) to block selected bonding sites before binding the active metal species (Figure~\ref{fig:ALE_GPClearnings}a2); 
(3) sometimes the reaction temperature can be used (Figure~\ref{fig:ALE_GPClearnings}a3); 
(4) use the size of the molecule or its chemical character (Figure~\ref{fig:ALE_GPClearnings}a4); and (5) grow compound layers (i.e., use repeated ALD cycles) to increase the metal concentration (Figure~\ref{fig:ALE_GPClearnings}a5).\cite{Lakomaa1994}
Note that for a constant GPC per unit area and linear increase of total areal number density of the metal (M/nm$^2$), the weight-based loading decreases in steps of decreasing size, because of the mass that is added on the surface in each ALD cycle (Figure~\ref{fig:ALE_GPClearnings}a5)\cite{2005Puurunen,Baumgarten2022}. 
Pre-blocking of reactive sites to reduce the GPC has become an active research field in area-selective ALD, where the aim is to completely block the reactions and to produce a "non-growth surface" next to a chemically dissimilar "growth surface"; 
similar molecules were initially used in the catalyst research from 1990s, such as hexamethyldisilazane (HMDS) \cite{1993Haukka} and acetylacetone (Hacac) \cite{Kytokivi1997a,Silvennoinen2007a}, and the selection has further expanded \cite{2013YanguasGil,2024Yu}. 

Another concept of general applicability stemming from ALD catalyst research is the geometric model to analyse GPC from size and number of ligands, and assumed reaction mechanisms (Figure~\ref{fig:ALE_GPClearnings}b).\cite{2003Puurunen,2003aPuurunen}
The model rationalizes how the same amount of ligands per unit area could be obtained on many different supports and reaction conditions, while at the same time the amount of metal adsorbed per unit area could vary according to the number of ligand-releasing reactive sites \cite{Haukka1998, 2002PuurunenDr, 2005Puurunen, 2005bPuurunen}. 

\subsubsection{Data-informed analysis}

Since approximately 2010, ALD for thermocatalysis has been the subject of intense global interest (Figure~\ref{fig:tc_timeline}).
While during the 1990s, four review type articles were published\cite{Lakomaa1994, 1995Haukka, 1996Malygin, Haukka1998} (and to our knowledge none in the first decade of 2000s), since 2010 the number of review articles published fully or partly about catalysis ALD is rapidly increasing and today counted in tens.\cite{2011Detavernier, 2012Stair, 2012Wegener, 2013Lakomaa, 2013Lu, 2013Zaera, ONeill2015a, 2015Munnik, Malygin2015,  2015Sobel, 2016Lu, 2017Bui, 2017Ramachandran, 2017Singh, 2018Onn, 2018Cao, 2018Zhang, 2018Wang, 2019Mackus, 2019Chen, 2020DeCoster, 2021Otroshchenko, 2021Lu, 2021Fonseca, 2021Plutnar, 2021Lin, 2021Huo, 2021Xu, 2021Zaera, 2021Sarnello, 2022Zaera, 2022Li, 2022Hu, 2023Lu, 2023Liu, 2023Zhou, 2023Dai, 2024Lausecker, 2024Olowoyo, 2024Zheng, 2024Abdelrahman, 2025Jung}.
While a few review articles deal with a specific historical branch of ALD (ML or ALE),\cite{Lakomaa1994, 1995Haukka, 1996Malygin, Haukka1998,2013Lakomaa, Malygin2015} several reviews have dealt with catalysis ALD in general.\cite{2011Detavernier, 2012Stair, 2013Lu, ONeill2015a, 2016Lu, 2017Ramachandran, 2017Singh, 2018Wang, 2021Lu} 
Additionally, there are reviews dealing with ALD among other catalyst preparation techniques;\cite{2012Wegener, 2013Zaera, 2015Munnik, 2020DeCoster, 2021Otroshchenko, 2021Lin, 2021Zaera, 2022Zaera, 2022Li, 2022Hu} 
reviews specific to ALD where catalysis among the  applications;\cite{2015Sobel, 2017Bui, 2018Onn, 2019Mackus, 2024Lausecker} and reviews dealing with specific type of ALD for catalysis or specific application by catalysis with focus on ALD, such as area-selective ALD,\cite{2018Cao} tailoring the metal--oxide interface by ALD,\cite{2018Zhang}  ALD-made catalysts for carbon dioxide reduction,\cite{2019Chen, 2024Olowoyo} ALD for single-atom catalysts,\cite{2021Fonseca} ALD for energy conversion,\cite{2021Plutnar, 2025Jung} ALD-based protective overcoats,\cite{2021Sarnello, 2023Lu} hydrothermal stability of catalysts,\cite{2021Huo} zeolite modification by ALD,\cite{2021Xu} catalytic and energy materials for eco-friendly vehicles,\cite{2023Liu} metal-organic frameworks,\cite{2023Zhou,2024Zheng} and recent research especially on heterogeneous catalysis, especially electrocatalysis.\cite{2023Dai} 
Also, one career perspective has been published, where ALD for catalysis is in focus.\cite{2024Abdelrahman}

According to the periodic table frequency map of Figure~\ref{fig:periodic},  the most popular elements grown by ALD for thermocatalysis are platinum, aluminum,  cobalt, titanium, nickel and palladium, other popular elements are zinc, iron, chromium, vanadium, zirconium, iridium and tungsten. Some examples of the use of these elements in thermocatalysis are as follows; note that an exhaustive review is out of the scope of this work. Platinum is a multipurpose catalyst, and ALD-made Pt-thermocatalysts have for example been applied for the liquid-phase reforming of 1-propanol, where Pt/titania was found especially active.\cite{Lobo2012} Aluminum can be used to make alumina as an overcoat to postpone Pt and Pd catalyst deactivation by sintering and coking \cite{Lu2020,Lu2012a}, and also applied on to increase the acidity of silica, zeolites and similar supports \cite{2011Sree,Vandegehuchte2014, Weng2019}. Growth of alumina takes place both on the support and on the metal, and selective growth on undercoordinated sites (edges, corners) sites has been suggested \cite{2010Lu}. In addition to toluene hydrogenation \cite{Backman1998,Backman2000,Backman2001,Backman2009}, cobalt has been used in metal-organic-frameworks for the oxidative dehydrogenation (ODH) of propane to propene under mild conditions,\cite{2017Li} and as bimetallic Co--Pt catalysts for selective hydrogenation optionally with ALD titania \cite{2019Zhang}. Titanium is used, for example, in protective titania coatings for single-atom palladium catalysts \cite{PiernaviejaHermida2016}. Nickel has been used in addition to the early application for toluene hydrogenation as dry reforming catalyst \cite{Shang2017}, zinc for promoting copper catalysts for carbon dioxide to methanol,\cite{Arandia2023} and vanadium for dehydrogenation and selective oxidation.\cite{Keranen2002, 2004Gervasini} Supported palladium nanoparticles have been prepared by the ABC-type scheme with protective ligands left in step A and alumina or titania grown in steps BC of the cycle (Figure \ref{fig:ALD_catalyst_preparation}e) \cite{2010Lu,Lu2010}, and also used as coking- and sintering-resistant catalysts with a protective overcoating in the oxidative dehydrogenation of ethane (Figure \ref{fig:ALD_catalyst_preparation}d) \cite{Lu2012a}. Bimetallic Ru--Pt nanoparticles grown through selective ALD of metal on metal (Figure \ref{fig:ALD_catalyst_preparation}c) are active e.g. in methanol decomposition \cite{Christensen2010,Lu2014}. The growth of iron oxide selectively through catalytic oxygen activation on Pt and not e.g. on silica, presents an elegant example of area-selective ALD (Figure \ref{fig:ALD_catalyst_preparation}c) \cite{2018Singh}. An example of changing the surface termination (Figure \ref{fig:ALD_catalyst_preparation}b) is the growth of niobia on mesoporous silica, used as a hydrothermally stable acidic mesoporous catalyst support for Pd catalysts for $\gamma$-valerolactone to pentanoic acid conversion \cite{PaganTorres2011}. 

Data-driven analysis in this work allows examination of typical features of an ALD experiment for thermocatalysis. Figure~\ref{fig:datadriven_thermocatalysis} shows a compilation of some trends, and further details are presented in the supplementary information. Most typical support materials are oxides, but also carbons and metals on a support are often used (Figure~\ref{fig:datadriven_thermocatalysis}c). Catalyst support particles with average diameter between 0.1 and 1 mm are typically used (Supporting Information Figure~S5). While most often, the supports are mesoporous oxides (average pore size 2--30\,nm; Figure~\ref{fig:datadriven_thermocatalysis}g) with specific surface area of some hundreds of square meters per gram (Figure~\ref{fig:datadriven_thermocatalysis}g), microporous materials (zeolites, metal-organic frameworks, carbons) are also frequently used. Because of the high aspect ratios (AR) given by the ratio of the particle radius to the pore size, typically in the range of thousands (e.g 100\,µm radius and 10\,nm pore size give AR of 10,000:1), thermocatalysis ALD almost exclusively operates under diffusion-limited conditions (Thiele modulus well over one \cite{2024Gonsalves}), Kundsen diffusion most often dominating (Figure~\ref{fig:pressuretransport}). Consequently, unless process conditions (exposure) are properly tuned, an egg-shell distribution of metal within the catalyst particle will necessarily result instead of the desired (and perhaps assumed) uniform distribution.\cite{2024Heikkinen,2024YimDr} The typical batch size varies from 0.1 to 10\,g (Figure~    \ref{fig:datadriven_thermocatalysis}d), and the total surface area coated at once is calculated as hundreds or thousands of square meters (Figure~\ref{fig:datadriven_thermocatalysis}e). Before ALD, the particles have been heat-treated (the heat-treatment temperature varies wildly and can be up to a thousand degrees Centigrade, Supporting Information Figure~S8);  the ALD temperature is mostly within 100--350\,$^\circ$C, although even temperatures above 800\,$^\circ$C were occasionally employed (Supporting Information Figure~S9). Most typically, just one ALD reaction cycle, or perhaps just the first reaction of a reaction cycle, is made (Supporting Information Figure~S10). In case more than one ALD reaction cycle is made, the number of cycles typically does not exceed ten (although there are examples up to some tens of cycles, Supporting Information Figure~S10). This is because catalysis is a surface phenomenon and modification of the surface beyond monolayer coverage is seldom useful (overcoats are an exception, Figure~\ref{fig:ALD_catalyst_preparation}d). One ALD reaction cycle takes typically minutes to complete (Supporting Information Figure~S11). The GPC in terms of areal number density (metal atoms per unit surface area; adsorption capacity) is no more than six per square nanometer, and typically between one and two (Figure~\ref{fig:datadriven_thermocatalysis}b).\cite{ThermocatalysisGPC} In terms of mass loading (wt\%), typical values are around one wt\% of metal, although there are reports of preparing several tens of wt\% as well (Supporting Information Figure~S12).   

\begin{figure}
    \centering
    \includegraphics[width=1\linewidth]{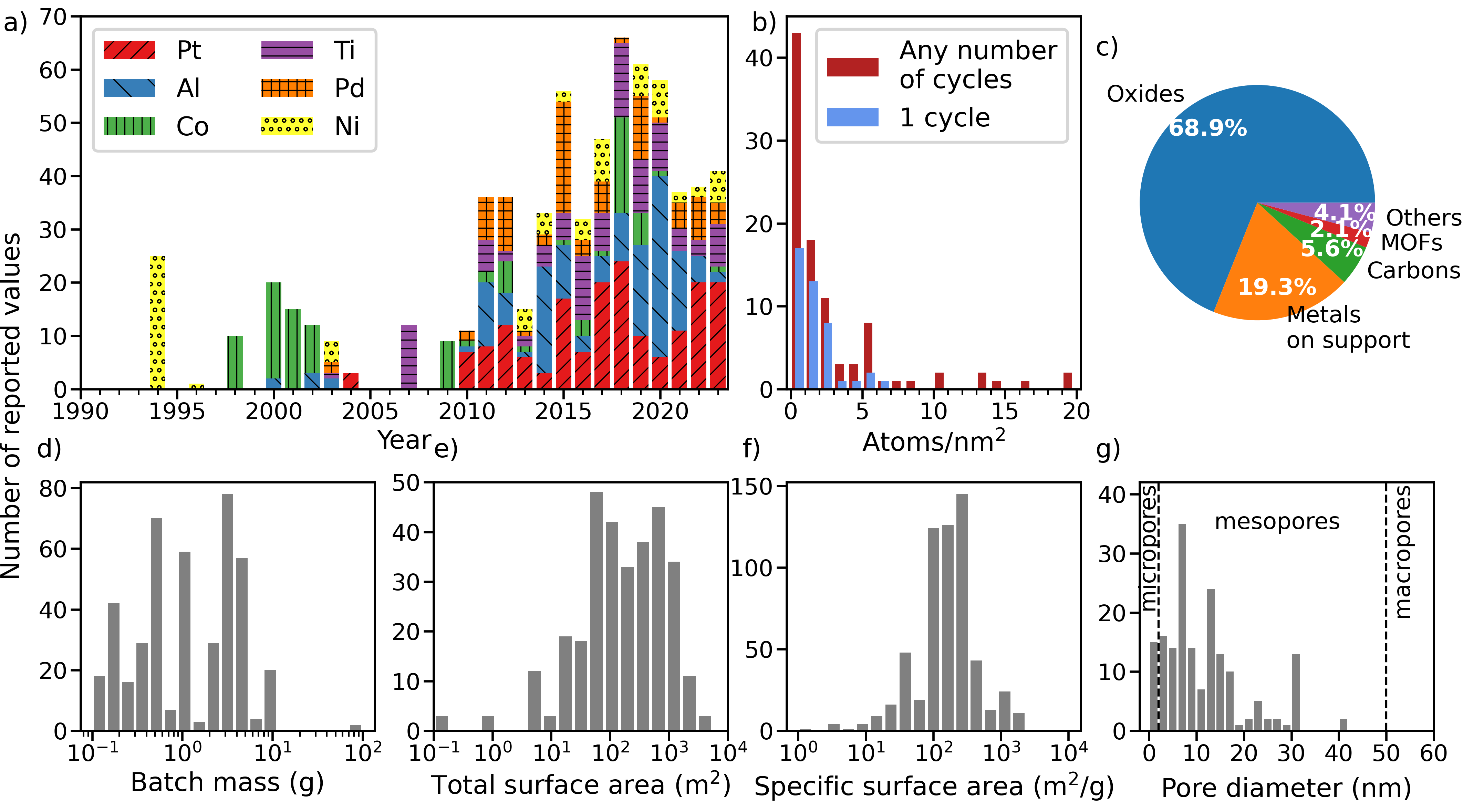}
    \caption{ALD on particulate materials for thermocatalysis. The histograms show the number of reported values, combinations thereof, or processes (more than one per publication possible, respectively). a) Popular coating elements trends; b) number of atoms/nm$^2$ (where reported); c) most common support materials; d) batch mass; e) total surface area of support material coated per ALD run; f) specific surface area of supports; g) pore diameters and classification according to IUPAC\cite{2025IUPACMesopore}.}
    \label{fig:datadriven_thermocatalysis}
\end{figure}

\subsection{Photocatalysis}

A photocatalyst is a light-absorbing catalyst that enhances the rate of a photocatalytic reaction by providing a more energy-efficient reaction pathway. 
A photocatalytic reaction starts when photons with sufficient energy---an energy greater than the band gap of the photocatalyst---excite electrons from the valence into the conduction band, thereby creating an electron-hole-pair in the photocatalyst. 
The generated electron–hole pair engages in a set of redox reactions at the surface of the photocatalyst, referred to as a photoreaction\cite{2013Motegh}. 
Photocatalysts have been studied for a range of reactions, but have been most successfully applied to degrade contaminants in water and air \cite{Li2017a,MartinSomer2020,Guo2020}.

Already in 1911---only 76 years after Berzelius proposed the term catalysis---the word photocatalysis appeared in several communications.\cite{2013Coronado} Compounds such as ZnO and {UO}$_{2}^{+}$ salts were used as photocatalyst \cite{1911Eibner, 1911Bruner,1913Landau}. Later, more and more multimaterial photocatalysts were studied \cite{Benz2020b}. As these are often systems with a nanoscale structure---coordinating the interaction between photons and atoms and/or having an active material on a carrying particle---ALD, with its nanoscale precision, is an attractive method to make these materials. The first report of using ALD to make photocatalysts is in a PhD-level thesis from the USSR \cite{1984Artemev} in which the deposition of iron and titanium on silica particles has been described. The photocatalytic activity in the water decomposition reaction was tested. 
In other countries, it took much longer before the use of ALD to make photocatalytic materials was reported. Within our dataset, the first articles on this topic (outside the USSR) appeared only in the early 2000s\cite{Hakim2007c, King2008b}.

\begin{figure}
    \centering
    \includegraphics[width=0.4\linewidth]{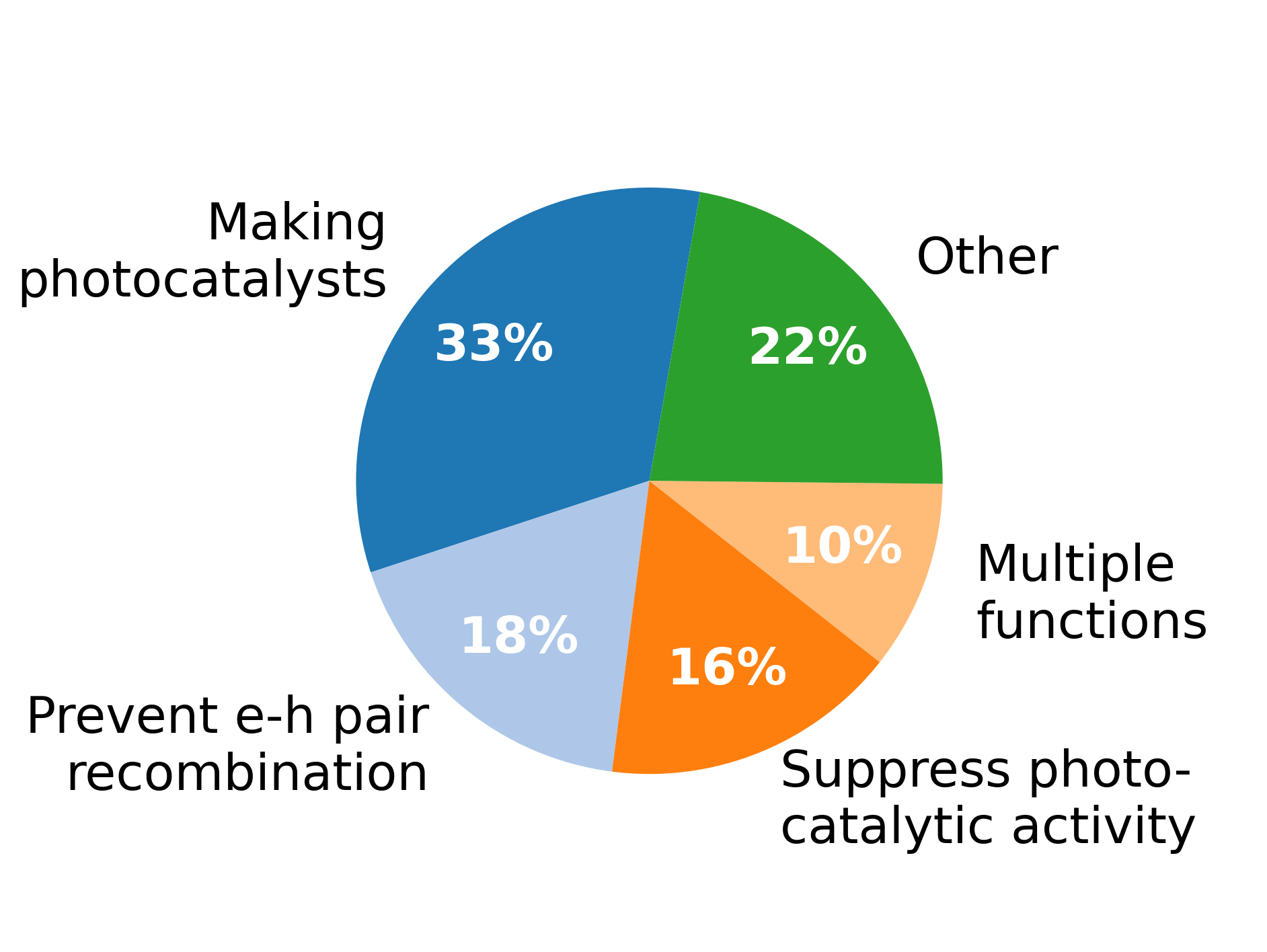}
    \caption{Functionalities that have been obtained using ALD from 67 articles in the application area of photocatalysis.}
    \label{fig:photocat_pie}
\end{figure}

From our dataset, 67 articles report the use of ALD to modify particles while aiming for a range of different photocatalytic functionalities (Figure~\ref{fig:photocat_pie}).
The most widely used approaches are (1) making a photocatalyst by depositing photocatalytic material using ALD, (2) increasing the activity of an existing photocatalyst by adding materials that reduce electron-hole pair recombination, and (3) suppressing photocatalytic activity.
In some cases, multiple functions are attained, or the function fits neither of these categories.

To make photocatalysts, the most widely used material is \ce{TiO2}; hence, Ti is also the most frequently reported coating element for this application (Figure~\ref{fig:periodic}).
A complication is, however, that the most active \ce{TiO2} photocatalysts are crystalline: typically anatase or a mixture of anatase and rutile, such as the widely used commercial photocatalyst P25\cite{2013bCoronado}. 
Since ALD of \ce{TiO2} below 200\,°C typically yields amorphous \ce{TiO2} with negligible photocatalytic properties – although the applied reactants also influence this \cite{2013Miikkulainen} – one either has to carry out the ALD at a higher temperature or anneal the obtained material afterwards. With this approach, photocatalytic material can be deposited on a range of supports. An interesting one is using magnetic material as the support, making a photocatalyst that can more easily be removed from water after cleaning. \cite{Zhou2010a}

A common strategy to increase the activity of a photocatalyst is to prevent electron-hole pair recombination by adding a small amount of a material to the particle surface.
Defects at the surface can provide energy levels within the bandgap and thereby facilitate recombination \cite{2013bCoronado}.
Once recombined, the electron-hole-pair is no longer available for the desired chemical reaction.
A thin coating layer on the particle surface can help reduce the density of surface defects on a particle, increase the lifetime of electron-hole-pairs, and hence increase the activity of the photocatalyst.
Pt has been the most widely deposited material to achieve this effect \cite{Zhou2010b,Zhang2017c,Benz2020a}, but also other materials, e.g. \ce{Au}, \ce{CeO2} and \ce{Cu2O} have been used for this purpose. \cite{Hashemi2020,Wang2017f,Benz2021}

There are also examples of coating a photocatalyst to suppress its photocatalytic activity. This is either done to demonstrate that ALD on particulate matter can make a pinhole-free film, or to make a pigment that is doing no harm to the layer in which it is incorporated (e.g., in paint ) \cite{King2008c,LaZara2020}.

Beyond making photocatalysts, enhancing electron-hole pair recombination, or suppressing photocatalytic activity, ALD has been used in multiple sophisticated ways to modify photocatalysts. 
Adding a thin ALD coating can be used to tune the bandgap, such that the photocatalyst can work with visible light, and no longer require high-energy UV-light\cite{Cao2020}. 
Further, ALD can be applied to tailor the catalyst surface (e.g., modify the zeta potential), such that reactant molecules or microbes to be converted more easily bind to it \cite{Benz2020b}. It is also possible to combine multiple functionalities by adding one material after another.  \cite{Benz2020b,Liu2022}
Here, ALD offers a large range of possibilities through sophisticated process sequences with multiple reactants. If carried out in the same ALD system, this can be implemented in a rather straightforward manner \cite{Benz2020b}.

\subsection{Electrocatalysis}\label{ch:electrocatalysis}

The challenges posed by global warming have driven the world to shift from conventional fossil fuels to clean, renewable energy sources in order to reduce \ce{CO2} emissions. This energy transition necessitates the use of advanced renewable energy conversion and storage devices, such as fuel cells and water-splitting systems\cite{2022aGupta}. These devices, however, rely heavily on the design and development of nanoscale electrocatalysts to achieve efficient product conversion at high rates. 

In most electrochemical processes, supported metal nanoparticles, particularly noble metals, are the key functional components to achieve satisfactory performance for a specific reaction, owing to their unique properties such as size, shape, and composition. 
Many studies have found that the size of metal nanoparticles significantly influences their catalytic activity \cite{2011Shao, 2020Klein,2012PerezAlonso}. For example, in the hydrogen evolution reaction (HER), Pt nanoclusters with a particle size of approximately 2 nm exhibit the highest activity \cite{2020Klein}. Therefore, precise control over the size and dispersion of metal nanoparticles on supports is crucial for maximizing their utilization efficiency within a limited surface area. This control is particularly important given the scarcity and high cost of noble metals.

To this end, ALD has received considerable attention for the synthesis of nanocatalysts used in electrochemical conversion reactions. In this section, we analyzed 52 publications from the field of ALD on particulate materials  for electrocatalyst preparations (Figure~\ref{fig:electrocatalysis}), with one exception that focuses on MLD of alucone. The first study using ALD on particulate materials to synthesize electrocatalyst was in 2011, in which Hsu et al.\ utilized ALD as a synthesis method to deposit Pt nanoparticles on tungsten monocarbide (WC) to study their activity for the oxygen evolution reaction (OER)\cite{Hsu2011}. Discrete Pt particles were formed over the WC support within 20 ALD cycles, and the performance of these Pt particles was comparable to a Pt thin film while having a lower loading. 
Following this first study, more and more attention has been paid to functionalize electrocatalysts by ALD for more than 10 different types of applications (Figure~\ref{fig:electrocatalysis}c). The main attention of using ALD on particles to functionalize electrocatalysts can be divided into two main purposes: 1) improved activity of noble metals (mainly Pt and Pd) with controlled particle size and distribution and 2) enhanced stability of metal catalysts through metal oxide/nitride layer coating. These two purposes are reflected in the coating material categories in Figure~\ref{fig:electrocatalysis}b. 

\begin{figure}
  \includegraphics[width=0.8\textwidth]{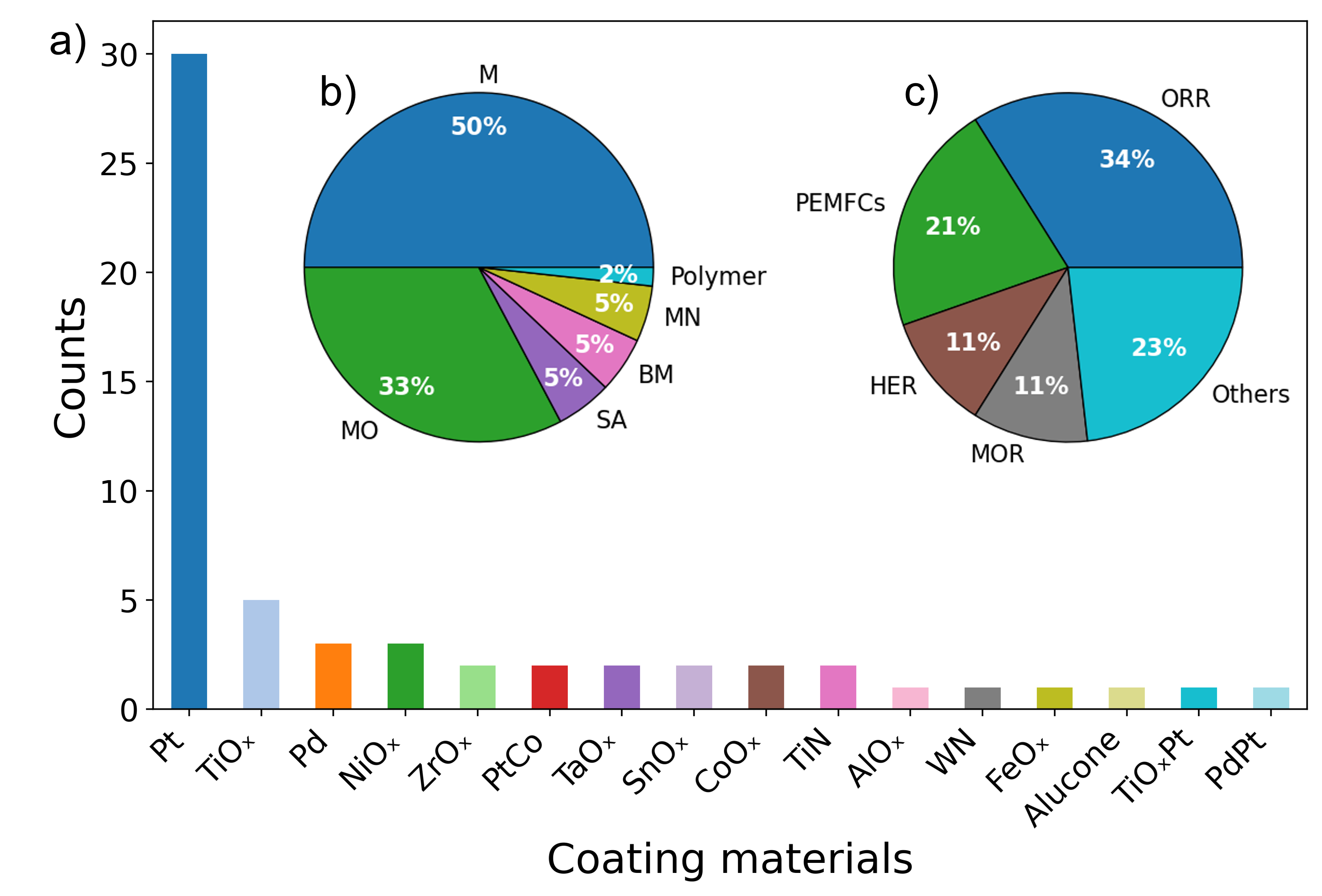}
  \caption{Overview of the coating materials using ALD on particulate materials for electrocatalytic applications. a) The number of counts of each coating material reported in the literature, b) Pie chart of the ALD coating materials for electrocatalysis applications. M: metal; MO: metal oxide; SA: single atom; MN: metal nitrides; BM: bimetallic catalyst, and c) Pie chart depicting four major and some minor (others) electrochemical applications of ALD on particulate materials. PEMFCs: proton exchange membrane fuel cells, ORR: oxygen reduction reaction, HER: hydrogen evolution reaction, MOR: methanol oxidation reaction. The others include OER: oxygen evolution reaction, FAOR: formic acid oxidation reaction, SOFC: solid oxide fuel cells, GOR: glycerol oxidation reaction, AOR: alcohol oxidation reaction, HP: hydrogen pumping, and CO2RR: \ce{CO2} reduction reaction.}
  \label{fig:electrocatalysis}
\end{figure}

As shown in Figure~\ref{fig:electrocatalysis}c, the majority of applications of ALD on particulate materials in electrocatalysis are in oxygen reduction reactions (ORR) and proton exchange membrane fuel cells (PEMFCs), which together account for more than 50\,\% of the total applications. Here, we separate the ORR and PEMFCs into two categories based on the title of the article, while acknowledging that ORR is the half-reaction of PEMFCs. The ORR at the cathode side of PEMFCs is limited by the slow kinetics, and Pt is currently the most efficient catalyst for such a reaction. However, the material cost of Pt is high, which accounts for half of the overall cost of fuel cells \cite{Zaccarine2022}. For this reason, half of the coating materials by ALD on particulate materials is about supported Pt nanoparticles (Figure~\ref{fig:electrocatalysis}a) with the aim to improve their activity and stability in ORR (and PEMFCs) via particle size and shape control and metal-support interactions. Compared to commercial Pt/C (e.g., Premetek, 60\,wt.\%), the ALD-synthesized Pt on the (modified) carbon support exhibited a uniform particle size with much less agglomeration, resulting in a higher electrochemical active surface area (ECSA) and thus a better performance in ORR \cite{Lee2019c, Lee2020a, Xu2018a, Song2018, Shi2019}. To reduce the usage of Pt, researchers have also explored the ALD synthesis of bi-metallic catalysts (PtCo) \cite{Sairanen2014} and non-nobel catalysts (\ce{CoOx}) \cite{Sun2019} for ORR chemical reactions. These catalysts have shown comparable electrocatalytic performance and durability as Pt.

The poor durability of commercial Pt/C during long-term operation is another key challenge for applications in ORR/PEMFCs. This challenge can be mitigated by replacing carbon with more corrosion-resistant support materials such as molybdenum carbide (\ce{Mo2C}) \cite{Saha2015}, WC \cite{Hsu2012,Hsu2015}, carbon nanotubes (CNT) \cite{Wang2016a,Sun2019,Gan2020}, zirconium carbide (ZrC) \cite{Cheng2015a}, and nickel nanowires \cite{Zaccarine2022}. To improve the stability of Pt nanoparticles, significant efforts have also been made to anchor Pt nanoparticles on carbon supports by coating them with protective layers, such as \ce{TaO}$_x$ \cite{Song2017} and \ce{ZrO2} \cite{Cheng2015b}. 

Following PEMFCs and ORR, the HER and methanol oxidation reaction (MOR) also significantly contribute to the application of electrocatalysts prepared by ALD on particulate materials, each accounting for 10.7\,\% of total usage. Pt remains the predominant catalyst for both HER and MOR. In the context of HER applications using ALD on particles, significant research efforts have focused on depositing Pt onto alternative support materials such as WC \cite{Hsu2015}, \ce{Mo2C} \cite{Saha2015}, and graphene \cite{Cheng2016} to enhance cost-efficiency and performance.   Regarding MOR, researchers have explored the ALD technique to synthesize more efficient Pt catalysts to reduce Pt loading for direct methanol fuel cells (DMFCs) \cite{Zhang2015a, Wang2016a, Kim2018b}.

Other, less researched electrocatalytic applications in the context of ALD on particles include the OER \cite{Hsu2011,Kim2015c,Palmer2018,Sun2019}, formic acid oxidation reaction (FAOR)\cite{Liu2016a,Shi2019}, solid oxide fuel cells (SOFC) \cite{Page2019,Jo2023}, glycerol oxidation reaction (GOR) \cite{Lee2020b,Han2020}, alcohol oxidation reaction (AOR) \cite{Rikkinen2011}, and \ce{CO2} reduction reaction (CO2RR) \cite{Li2022b}. The motivation for using ALD in these applications is similar to the applications discussed above.

\subsection{Batteries}

An application of ALD on particulate materials that has obtained increasing attention in recent years is the synthesis of battery materials. The first report (within the scope of this review) of ALD on particles for batteries by Snyder et al.\ dates back to 2007\cite{Snyder2007}, but only from 2010 a significant number of articles has been published about this application (see Figure~\ref{fig:application}). 
Rather recently, an extensive review on this specific topic has been written by Lee et al.\cite{Lee2022i}. Hence, we will keep this section very brief and refer the reader to that review for more detailed information. 
Note, however, that Lee et al.\ include coating of immobilized particles (i.e.\, in an electrode), whereas here we focus on particles in powder form. 

Our data-driven analysis shows that 82\% of the battery articles use supports composed of metal (or metalloid) compounds (mainly oxide), which in the vast majority will be cathode materials. 17\% concerns typical anode materials such as carbon-based supports and silicon. Only 1\% concerns other materials. Among these categories, cathode materials have received the most attention in research. 
A critical reason for this is that they are typically more expensive than the anode material. 

\begin{figure}
    \centering
    \includegraphics[width=0.4\linewidth]{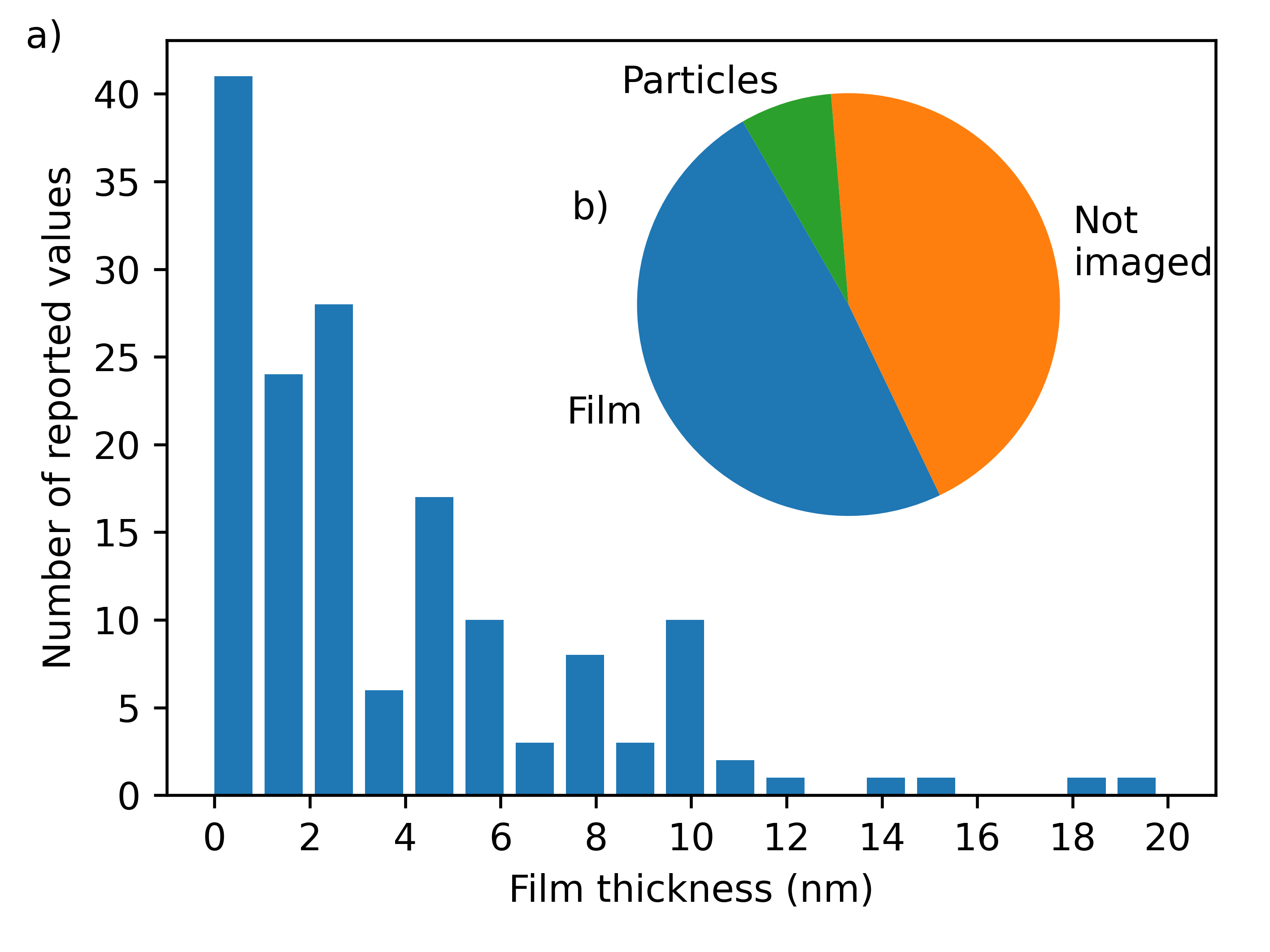}
    \caption{a) Reported film thicknesses for application in batteries (up to 20\,nm, only few reports of thicker layers\cite{Liu2015b, Luan2012}). b) Structure of deposited material where proof was provided in the respective report through imaging.}
    \label{fig:batteries}
\end{figure}

In many battery types, the solid/electrolyte interphase (SEI) layer may show instabilities due to a reaction between the electrode and the electrolyte. By coating the electrode surface, this can be partly or entirely prevented. Often, coating methods such as sol--gel synthesis and chemical vapor deposition have been used to this end. However, these methods may give coatings that are inhomogeneous and/or too thick, leading to imperfect protection and/or a decrease in the battery capacity. The ultrathin films that can be made by ALD (see Figure~\ref{fig:batteries}a) have been shown to strongly extend their lifetime while hardly reducing the battery capacity \cite{Zhao2013b,Shapira2018}. This is effective both for cathode and anode particles \cite{Lee2022i}.

Most coatings for battery applications grow in film structure (see Figure~\ref{fig:batteries}b), and those are primarily metal oxides (see Figure~\ref{fig:periodic} for an overview of coating elements).
A small minority of coatings grow in particle form, and these are mostly on carbon-based support materials\cite{Azaceta2020, Lu2013b, Lu2017, Luo2015, Lv2017, Meng2015, Sun2014} (see also Section \ref{sec:supportschem}) and/or noble metal coatings\cite{Li2022d, Lu2013b, Luo2015}.

\subsection{Luminescent phosphors}

Luminescent phosphors are materials that can convert light of shorter wavelengths to longer wavelengths through electronic transitions, e.g., convert the blue light of an LED to red light.\cite{2021Gupta}
These materials are an essential component in solid-state lighting applications or displays.
However, the adoption of luminescent phosphor particulate materials in display applications can be undermined by the degradation of their luminescent properties due to thermal and chemical instability.
To improve their stability, phosphor particles can be coated with a thin protective barrier layer;
here, ALD offers an attractive alternative to conventional coating techniques (e.g., sol--gel, CVD), providing conformal and ultra-thin protective features, without compromising the bulk material's optical characteristics.
This has been demonstrated on phosphor powders of different colors (blue, blue-green, yellow, and red), with particle sizes typically spanning 1--100\,μm.

The most common protective coatings, typically deposited to serve a function as a moisture barrier, are thin metal oxides and metalloid oxides.
The optimal coating thickness is reported to be between 2-50\,nm, as illustrated in Table~\ref{tab:phosphors}.
Nearly half of the reported studies focus on the deposition of \ce{Al2O3} films\cite{Kim2007b, Zhou2016, Zhang2018a, tenKate2019, Verstraete2019, Zhao2019b, Zhao2020, Karacaoglu2020, Zhao2022, Karacaoglu2023, Zhao2023} due to their favorable barrier properties and the availability of well-defined ALD processes.
Additionally, several other oxide-based coatings, such as \ce{TiO2} and \ce{TiO2}-\ce{Al2O3} composites\cite{Yoon2011} (which enhance photocatalytic reactivity\cite{Kim2009b}, in addition to their protective function\cite{Verstraete2019, Karacaoglu2020}), \ce{SiO2}\cite{Kim2007a, Jeong2009, Huang2022} and \ce{MgO}\cite{Kim2011d} have been demonstrated.
A \ce{TiN} coating, applied by one or three ALD cycles\cite{Sosnov2012}, is the only reported non-oxide material implemented onto phosphor materials.

Besides increasing the stability of phosphors, ALD coatings can positively affect the phosphors' luminescent properties.
Studies have shown that photoluminescence intensity can even be improved for up to a certain coating thickness, possibly due to the diminishing of surface defects and the prevention of particle aggregation.\cite{Kim2007a, Kim2007b, Kim2009b, Jeong2009}
In one instance, the fabrication of a phosphor material exclusively by ALD was demonstrated.
The process involved the conversion of $\gamma$-\ce{Al2O3} particles into spinel \ce{ZnAl2O4} by \ce{ZnO} deposition and thermal annealing, followed by the consecutive \ce{Eu} doping.\cite{Rauwel2011, Rauwel2012a}

To prevent negative effects of the ALD coating, and maintain performance, particle agglomeration and interaction of the coating with the phosphor’s dopant ions\cite{tenKate2019} should be minimized.
Additionally, introduction of porosity to the grown film, and deposition proceeding via island-growth mode rather than layer-by-layer growth (see Section~\ref{sec:GPC}) can also be detrimental for the phosphor performance.\cite{Verstraete2019}

With respect to processing conditions, care must be taken, especially for temperature-sensitive phosphors.
The thermal stability of phosphors can limit the use of \ce{H2O} as a counter-reactant due to the relatively high temperature requirements for effective purging.
For coating of highly sensitive phosphors, there is a trend towards the use of less surface-adhesive counter-reactants (\ce{O2}, \ce{O2}* and \ce{O3}) and the lowering of the reaction temperature (typically below 200$^{\circ}$C, down to room temperature).

\begin{table}
\caption{ALD phosphor coating studies (the reported thickness values are derived by TEM; optimal thickness is reported in nm, when available, or as the corresponding number of ALD cycles performed; "c" stands for ALD cycle).}
\label{tab:phosphors}

\scriptsize
\begin{tabular}{lcccccc}
\thead
{Phosphor}
& \thead{Color}
& \thead{ALD scheme}
& \thead{Temperature
\\ {[}$^{\circ}$C{]}}
& \thead{Coating}
& \thead{\makecell[c]{Optimal
\\ thickness 
\\ {[}nm{]}}}
& \thead{Ref.}
\\
\midrule
\ce{BaMgAl10O17}:Eu$^{2+}$
& blue
& TEOS/\ce{H2O}
& RT
& \ce{SiO2}
& 14 (200c)
& \citenum{Kim2007a}
\\
\ce{BaMgAl10O17}:Eu$^{2+}$
& blue
& TMA/\ce{H2O}
& 350
& \ce{Al2O3}
& (500c)
& \citenum{Kim2007b}
\\
\ce{CaAl2O4}:Eu$^{2+}$,Nd$^{3+}$
& blue
& \ce{TiCl4}/\ce{H2O}
& 240
& \ce{TiO2}
&  
& \citenum{Kim2009b}
\\
\ce{BaMgAl10O17}:Eu$^{2+}$
& blue
& TEOS/\ce{H2O}
& 70
& \ce{SiO2}
& 20 (300c)
& \citenum{Jeong2009}
\\
\ce{ZnAl2O4}:Eu$^{2+}$
& red
& DEZ/\ce{H2O}
& 175
& \ce{ZnO}
& 
& \citenum{Rauwel2011}
\\ 
\ce{BaMgAl10O17}:Eu$^{2+}$
& blue
& \ce{Mg(CpEt)2}/\ce{H2O}
& 120
& \ce{MgO}
& 12 (400c)
& \citenum{Kim2011d}
\\ 
\ce{CaAl2O4}:Eu$^{2+}$,Nd$^{3+}$
& blue-green
& \makecell[c]{TTIP/\ce{H2O}
\\ TMA/\ce{H2O}}
& \makecell[c]{270
\\ 270}
& \makecell[c]{\ce{TiO2}
\\ \ce{Al2O3}}
& 
& \citenum{Yoon2011}
\\ 
\ce{ZnAl2O4}:Eu$^{2+}$
& red
& DEZ/\ce{H2O}
& 175
& \ce{ZnO}
& 
& \citenum{Rauwel2012a}
\\
\ce{Zn2SiO4}:Mn$^{2+}$
& green
& \ce{TiCl4}/\ce{NH3}
& 340–420
& \ce{TiN}
& (1c)
& \citenum{Sosnov2012}
\\ 
\ce{Y3Al5O12}:Ce$^{3+}$
& yellow
& TMA/\ce{O2}
& 25
& \ce{Al2O3}
& 2 (15c)
& \citenum{Zhou2016}
\\ 
\ce{Ca9K(PO4)7}:Eu$^{2+}$,Mn$^{4+}$,Dy$^{3+}$
& red
& TMA/\ce{H2O}
& 75
& \ce{Al2O3}
& 50 (500c)
& \citenum{Zhang2018a}
\\ 
\ce{K2SiF6}:Mn$^{4+}$
& red
& TMA/\ce{O3}
& 50; 120
& \ce{Al2O3}
& 
& \citenum{tenKate2019}
\\ 
\ce{K2SiF6}:Mn$^{4+}$
& red
& \makecell[c]{TMA/\ce{H2O}; TMA/\ce{O2}*
\\ TDMAT/\ce{H2O}}
& \makecell[c]{100
\\ 100}
& \makecell[c]{\ce{Al2O3}
\\ \ce{TiO2}}
& 
& \citenum{Verstraete2019}
\\ 
\ce{Sr2Si5N8}:Eu$^{2+}$
& red
& TMA/\ce{H2O}; TMA/\ce{O3}
& 30–200
& \ce{Al2O3}
& 8–10 (15–30c)
& \citenum{Zhao2019b}
\\
\ce{RbLi(Li3SiO4)2}:Eu$^{2+}$
& green
& TMA/\ce{H2O}
& 60
& \ce{Al2O3}
& 
& \citenum{Zhao2020}
\\
\ce{SrAl2O4}:Eu$^{2+}$,Dy$^{3+}$
& green
& \makecell[c]{TMA/\ce{H2O}
\\ \ce{TiCl4}/\ce{H2O}}
& \makecell[c]{120
\\ 120}
& \makecell[c]{\ce{Al2O3}
\\ \ce{TiO2}}
& 
& \citenum{Karacaoglu2020}
\\ 
\ce{Sr[LiAl3N4]}:Eu$^{2+}$
& red
& TDMASi/\ce{O3}
& 150
& \ce{SiO2}
& 
& \citenum{Huang2022}
\\
\ce{K2GeF6}:Mn$^{4+}$
& red
& TMA/\ce{O3}
& 90
& \ce{Al2O3}
& (30c)
& \citenum{Zhao2022}
\\
\ce{BaAl2O4}:Eu$^{2+}$,Dy$^{3+}$
& blue-green
& TMA/\ce{H2O}
& 150
& \ce{Al2O3}
& 
& \citenum{Karacaoglu2023}
\\ 
\ce{BaSi2O2N2}:Eu$^{2+}$
& blue-green
& TMA/\ce{O3}
& 100
& \ce{Al2O3}
& 11±1 (20c)
& \citenum{Zhao2023}
\\
\end{tabular}   
\end{table}

\subsection{Healthcare}

\subsubsection{Pharmaceuticals}\label{ch:pharma}

Solid organic powders, such as active pharmaceutical ingredients (APIs) and excipients, are challenging materials to process by ALD due to their complex surface functionalities, thermal stability requirements, and size-related flow characteristics.
Certain ALD processes that can be operated at low temperatures, even close to room temperature, can be harnessed for coating of pharmaceutical powders to enhance their performance.

The surface modification of acetaminophen, by means of metal oxide ALD (\ce{Al2O3}, \ce{TiO2} and \ce{ZnO}) was the first case of pharmaceutical coating reported.\cite{Kaariainen2017}
No degradation or change of the API form due to the applied ALD temperature (100\,$^{\circ}$C) or the use of ALD reactants (TMA, \ce{TiCl4} or TTIP, and DEZ respectively) was observed, and a delayed release behavior was verified.
The choice of reactants can influence the growth mode, as illustrated for 5-aminosalicylic acid particles, where the coating was diffused inside the API particle for the case of \ce{ZnO} while the \ce{Al2O3} coating formed a sharp interface with the particles.\cite{Sosa2023}
A similar approach was followed for the modification of the injectable drug, indomethacin, drastically altering its release profile after the coating with a relatively thick (30--35 nm) \ce{Al2O3} film.\cite{Hellrup2019}
In addition, ALD has been used to provide thermostable, single-dose vaccines, as reported in a study on coating HPV 16 L1 capsomer-based particles by up to 500 cycles of \ce{Al2O3} ALD.\cite{Garcea2020}
A relatively high number of ALD cycles (250, corresponding to a film thickness of 50--60\,nm) was required, as demonstrated in the coating studies of Lys-phage-like particles\cite{Witeof2022} and other HPV variants (HPV 18 and HPV 31) of the capsomer-based particles.\cite{Witeof2023}
Additionally, the effect of coating a model injectable protein, myoglobin, and myoglobin-mannitol with alumina or silica ALD was probed, showing improved stability during storage.\cite{Barros2023}

A large number of \ce{Al2O3} ALD cycles (up to 500) at a low temperature (50\,$^{\circ}$C) was applied to prevent moisture absorption of a hygroscopic amorphous lactose powder.\cite{Hellrup2017}
The possibility to effectively alter the flowability characteristics of commonly used pharmaceutical excipients was demonstrated by \ce{TiO2} coating of croscarmellose-sodium, microcrystalline cellulose, $\alpha$-lactose monohydrate, sorbitol, and dibasic calcium phosphate dihydrate.\cite{Hirschberg2019}
However, an attempt to implement \ce{TiO2} ALD coatings directly on the drug dosage form level (minitablets of denatonium benzoate ) to provide taste masking was not deemed successful.\cite{Hautala2017}

Coating of budesonide API and lactose with alumina under ambient processing conditions (30\,$^{\circ}$C, atmospheric pressure)\cite{Zhang2017b} in a fluidized bed reactor, showed uniform coating characteristics for the case of budesonide.
The coating of lactose resulted in films of varying thickness from particle to particle, as evidenced by TEM.\cite{Zhang2017b}
The same behavior was also observed for an ALD process utilizing \ce{O3} instead of \ce{H2O} vapor as the counter-reactant, on two different sizes of lactose particles.\cite{Zhang2019d}
Nevertheless, the ability to enhance the processability of the powders, while also controlling the drug release characteristics, was demonstrated.\cite{Zhang2019d}
To explore the potential of ALD for inhalation drug dosage forms, the research was expanded to other ALD, MLD and hybrid ALD/MLD schemes, namely for deposition of \ce{Al2O3}, \ce{TiO2}, \ce{SiO2}, PET and titanicone coatings.\cite{LaZara2021a, LaZara2021b}
Consequently, it was shown that the wettability of the API could be tuned from highly hydrophilic to superhydrophobic, presenting an opportunity to alter the dissolution rate, dispersion, flowability and the solid-state stability of budesonide.\cite{LaZara2021a, LaZara2021b}
Inorganic nanofilms showed a clear benefit over their organic and hybrid counterparts in modifying the flowability of budesonide.\cite{Zhang2023}

ALD of \ce{Al2O3} and \ce{TiO2} on palbociclib and pazopanib \ce{HCl} has been implemented to reduce electrostatic charge build-up.\cite{Gupta2022}
A rotary drum reactor was used to coat the API powders with 50 or 100 cycles respectively.\cite{Gupta2022}
Raman spectroscopy was the first implementation of process analytical technology (PAT) reported to study the extent of ALD reactions.\cite{Mittal2022}
The potential for process scale-up was effectively demonstrated by the use of a double cone reactor that can handle larger powder batches ($\geq$1000\,g), for the coating of several model APIs (acetaminophen, ibuprofen, theophylline, metformin \ce{HCl}) with metal oxide coatings (Al-, Ti- and Zn-based).\cite{Swaminathan2023} 

The term ALC has been introduced to explicitly describe ALD processing, when applied to powders/particles of pharmaceutical relevance.\cite{Moseson2022}
It is characterized by low deposition temperatures, below 100\,$^{\circ}$C (and usually, as low as 30--35\,$^{\circ}$C).\cite{Swaminathan2023}
When applied to amorphous solid dispersions (ASDs) ALD has the potential to stabilize them, avoiding crystallization and enabling higher drug loadings as showcased in the cases of ezetimibe\cite{Duong2022}, erlotinib, naproxen, lumefantrine\cite{Moseson2022} and posaconazole.\cite{Moseson2023a}
The coatings were predominantly based on \ce{Al2O3} (one case of \ce{ZnO} reported) and required a thickness in the order of 4-40\,nm to demonstrate the observed benefits.
Apart from crystallization inhibition, the coatings also improved the flowability of the powders.
The inherent ALD characteristics of pinhole-free coatings with precisely controlled thickness are considered of high relevance for the realization of ASDs with high drug loadings.\cite{Mamidi2023, Moseson2023c}

The advantages that ALD impacts on pharmaceutical particulate materials, either in terms of protection and stabilization (moisture, oxidation, light), release control of the active ingredients and processability improvements highlight the emerging possibilities.\cite{2024Schenck}
Implementation of ALD for enhancing the performance of APIs has already been demonstrated for oral, injectable, and inhalation drug dosage forms.
Compatibility of the ALD reactants and coatings with the requirements set by the regulatory authorities will have to be demonstrated, in order to allow the establishment of ALD as a technology platform.
ALD process schemes operated at atmospheric pressure seem to result typically to higher growth rates (Table~\ref{tab:pharma}).
A differentiation on the application of ALD should be made with regard to the coating characteristics.
For protection, typically thicker (up to 60\,nm) and conformal coatings are required.
In terms of dissolution, thicknesses below 20\,nm with a sufficient level of surface defects have been reported as optimal.\cite{Moseson2023a}
Finally, in view of processability, surface passivation suffices, and it is envisioned that even a small number of ALD cycles can have a major impact on powder flow behavior.

\begin{table}
\footnotesize
\begin{tabular}{llllll}
\thead
{Support}
& \thead{Reactants}
& \thead{Temperature
\\ {[}$^{\circ}$C{]}}
& \thead{Thickness 
\\ {[}nm{]}}
& \thead{GPC 
\\ {[}nm{]}}
& \thead{References}
\\
\toprule
lactose
& \makecell[l]{TMA/\ce{H2O} 
\\ TMA/\ce{O3}}
& \makecell[l]{30 
\\ 30}
& \makecell[l]{2.5-11.0 
\\ 3.8-10.3}
& \makecell[l]{0.47-0.73 
\\ 0.47-1.00}
& \makecell[l]{\citenum{Zhang2017b,Zhang2019d} 
\\ \citenum{Zhang2019d}}
\\
\midrule
budesonide
& \makecell[l]{TMA/\ce{H2O} 
\\ TMA/\ce{O3} 
\\ \ce{TiCl4}/\ce{H2O}
\\ \ce{SiCl4}/\ce{H2O}
\\ \ce{TiCl4}/EG
\\ TC/EG}
& \makecell[l]{40
\\ 40
\\ 40
\\ 40 
\\ 120 
\\ 150}
& \makecell[l]{0.8-2.0
\\ 20.0-50.0
\\ 3.0-15.0
\\ 1.5-10.0
\\ 4.5-23.0
\\ 1.5}
& \makecell[l]{0.20-0.33 
\\ 1.00-2.00 
\\ 0.30-0.32 
\\ 0.10-0.15 
\\ 0.45-0.46 
\\ 0.03}
& \makecell[l]{\citenum{Zhang2017b} 
\\ \citenum{LaZara2021a,LaZara2021b,Zhang2023}
\\ \citenum{LaZara2021a,LaZara2021b,Zhang2023}
\\ \citenum{LaZara2021a,LaZara2021b,Zhang2023}
\\ \citenum{LaZara2021a,Zhang2023}
\\ \citenum{LaZara2021a,Zhang2023}}
\\
\midrule
indomethacin
& TMA/\ce{H2O}
& 50
& 30.0-35.0
& 0.26-0.30
& \citenum{Hellrup2019}
\\
\midrule
HPV 16 L1 capsomer
& TMA/\ce{H2O}
& 70
& 40.0-59.9
& 0.16-0.24
& \citenum{Garcea2020}
\\
\midrule
posaconazole/HPMCAS
& TMA/\ce{H2O}
& 35
& \makecell[l]{-
\\ 12.0
\\ 8.0-40.0} 
& \makecell[l]{\textgreater{}1.00
\\ 0.40-1.00
\\ \textless{}0.40}
& \citenum{Moseson2023a}
\\
\midrule
5-aminosalicylic acid
& \makecell[l]{TMA/\ce{H2O}
\\ DEZ/\ce{H2O}}
& \makecell[l]{120
\\ 120}
& \makecell[l]{38.0
\\ 24.7}
& \makecell[l]{0.13
\\ 0.12}
& \citenum{Sosa2023}
\\
\midrule
myoglobin
& TMA/\ce{H2O}
& 30
& 8.3
& 0.21
& \citenum{Barros2023}                                                                                         
\end{tabular}
\caption{Reported thickness range and GPC values (obtained by microscopy) for ALD coatings onto particulate materials of pharmaceutical relevance (temperature classification as “low” for $<70\,^{\circ}$C and “high” for $>70\,^{\circ}$C; coating thickness split to “low” for $<20$ nm and “high” for $>20$ nm; GPC of \ce{Al2O3} indicated as “slow” for $<0.4$\,nm, “average” for 0.4-1\,nm and “fast” for $>1$\,nm according to the work of Moseson et al.\ \cite{Moseson2023a}; the typically reported GPC value for \ce{Al2O3} on planar substrates is 0.1--0.2\,nm \cite{Zhang2017b}).}
\label{tab:pharma}

\end{table}

\subsubsection{Other healthcare}

The inherent capability of ALD to precisely modify the interface of particulate materials was the driver of the initial attempts to implement the technique in healthcare applications.
Enhancement of bio-compatibility, i.e. for tissue engineering\cite{Liang2009b} and magnetism-mediated therapeutics\cite{Uudekull2017, Seong2019b} were the first explored applications.
Also, in the field of nutraceuticals, the deposition of a bio-compatible (\ce{TiO2}) coating for the prevention of vitamin C oxidation was reported.\cite{HaghshenasLari2018}
Coatings by ALD have also been implemented to screen the toxicity of novel functional nanomaterials and drug delivery systems\cite{Hilton2017, Mestres2016} and to enhance the antimicrobial activity of porous nanoparticles.\cite{LopezdeDicastillo2019}
Additionally, niche applications exploring the potential of X-ray mediated photodynamic therapy\cite{Sengar2019} and sorbent materials for medical radionuclides\cite{Moret2020a} have been reported.

In the aforementioned cases, the reported thickness of the ceramic oxide coatings (\ce{Al2O3}, \ce{TiO2} and \ce{ZnO}) ranged from 1 to 30\,nm, thus requiring the application of (up to) a few hundred ALD cycles.
The choice of deposited materials was based on their biocompatibility characteristics and the availability of well-defined ALD process schemes, capable of deposition at relatively low temperatures.

\subsection{Other applications}\label{ch:otherapps}

Particles that are protected or functionalized using ALD have a wide range of applications, of which we have thus far only discussed the ones that received the most attention; the various forms of catalysis clearly stand out. However, many more applications (see Supporting Information Table~S2) have only been investigated in 10\,or fewer articles from the set we analysed.  
Coating particulate material using ALD can help make better thermoelectric materials: materials that can convert waste heat into electricity\cite{He2022,Jung2022}. ALD can also be used to functionalize particles for separation, e.g., for \ce{CO2} capture\cite{Armutlulu2017} or for adsorption of organic pollutants\cite{Wang2017d}. Another application is to modify particles that are applied as colloids (dispersed in a liquid), e.g.\, to study their electrosurface characteristics\cite{Ermakova2002} or give them better dispersibility\cite{Kim2011b}.

\section{Outlook \& future developments}\label{ch:outlook}
ALD on particulate materials as a research field has received fluctuating levels of attention over the years, in both absolute and relative terms, compared to the broader field of ALD.
In the early years (1960s--1990s), particles were regularly used as substrates.
However, the relative attention for particles decreased with the substantial increase in planar substrate ALD research driven by the semiconductor industry in the early 2000s (see Supporting Information Figure~S2). 
In recent years, the interest in ALD on particulate materials has again been rising, as is visible from the growing number of articles (Figure~\ref{fig:vstime}) and the increasing number of sessions on particles at ALD conferences. This is not surprising, given the rise of current large scale particle-based applications (e.g.\ batteries), and the unique features that ALD has to offer for them. 
Among those features are nano-scale precision of coatings and the potential to minimize the required amount of materials, thereby addressing a growing concern for critical materials. 

For part of the expected future developments, we can look at recent developments for ALD on planar substrates, where atomic layer etching and area-selective ALD (ASALD) have received much attention in recent years.
In atomic layer etching, the substrate or earlier deposited materials are removed layer by layer in a self-limiting set of reactions similar to ALD. 
In the context of particles, atomic layer etching is still rare\cite{2023Partridge,2021LiiRosales,2020Clancey}, and convincing applications are yet to be found.
In ASALD, selective deposition on parts of the substrate takes place, depending on their surface chemistry. In some cases, this occurs without further process steps, but can nevertheless be useful to make, e.g., bimetallic electrocatalysts\cite{Christensen2010,Li2022b,Jang2020,Wang2018,Wang2016b,Wang2015,Lu2014}.
In other cases, material is deliberately deposited around a catalytically active metal, thereby increasing the stability of the electrocatalyst\cite{Song2017,Cheng2015b}.
These examples show that ASALD enables new functionalities that would otherwise not be attainable.
However, the plethora of available planar substrate processes suggests that the potential for ASALD on particles is far from exhausted. 

The trend from temporal (ALD cycles take place in time) to spatial (ALD cycles take place in space) reactors is observed in both planar and particle substrate ALD processes.
This shift allows converting a batch process into a continuous process.
The latter typically leads to more constant product quality, smaller equipment volumes, and easier integration in the product chain.

There is also room for more research on the optimization of purging. 
Particulate material typically has a much larger surface (even more when porous) than planar substrates, and thus larger reactant amounts are used. 
Also, the porosity implies different mass transport regimes compared to flat substrates (see section~\ref{sec:pressure}).
Hence, the purging step demands more attention compared to planar substrate ALD, and typically much longer purge times, especially for basic research of ALD processes, where any potential CVD-like behavior must be ruled out.
In contrast, for the development of production processes, one will typically try to minimize the purge time in order to reduce the costs. 
Very few articles in the academic literature specifically probe the requirements of purge times for ALD on particulate materials. 

Forming a comprehensive picture of an entire research field remains challenging.
The present review is based on a systematic literature search using a defined set of search terms (keywords), which led to the quantitative literature dataset of 799 articles (see section~\ref{ch:method}).
The discussion in text form, e.g., in the thermocatalysis section, was enriched with individual references to articles from outside of this dataset.
It turns out that some of these articles would have indeed qualified to be included in the quantitative dataset\cite{2008Plomp,2004Buijink,2000Ferguson,1993Haukka,2023Zhou}, but only showed up in a context-oriented search. 
Hence, future review work could additionally include semantic search methods, using AI and natural language processing, to uncover even those articles with deviating terminology that did not show up in the keyword-based search, thereby providing an even more comprehensive overview of the field.

With regard to obtaining specific information from articles through large language models (LLMs), our current experience is ambivalent, but AI tools certainly bear potential for the future.
Though LLMs can provide a qualitative overview of a specific article, filtering out reliable quantitative information remains challenging, as LLMs tend to interpolate missing information from context.
In this respect, specialized AI-models that sharply distinguish factual data from context would be required. 
Curated datasets, like the one provided through our review on an open-access platform\cite{ALDpmZenodo}, can serve as a basis for training such models. 
Ultimately, scientific literature mining, content relevance screening and data extraction are expected to become increasingly automated.
This route can help transition scientific literature reviewing from expert-driven narratives to data-centric meta-analysis, especially in the field of material synthesis.\cite{2025bYanguasGil}

The intrinsic heterogeneity and complexity of particulate materials---varying in size, size distribution, porosity, and surface functionalities---is expected to continue challenging the applicability of high-throughput experimentation methods\cite{CamachoBunquin2015, Knemeyer2021a}.
These limitations will reinforce the value of predictive modeling as a means to reduce reliance on resource-intensive iterative experimentation.
Two key directions for future development are the use of machine learning algorithms for ALD parameter prediction\cite{2022bKim, Yoon2023a}, and the deployment of machine learning for ALD process optimization \cite{2022YanguasGil}, ultimately contributing to the development of reactor digital twins. 
The development and training of such models will depend critically on curated datasets of the kind presented herein, and on fundamental information obtained on the kinetics of ALD processes obtained through saturation profile analysis in well-defined high-aspect-ratio test structures such as PillarHall \cite{2015Gao, 2018Ylilammi, 2019Arts, 2020Yim, PillarHallWebNOTE, PillarHallZoteroNOTE} or, e.g., porous particles of well-defined geometry \cite{2021Gayle, 2024Heikkinen}. 
Preferably, Open Science principles will be followed and data and research software made openly available; such examples already exist in the ALD field \cite{ZenodoNOTE, GitHubNOTEmachball, GitHubNOTEDReaMALD, GitHubNOTEdrmaldarts}.

\section{Conclusion}

Although most publications on ALD are focused on substrates with a planar nature, there is a significant body of literature describing ALD on particulate materials.
In the early days of ALD, particulate substrates were regularly used, often for fundamental studies or to make catalysts. In the early 2000s, with the strong interest in ALD for semiconductor applications, the relative share of particulate substrates in the ALD literature declined, but more recently started rising again.
In this review, we quantitatively analyzed 799 indexed articles on this topic, published from 1988 through 2023. 

Historically, ALD research emerged earlier, and appeared in different places independently.
To our knowledge, the earliest investigations of ALD on particles started in the 1960s in the USSR, and independently later in Finland. We included the early USSR works in a qualitative recapitulation of the history of ALD on particles, thereby clarifying some of the misconceptions around this topic. 

Though the fundamental principles of ALD remain the same, our review of equipment, particulate substrates (supports) and processing, shows that a clear distinction between ALD on planar substrates (e.g., wafers) and on particulate materials must be made. 
The particulate nature of the supports requires dedicated reactors, especially when processing larger amounts of particles. 
Moreover, the typical chemical composition of the support material differs from typical planar substrates, and the surface areas to be coated can easily be several orders of magnitude larger, especially for porous or sub-micron scale particles. 
With respect to processing conditions, this larger surface area and the often porous nature of the support material requires long pulse and purge times. 
Also, typically a higher process pressure, up to atmospheric pressure, is used for ALD on particles, with implications for the mass transport phenomena of reactants. 
Finally, we highlighted a selection of characterization methods specifically aimed at  particles and coatings in the context of ALD.

With respect to applications, thermocatalysis has been the driving force of technology development in the early days of ALD on particulate materials, and still is a major application today. We therefore placed a special emphasis on this topic in the present review, where we find that many of the basic concepts of material design and processing have been developed to make thermocatalysts, but have nonetheless proven valuable to other applications. Among the plethora of reported applications, we highlighted the most popular ones, i.e., photocatalysis, electrocatalysis, batteries, luminescent phosphors, and healthcare.

Electrocatalysis and batteries play an important role in the ongoing energy transition, and thereby address a most pressing global challenge.
In these highly relevant and large-scale applications, ALD on particulate materials could potentially take a key role due to the unique materials properties attainable via the technology, similar to the key role that ALD on planar substrates has for semiconductor manufacturing today.

\section{Appendix}

Some of the articles that were not explicitly discussed were nevertheless used for the statistical figures.
Table~\ref{tab:allreferences} shows a complete list of all 799 quantitatively evaluated articles, grouped by applications.

\begin{table}[!htbp]
\footnotesize
   \centering
    \begin{tabularx}{\textwidth}{l X}
    \toprule
    \textbf{Application} & \textbf{References} \\
    \midrule
Thermocatalysis & \citenum{Adebayo2022, Afzal2020, AlbaRubio2014, Arandia2023, Arifin2012, Asakura1988b, Asundi2021, Backman1998, Backman2000, Backman2001, Backman2009, Bahrani2023, Baltes2001, Bueno2019, CamachoBunquin2015, CamachoBunquin2017, CamachoBunquin2018a, CamachoBunquin2018b, Caner2017, Canlas2022, Cao2019, Chen2006, Chen2008, Chen2009, Chen2010a, Chen2010b, Chen2023, Christensen2010, Cronauer2011, Cronauer2012, Daresibi2022, Daresibi2023, DePrins2019, Deng2011, Deng2022, DiazLopez2021, Ding2016, Du2017, Duan2019, Enterkin2011b, Fan2023, Feng2010, Feng2011a, Feng2011b, Feng2015, Feng2020, Fu2013, Gao2017a, Gao2019a, Gong2015, Gong2016, Gong2019, Gould2013, Gould2014, Gould2015a, Gould2015b, Hackler2020, Hakuli2000, Halder2020, Han2016a, Han2016b, Han2016c, He2018, Hedlund2016, Heikkinen2021, Hirva1994, Hu2019, Huang2017, Huang2021, Iiskola1997a, Iiskola1997b, Ikuno2017, Ingale2021, Jackson2013, Jackson2015, Jackson2019, Jacobs1994a, Jang2020, Jeong2014a, Jeong2016a, Jeong2016b, Jeong2016c, Jiang2014, Jin2023, Juvaste2000a, Juvaste2000b, Kanervo2001, Karinen2011, Kennedy2018, Keranen2002, Keranen2003, Kim2011a, Kim2011c, Kim2013, Kim2014a, Kim2015b, Kim2016, Kim2018a, Knemeyer2021b, Korhonen2007, Krishna2020, Kwon2021, Kytokivi1996a, Lashdaf2003a, Lashdaf2004a, Lashdaf2004b, Lee2014, Lee2021, Lee2022g, Lee2022h, Lei2012, Lei2014, Li2010, Li2015, Li2017b, Li2018a, Liang2011b, Liang2012b, Liang2013b, Liang2017, Lichty2012, Lin2018, Lin2019b, Lin2020a, Lin2020b, Lindblad1994, Lindfors1994, Littlewood2019, Liu2017a, Liu2018a, Liu2018b, Liu2020b, Liu2023b, Lobo2012, Lu2012a, Lu2012b, Lu2016a, Lu2018, Lu2019, Lu2020, Lu2022a, MakiArvela2003, Mao2019a, Mao2019b, Mao2020a, Mao2020b, McNeary2021, Meng2023, Milt2000, Milt2001, Milt2002, Moulijn2023, Muntean2013, Nadjafi2021, Najafabadi2016, Nasriddinov2019, ONeill2014, Okumura1997a, Okumura1997b, Onn2015, Onn2016, Onn2017a, Onn2017b, Onn2017c, Onn2018, PaganTorres2011, Peters2015, PiernaviejaHermida2016, PlateroPrats2017, Puurunen2002a, Puurunen2003, Qin2023, Ro2016, Ruff2019, Sairanen2012, Scheffe2010, Scheffe2011, Scheffe2013, Seo2013, Setthapun2010, Settle2019, Shang2013, Shang2017, Shen2020, Shen2023b, Shen2023c, Shi2023, Silvennoinen2007b, Smeds1996, Sree2016, Srinath2021, Strempel2016, Tan2017, Tang2021, Tian2022, Tian2023, Timonen1999, Uglietti2023, Uusitalo2000a, Uusitalo2000b, VanBui2017, Vandegehuchte2014, Verheyen2014, Voigt2019, Vuori2009, Vuori2011, Wang2014, Wang2015, Wang2016b, Wang2016c, Wang2016e, Wang2016g, Wang2017a, Wang2017b, Wang2018, Wang2019a, Wang2019b, Wang2020, Wang2021, Wang2022d, Wang2023a, Weng2018b, Weng2019, Wu2017, Xie2013, Xing2023, Xu2018b, Yan2015, Yan2017b, Yan2018a, Yan2018b, Yang2017a, Yang2017c, Yang2018, Yang2019b, Yang2020, Yang2023, Yao2016, Yi2015, Yi2017, Yoon2022b, Zhang2014a, Zhang2014b, Zhang2015b, Zhang2017a, Zhang2018b, Zhang2019a, Zhang2021, Zhao2017c, Zhao2019a, Zhilyaeva2018, Zhong2023, Zuo2022, deSouza2007, vanNorman2015}\\ 
 \midrule
Not applied & \citenum{Asakura1992, Baumgarten2022, Beetstra2009, Bloch2014, Bodalyov2019, Cai2021, Cavanagh2009, Chen2011, Christensen2009, Clancey2015, Clary2020, Coile2020, Didden2014, Didden2016, Duan2015, Duan2016a, Duan2017, Duran2014, Ek2003a, Ek2003b, Ek2004, Elam2007, Elam2010, Fu2014, GarciaGarcia2018, Gertsch2023, Goldstein2008, Goulas2013, Goulas2014, Greenberg2020, Grillo2017, Grillo2018a, Grillo2018b, Hakim2005a, Hakim2005b, Hakim2006, Hashemi2019, Haukka1993a, Haukka1993b, Haukka1994a, Haukka1994b, Haukka1994c, Haukka1995a, Haukka1996, Haukka1997a, Heikkinen2022, Ingale2020, Jackson2014, Jacobs1994b, Jaggernauth2016, Juvaste1999a, Juvaste1999b, Juvaste2000c, Kaushik2021, Ke2023, Kikuchi2016, King2007, King2008d, King2009a, Knemeyer2020, Knemeyer2021a, Koshtyal2014, KrogerLaukkanen2001, Kytokivi1996b, Kytokivi1997a, Kytokivi1997b, Lakomaa1992, Lakomaa1996, Lashdaf2003b, Lee2022b, Lei2013a, Li2016b, Li2022a, Liang2008a, Liang2009c, Liang2009d, Liang2011a, Liang2012a, Liang2012c, Liang2013a, Libera2008, Lindblad1993, Lindblad1997, Lindblad1998, Liu2013, Liu2017b, Longrie2012, Longrie2014a, Lu2007, Lu2009, Lu2010, Lu2014, Lubers2015, Mahtabani2021, Mahtabani2023, Malkov2010, Malygin1997, Malygin2002, Malygin2012, Manandhar2016, Manandhar2017, Marquez2022, Masango2014, McCormick2007a, McCormick2007b, Meledina2016, Meng2010, Molenbroek1998, Moret2020b, Myers2021, Nguyen2019, Nieminen1999, OToole2021b, Okumura1998a, Okumura1998b, Pallister2014, Park2014, Patel2015b, Puurunen2000a, Puurunen2000b, Puurunen2001, Puurunen2002b, Ramachandran2016, Rampelberg2014, Rasteiro2023, Rautiainen2002, Rauwel2012b, Scheffe2009, Shah2021, Shen2023a, Silvennoinen2007a, Sosnov2010, Sosnov2011, Sosnov2017, Sree2012, Strempel2017, Strempel2018, Suvanto1997, Suvanto1998, Tiznado2014, Tsyganenko2000, Valdesueiro2015, Wang2017c, Wang2017e, Wank2004a, Wank2004b, Weng2018a, Wiedmann2012, Wilson2008, Wolff2023, Yan2019, Yim2023, Yoon2023a, Younes2022, Zang2006, Zemtsova2020, Zhan2008, Zhang2023b, Zhou2012}\\ 
 \midrule
Batteries & \citenum{Azaceta2020, Aziz2014, Bai2016, Bao2020, Basak2022, Cai2023, Cao2021, Chae2018, Chen2019, Cheng2019, Chu2019, Dai2016, Dannehl2018, Deng2019, Fang2018b, Gao2017b, Gao2018, Gao2019b, Gao2020b, Gao2020c, Gong2018, Guan2013, Guo2022, Hall2019, Han2019, Han2022, He2014, He2020b, Hood2023, Hoskins2019b, Huang2019, Jackson2016, Jin2019, Jin2022, Jung2010a, Jung2010b, Kaliyappan2015, Kang2011, Kang2022, Kim2014b, Kim2015a, Kim2015d, Kim2021, Kim2023, Kong2016, Kong2019, Kraytsberg2015, Laskar2017, Lee2022a, Lee2022f, Lei2013b, Li2012, Li2018e, Li2019b, Li2022c, Li2022d, Li2023, Liang2020, Lin2023, Liu2015a, Liu2015b, Liu2021a, Liu2021b, Liu2023a, Lu2013b, Lu2017, Luan2012, Luo2015, Lv2016, Lv2017, Mao2020d, Meng2015, Mohanty2016, Moryson2021, Nizami2023, Ostli2021, Ostli2022, Palaparty2016, Park2017, Patel2015a, Patel2016a, Patel2016b, Patel2017, Qin2016, Riley2010, Riley2011, Saha2023, Sarkar2017a, Sarkar2017b, Shapira2018, Snyder2007, Sun2014, Sun2023, Tesfamhret2021, Tiurin2020, Vinado2018, Wang2016d, Wang2022a, Wise2015, Woo2015, Xiao2015, Xiao2017, Xie2015a, Xie2015b, Yang2022a, Yang2022c, Yu2021, Yu2022a, Yu2022b, Zhang2013, Zhang2020, Zhao2012, Zhao2013a, Zhao2013b, Zhao2019c, Zhou2020, Zhu2019, urrehman2018}\\ 
 \midrule
Photocatalysis & \citenum{Ahmad2020, Azizpour2017, Benz2020a, Benz2020b, Benz2021, Benz2022, Cao2017, Cao2018, Cao2020, Choi2013, DiMauro2017, Dominguez2018, Feng2018, Guo2018, Guo2020, Hakim2007c, Hashemi2020, Hussain2022, Jang2016, Jang2019a, Jang2019b, Jeong2014b, Justh2018, Keri2019, King2008a, King2008b, King2008c, LaZara2020, Lange2021, Lee2016, Lee2019a, Li2017a, Li2018b, Li2021, Liang2009a, Liang2010a, Liang2010b, Liang2014, Liu2016b, Liu2022, Lopez2021, MartinSomer2020, Nagy2016, Pan2020, Podurets2020, Rihova2021, Scott2019, Seong2019a, Shin2020, Singh2016, Sridharan2015, Wang2017f, Williams2012, Xie2022, Yang2017b, Yang2022b, Zhang2017c, Zhang2019b, Zhao2017a, Zhao2017b, Zhao2018, Zhou2010a, Zhou2010b, Zi2021, vanOmmen2015}\\ 
 \midrule
Electrocatalysis & \citenum{Chen2016, Cheng2014, Cheng2015a, Cheng2015b, Cheng2016, Gan2020, Godoy2021, Han2020, He2020a, Hsu2011, Hsu2012, Hsu2015, Jo2023, Kamitaka2021, Kim2015c, Kim2018b, Lee2019c, Lee2020a, Lee2020b, Lee2022c, Lee2022d, Li2022b, Liu2016a, Liu2023c, Lubers2016, Lubers2017, Luo2019, McNeary2018, McNeary2019, McNeary2020, Page2019, Palmer2018, Ray2012, Rikkinen2011, Saha2015, Saha2016, Sairanen2014, Shi2019, Song2017, Song2018, Song2020, Sun2013b, Sun2019, Tsai2020, Wang2016a, Wang2016f, Wang2019c, Xu2018a, Zaccarine2022, Zhang2015a, Zhang2019c, Zhang2019e}\\ 
 \midrule
Healthcare & \citenum{Barros2023, Duong2022, Garcea2020, Gupta2022, HaghshenasLari2018, Hautala2017, Hellrup2017, Hellrup2019, Hilton2017, Hirschberg2019, Kaariainen2017, LaZara2021a, LaZara2021b, Liang2009b, LopezdeDicastillo2019, Mestres2016, Mittal2022, Moret2020a, Moseson2022, Moseson2023a, Sengar2019, Seong2016, Seong2019b, Sosa2023, Swaminathan2023, Uudekull2017, Vasudevan2015, Witeof2022, Witeof2023, Zhang2017b, Zhang2019d, Zhang2023}\\ 
 \midrule
Luminescent Phosphors & \citenum{Huang2022, Jeong2009, Karacaoglu2020, Karacaoglu2023, Kim2007a, Kim2007b, Kim2009b, Kim2011d, Rauwel2011, Rauwel2012a, Sosnov2012, Verstraete2019, Yoon2011, Zhang2018a, Zhao2019b, Zhao2020, Zhao2022, Zhao2023, Zhou2016, tenKate2019}\\ 
 \midrule
Other application & \citenum{He2022, Jung2022, Kim2020, Lee2022e, Lee2023, Lehmann2023, Li2018d, Li2020, Lim2020b, Zhang2019f, Armutlulu2017, Dong2019, Kim2019a, Lai2021, Li2018c, Liang2008b, Wang2017d, Eom2015, Ermakova2002, Hakim2007b, Kim2011b, Ok2017, Ugur2021, Xiang2022, BorbonNunez2017, Lin2013, Meng2011, MunozMunoz2015, Sree2017, Zhao2015, Cremers2018, Hakim2007a, Hoskins2018, Hoskins2019a, Nam2012, Qin2019, Li2019a, Liu2020a, Sun2012, Sun2013a, Wang2017g, Hering2021, Lichty2013, OToole2019, OToole2020, Ferguson2005, Qin2013, Qin2017, Yan2017a, He2021a, Kim2019c, Lim2020a, Sharma2023, Bull2020, Chen2017, Leick2021, Bohus2022, GilFont2020, Varady2022, Choi2011, Duan2016b, Smirnov2008, Liang2007a, Liang2007b, Nevalainen2012, King2009b, King2009c, King2009d, Wang2022b, Wang2022c, Wang2023b, Bhattacharya2019, Zhang2018c, Cremers2019, Gao2020a, Miller2020, Chazot2022, OToole2021a, Moghtaderi2006, Kim2022a, Devine2011, Housauer2020, Kilbury2012, Duran2016, Bull2021, He2021b, FornerEscrig2021, Park2012, Nevalainen2009, Valdesueiro2017, Valdesueiro2016, Zhang2015c, Qin2018, Weimer2008, Qi2022, Navarrete2020, Shvareva2008, Fang2020}
\\ 
\bottomrule
  
\end{tabularx}
\caption{Literature references to articles used in quantitative analyses throughout this review; grouped by their application areas.}
\label{tab:allreferences}
\end{table}

\begin{acknowledgement}

We thank Rens Kamphorst and Kalani Ostermeijer for proof-reading and feedback on the manuscript. RLP thanks Outi Krause for valuable discussions related to the history of catalysis ALD.  We acknowledge the Virtual Project on the History of ALD (VPHA) as the source of translations of article contents originally written in Cyrillic letters (Russian, Bulgarian). 

\end{acknowledgement}

\begin{suppinfo}

Supplementary information with additional quantitative analyses is available in the provided PDF file.

The tabulated data (which is the basis of the graphs in this review) is accessible through a public repository\cite{ALDpmZenodo}, along with instructions on how to efficiently access this dataset via Python scripts\cite{2025GithubALDpm}.

\end{suppinfo}


\clearpage
\bibliography{ALDpmBibliography.bib}

\begin{mcitethebibliography}{1047}
\bibitem[MLa()]{MLalsoInEnglishNOTE}
Occasionally, it has been claimed in the scientific literature, that publications on ML-ALD were available only in Russian. Such statements are demonstrably incorrect.\relax
\mciteBstWouldAddEndPunctfalse
\mciteSetBstMidEndSepPunct{\mcitedefaultmidpunct}
{}{\mcitedefaultseppunct}\relax
\EndOfBibitem
\bibitem[Cat()]{CatALDcitationJAPANnote}
The current authors are not aware of evidence that the Japanese researchers would have been aware of the ML--ALD or ALE--ALD works.\relax
\mciteBstWouldAddEndPunctfalse
\mciteSetBstMidEndSepPunct{\mcitedefaultmidpunct}
{}{\mcitedefaultseppunct}\relax
\EndOfBibitem
\end{mcitethebibliography}

\end{document}










\begin{figure}
    \centering
    \includegraphics[width=0.75\linewidth]{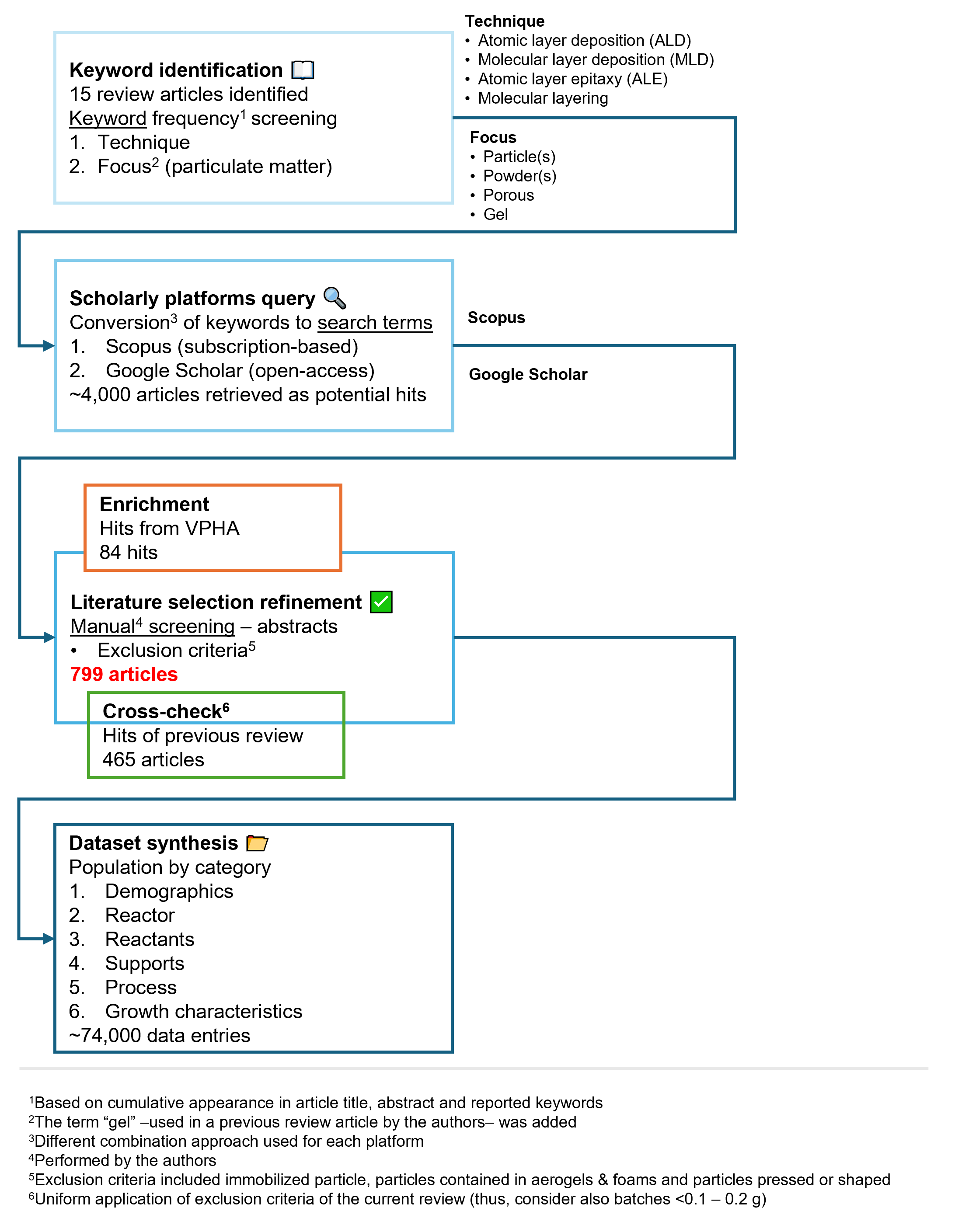}
    \caption{Search methodology applied in the current review describing the main steps involved to enable data-driven analytic insights.}
    \label{fig:method_vert}
\end{figure}

\begin{figure}
    \centering
    \includegraphics[width=0.5\linewidth]{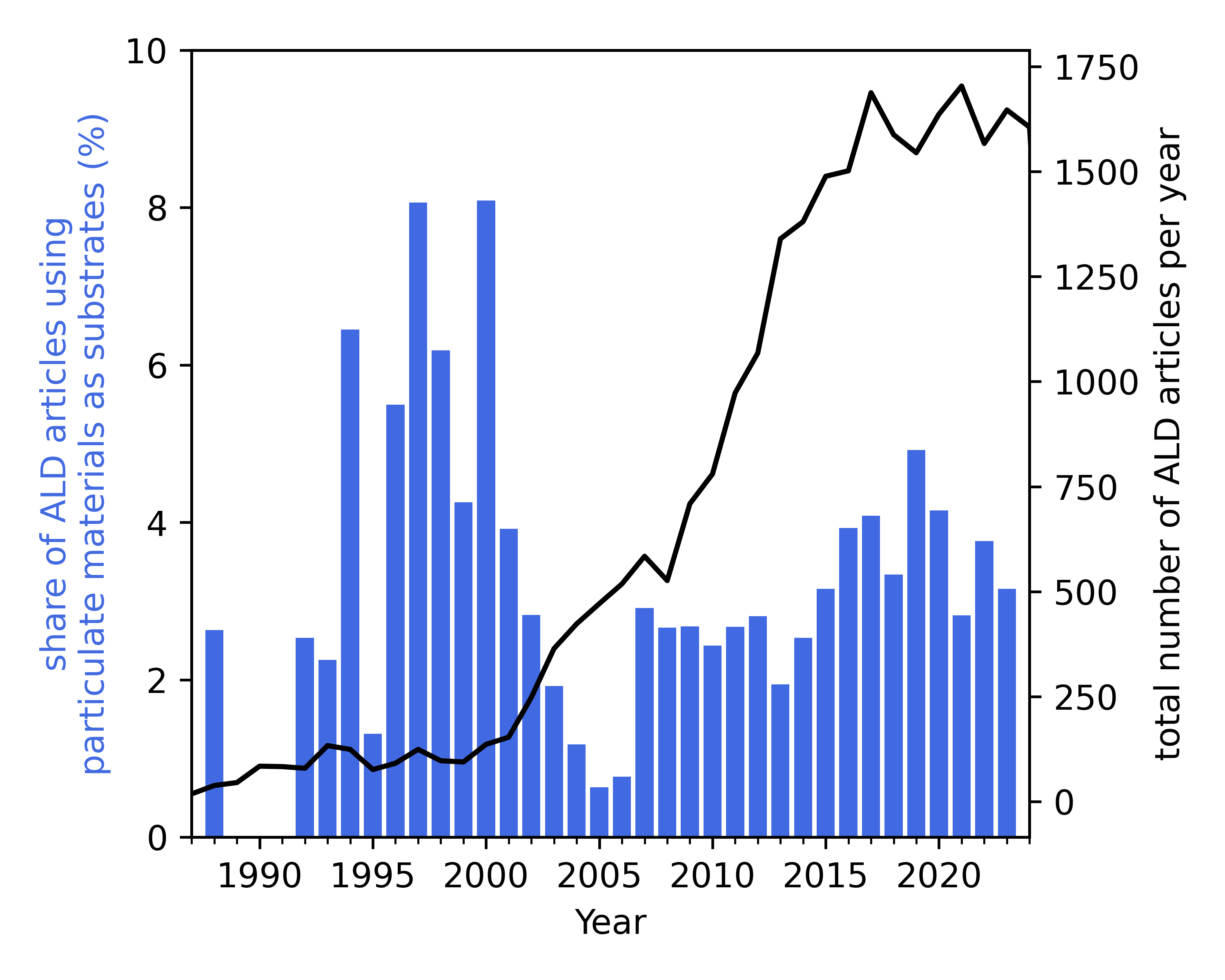}
    \caption{Share of articles from the field of ALD that use particulate materials as substrate. The number of ALD articles using particles as substrate per year is taken from our curated list of articles from this subfield of ALD (i.e. the main dataset used throughout the review). The number of articles from the field of ALD in general was obtained from a search run on Scopus using the five most common technology names for ALD that we identified from the dataset, i.e., searching "atomic layer epitaxy" OR "atomic layer deposition" OR "atomic layer chemical vapor deposition" OR "molecular layer deposition" OR "molecular layering" for the fields of title, abstract and keywords.}
    \label{fig:enter-label}
\end{figure}

\begin{figure}
  \includegraphics[width=0.75\textwidth]{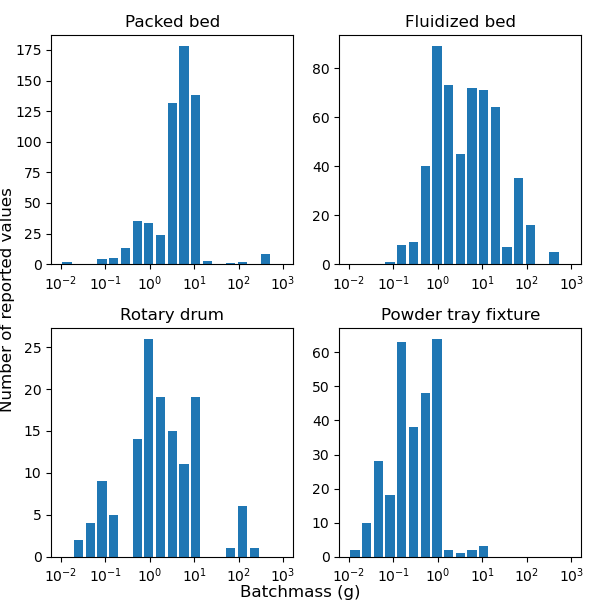}
  \caption{Reported batch masses for reactor types discussed in the main manuscript.}
  \label{SI:histobatchreactor}
\end{figure}

\begin{figure}
  \includegraphics[width=0.75\textwidth]{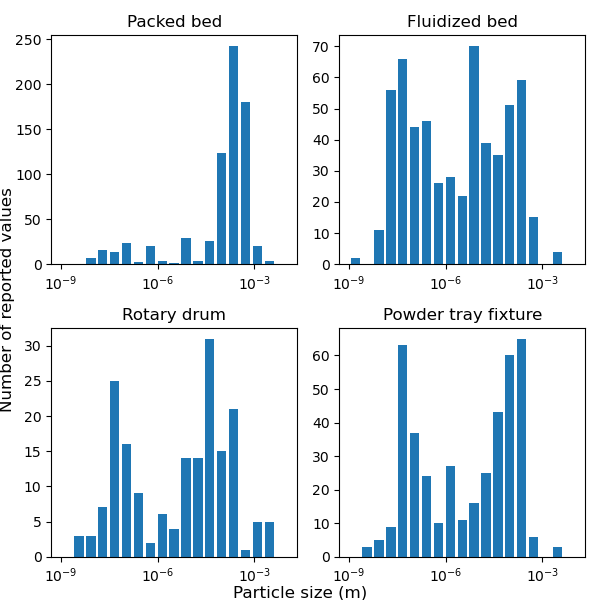}
  \caption{Reported particle size according to reactor type.}
  \label{SI:histosizereactor}
\end{figure}

\begin{figure}
  \includegraphics[width=0.75\textwidth]{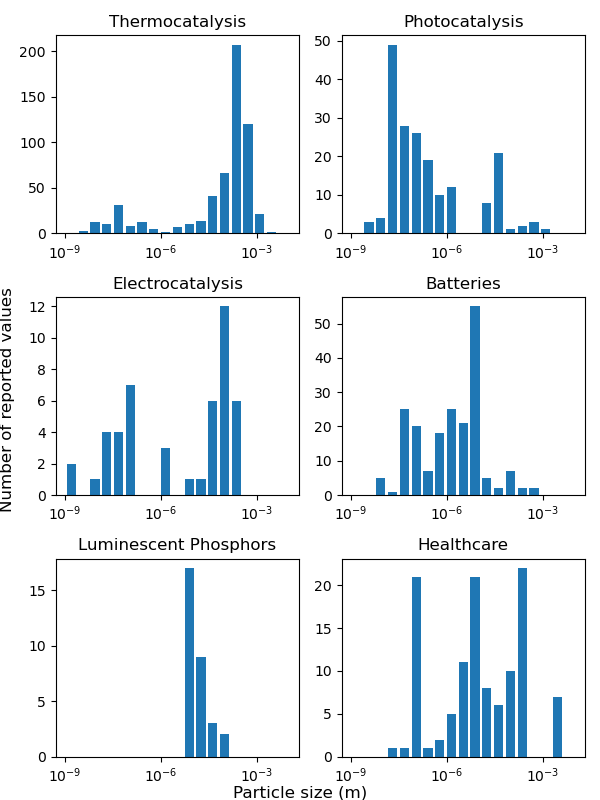}
  \caption{Reported particle diameter according to application.}
  \label{SI:histosizeapplication}
\end{figure}

\begin{figure}
    \centering
    \includegraphics[width=0.75\linewidth]{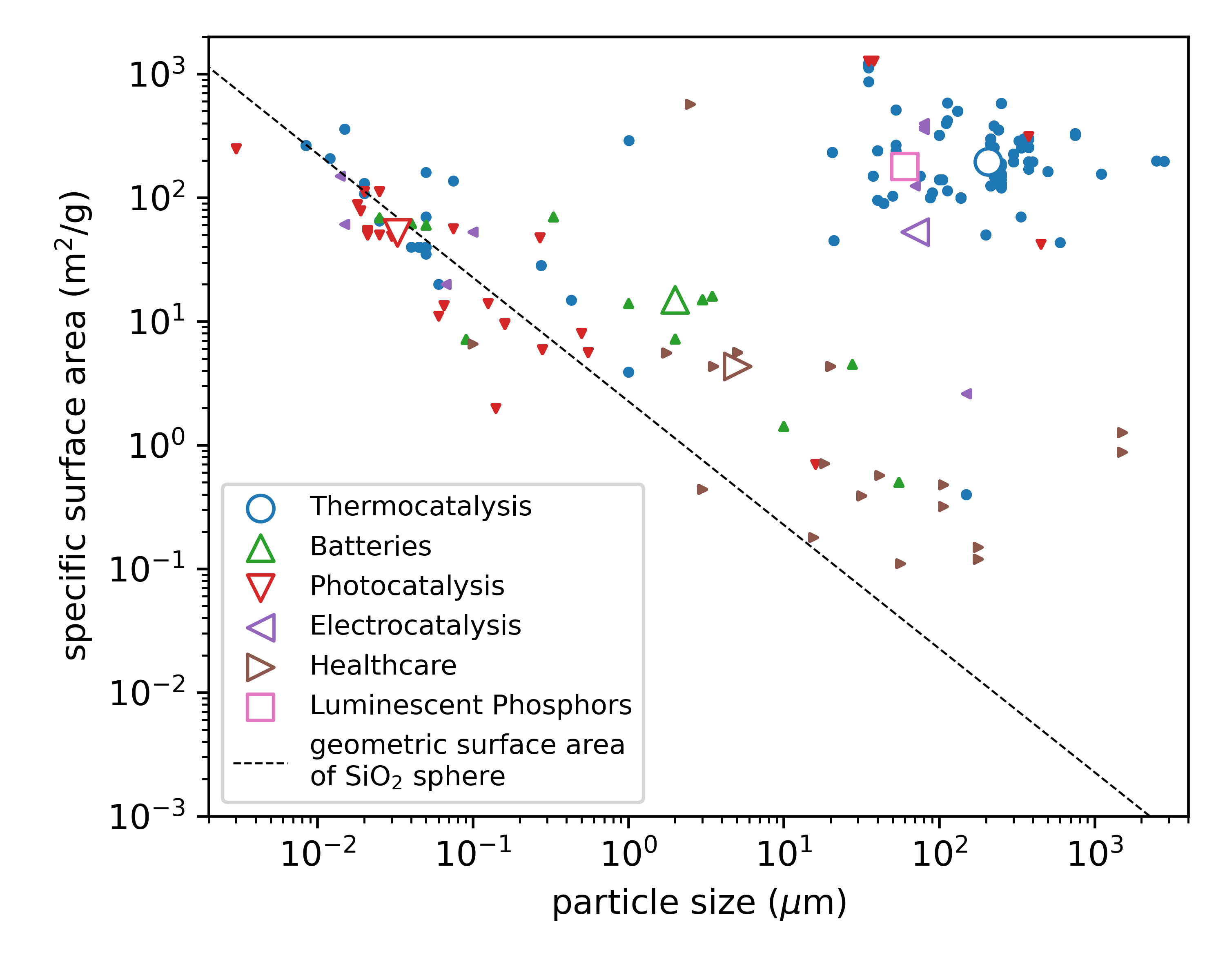}
    \caption{Specific surface area versus particle diameter per type of support material (where both values where reported). The large open symbols (see legend) mark the median values of the smaller filled symbols of the same color and shape.}
    \label{fig:SSA_appl}
\end{figure}

\begin{figure}
    \centering
    \includegraphics[width=1\linewidth]{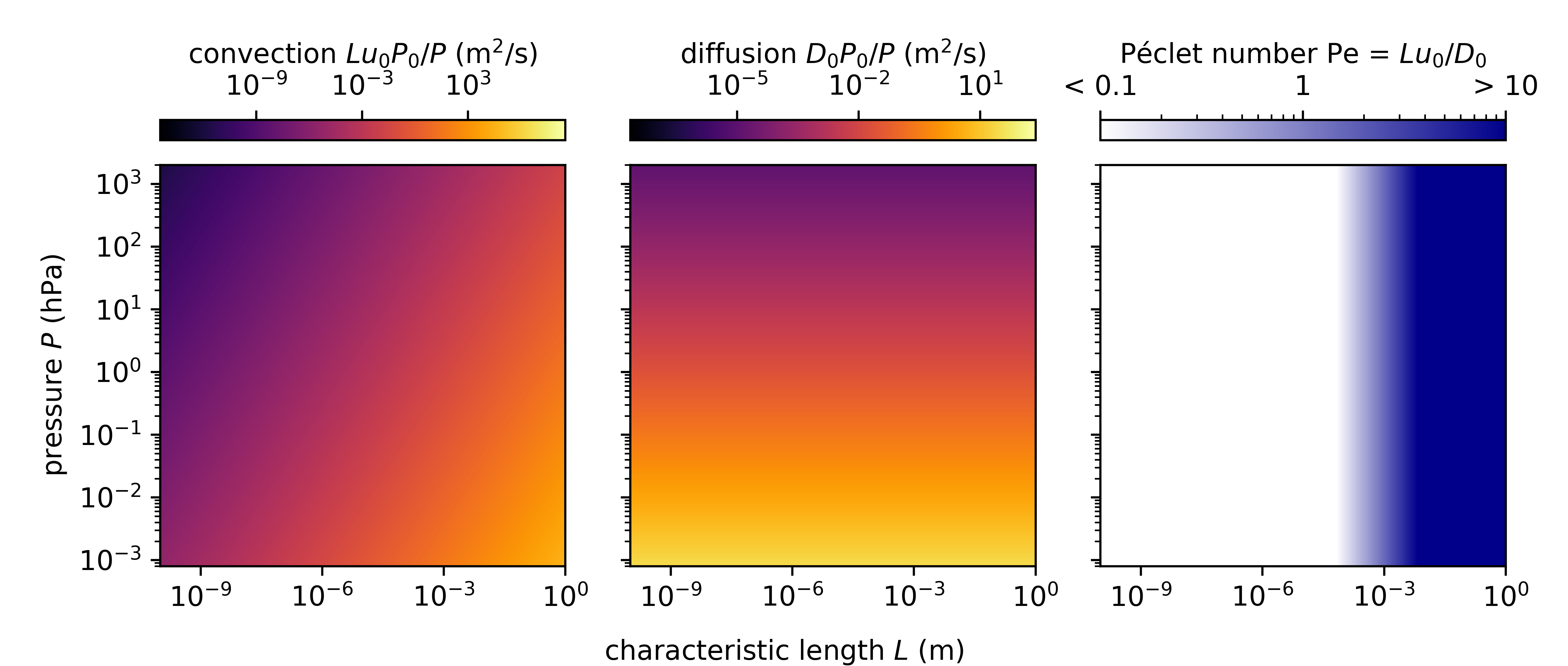}
    \caption{Mass transport of TMA through convection and diffusion, and Péclet number (ratio of the two), depending on pressure and characteristic length scale; calculated as outlined in the main manuscript, however, at constant mass flow of 1\,slm instead of constant gas velocity.}
    \label{fig:SI_masstransport}
\end{figure}

\begin{table}
    \centering
    \begin{tabular}{cc}
           Technique& Count\\
           \hline
           TEM (incl. special variants)& 891\\
           ICP-OES& 652\\
           INAA& 189\\
           STEM (incl. special variants)& 148\\
           XRF& 114\\
           EDX& 90\\
           TGA& 76\\
           AAS& 73\\
           LECO& 63\\
           ICP-MS& 60\\
           SEM (incl. special variants)&54\\
           XPS&32\\
    \end{tabular}
    \caption{Top 12 reported characterization techniques to for imaging of ALD deposited films or particles, or amount of deposited material. See main manuscript for explanation of abbreviations.}
    \label{tab:characaterization_ranking}
\end{table}

\clearpage

\begin{footnotesize}
\begin{longtable}{ccc}
\caption{List of application areas covered by articles used in quantitative analyses throughout this review.}\label{tab:allapplications}\\
\toprule
\textbf{Application} & \textbf{Number of publications} & \textbf{References} \\
\midrule
\endfirsthead

\toprule
\textbf{Application} & \textbf{Number of publications} & \textbf{References} \\
\midrule

\endhead

\midrule
\endfoot

\bottomrule
\endlastfoot
Thermocatalysis & 255 & \citenum{Adebayo2022, Afzal2020, AlbaRubio2014, Arandia2023, Arifin2012, Asakura1988b, Asundi2021, Backman1998, Backman2000, Backman2001, Backman2009, Bahrani2023, Baltes2001, Bueno2019, CamachoBunquin2015, CamachoBunquin2017, CamachoBunquin2018a, CamachoBunquin2018b, Caner2017, Canlas2022, Cao2019, Chen2006, Chen2008, Chen2009, Chen2010a, Chen2010b, Chen2023, Christensen2010, Cronauer2011, Cronauer2012, Daresibi2022, Daresibi2023, DePrins2019, Deng2011, Deng2022, DiazLopez2021, Ding2016, Du2017, Duan2019, Enterkin2011b, Fan2023, Feng2010, Feng2011a, Feng2011b, Feng2015, Feng2020, Fu2013, Gao2017a, Gao2019a, Gong2015, Gong2016, Gong2019, Gould2013, Gould2014, Gould2015a, Gould2015b, Hackler2020, Hakuli2000, Halder2020, Han2016a, Han2016b, Han2016c, He2018, Hedlund2016, Heikkinen2021, Hirva1994, Hu2019, Huang2017, Huang2021, Iiskola1997a, Iiskola1997b, Ikuno2017, Ingale2021, Jackson2013, Jackson2015, Jackson2019, Jacobs1994a, Jang2020, Jeong2014a, Jeong2016a, Jeong2016b, Jeong2016c, Jiang2014, Jin2023, Juvaste2000a, Juvaste2000b, Kanervo2001, Karinen2011, Kennedy2018, Keranen2002, Keranen2003, Kim2011a, Kim2011c, Kim2013, Kim2014a, Kim2015b, Kim2016, Kim2018a, Knemeyer2021b, Korhonen2007, Krishna2020, Kwon2021, Kytokivi1996a, Lashdaf2003a, Lashdaf2004a, Lashdaf2004b, Lee2014, Lee2021, Lee2022g, Lee2022h, Lei2012, Lei2014, Li2010, Li2015, Li2017b, Li2018a, Liang2011b, Liang2012b, Liang2013b, Liang2017, Lichty2012, Lin2018, Lin2019b, Lin2020a, Lin2020b, Lindblad1994, Lindfors1994, Littlewood2019, Liu2017a, Liu2018a, Liu2018b, Liu2020b, Liu2023b, Lobo2012, Lu2012a, Lu2012b, Lu2016a, Lu2018, Lu2019, Lu2020, Lu2022a, MakiArvela2003, Mao2019a, Mao2019b, Mao2020a, Mao2020b, McNeary2021, Meng2023, Milt2000, Milt2001, Milt2002, Moulijn2023, Muntean2013, Nadjafi2021, Najafabadi2016, Nasriddinov2019, ONeill2014, Okumura1997a, Okumura1997b, Onn2015, Onn2016, Onn2017a, Onn2017b, Onn2017c, Onn2018, PaganTorres2011, Peters2015, PiernaviejaHermida2016, PlateroPrats2017, Puurunen2002a, Puurunen2003, Qin2023, Ro2016, Ruff2019, Sairanen2012, Scheffe2010, Scheffe2011, Scheffe2013, Seo2013, Setthapun2010, Settle2019, Shang2013, Shang2017, Shen2020, Shen2023b, Shen2023c, Shi2023, Silvennoinen2007b, Smeds1996, Sree2016, Srinath2021, Strempel2016, Tan2017, Tang2021, Tian2022, Tian2023, Timonen1999, Uglietti2023, Uusitalo2000a, Uusitalo2000b, VanBui2017, Vandegehuchte2014, Verheyen2014, Voigt2019, Vuori2009, Vuori2011, Wang2014, Wang2015, Wang2016b, Wang2016c, Wang2016e, Wang2016g, Wang2017a, Wang2017b, Wang2018, Wang2019a, Wang2019b, Wang2020, Wang2021, Wang2022d, Wang2023a, Weng2018b, Weng2019, Wu2017, Xie2013, Xing2023, Xu2018b, Yan2015, Yan2017b, Yan2018a, Yan2018b, Yang2017a, Yang2017c, Yang2018, Yang2019b, Yang2020, Yang2023, Yao2016, Yi2015, Yi2017, Yoon2022b, Zhang2014a, Zhang2014b, Zhang2015b, Zhang2017a, Zhang2018b, Zhang2019a, Zhang2021, Zhao2017c, Zhao2019a, Zhilyaeva2018, Zhong2023, Zuo2022, deSouza2007, vanNorman2015}\\ 
 \midrule
Not applied & 159 & \citenum{Asakura1992, Baumgarten2022, Beetstra2009, Bloch2014, Bodalyov2019, Cai2021, Cavanagh2009, Chen2011, Christensen2009, Clancey2015, Clary2020, Coile2020, Didden2014, Didden2016, Duan2015, Duan2016a, Duan2017, Duran2014, Ek2003a, Ek2003b, Ek2004, Elam2007, Elam2010, Fu2014, GarciaGarcia2018, Gertsch2023, Goldstein2008, Goulas2013, Goulas2014, Greenberg2020, Grillo2017, Grillo2018a, Grillo2018b, Hakim2005a, Hakim2005b, Hakim2006, Hashemi2019, Haukka1993a, Haukka1993b, Haukka1994a, Haukka1994b, Haukka1994c, Haukka1995a, Haukka1996, Haukka1997a, Heikkinen2022, Ingale2020, Jackson2014, Jacobs1994b, Jaggernauth2016, Juvaste1999a, Juvaste1999b, Juvaste2000c, Kaushik2021, Ke2023, Kikuchi2016, King2007, King2008d, King2009a, Knemeyer2020, Knemeyer2021a, Koshtyal2014, KrogerLaukkanen2001, Kytokivi1996b, Kytokivi1997a, Kytokivi1997b, Lakomaa1992, Lakomaa1996, Lashdaf2003b, Lee2022b, Lei2013a, Li2016b, Li2022a, Liang2008a, Liang2009c, Liang2009d, Liang2011a, Liang2012a, Liang2012c, Liang2013a, Libera2008, Lindblad1993, Lindblad1997, Lindblad1998, Liu2013, Liu2017b, Longrie2012, Longrie2014a, Lu2007, Lu2009, Lu2010, Lu2014, Lubers2015, Mahtabani2021, Mahtabani2023, Malkov2010, Malygin1997, Malygin2002, Malygin2012, Manandhar2016, Manandhar2017, Marquez2022, Masango2014, McCormick2007a, McCormick2007b, Meledina2016, Meng2010, Molenbroek1998, Moret2020b, Myers2021, Nguyen2019, Nieminen1999, OToole2021b, Okumura1998a, Okumura1998b, Pallister2014, Park2014, Patel2015b, Puurunen2000a, Puurunen2000b, Puurunen2001, Puurunen2002b, Ramachandran2016, Rampelberg2014, Rasteiro2023, Rautiainen2002, Rauwel2012b, Scheffe2009, Shah2021, Shen2023a, Silvennoinen2007a, Sosnov2010, Sosnov2011, Sosnov2017, Sree2012, Strempel2017, Strempel2018, Suvanto1997, Suvanto1998, Tiznado2014, Tsyganenko2000, Valdesueiro2015, Wang2017c, Wang2017e, Wank2004a, Wank2004b, Weng2018a, Wiedmann2012, Wilson2008, Wolff2023, Yan2019, Yim2023, Yoon2023a, Younes2022, Zang2006, Zemtsova2020, Zhan2008, Zhang2023b, Zhou2012}\\ 
 \midrule
Batteries & 118 & \citenum{Azaceta2020, Aziz2014, Bai2016, Bao2020, Basak2022, Cai2023, Cao2021, Chae2018, Chen2019, Cheng2019, Chu2019, Dai2016, Dannehl2018, Deng2019, Fang2018b, Gao2017b, Gao2018, Gao2019b, Gao2020b, Gao2020c, Gong2018, Guan2013, Guo2022, Hall2019, Han2019, Han2022, He2014, He2020b, Hood2023, Hoskins2019b, Huang2019, Jackson2016, Jin2019, Jin2022, Jung2010a, Jung2010b, Kaliyappan2015, Kang2011, Kang2022, Kim2014b, Kim2015a, Kim2015d, Kim2021, Kim2023, Kong2016, Kong2019, Kraytsberg2015, Laskar2017, Lee2022a, Lee2022f, Lei2013b, Li2012, Li2018e, Li2019b, Li2022c, Li2022d, Li2023, Liang2020, Lin2023, Liu2015a, Liu2015b, Liu2021a, Liu2021b, Liu2023a, Lu2013b, Lu2017, Luan2012, Luo2015, Lv2016, Lv2017, Mao2020d, Meng2015, Mohanty2016, Moryson2021, Nizami2023, Ostli2021, Ostli2022, Palaparty2016, Park2017, Patel2015a, Patel2016a, Patel2016b, Patel2017, Qin2016, Riley2010, Riley2011, Saha2023, Sarkar2017a, Sarkar2017b, Shapira2018, Snyder2007, Sun2014, Sun2023, Tesfamhret2021, Tiurin2020, Vinado2018, Wang2016d, Wang2022a, Wise2015, Woo2015, Xiao2015, Xiao2017, Xie2015a, Xie2015b, Yang2022a, Yang2022c, Yu2021, Yu2022a, Yu2022b, Zhang2013, Zhang2020, Zhao2012, Zhao2013a, Zhao2013b, Zhao2019c, Zhou2020, Zhu2019, urrehman2018}\\ 
 \midrule
Photocatalysis & 65 & \citenum{Ahmad2020, Azizpour2017, Benz2020a, Benz2020b, Benz2021, Benz2022, Cao2017, Cao2018, Cao2020, Choi2013, DiMauro2017, Dominguez2018, Feng2018, Guo2018, Guo2020, Hakim2007c, Hashemi2020, Hussain2022, Jang2016, Jang2019a, Jang2019b, Jeong2014b, Justh2018, Keri2019, King2008a, King2008b, King2008c, LaZara2020, Lange2021, Lee2016, Lee2019a, Li2017a, Li2018b, Li2021, Liang2009a, Liang2010a, Liang2010b, Liang2014, Liu2016b, Liu2022, Lopez2021, MartinSomer2020, Nagy2016, Pan2020, Podurets2020, Rihova2021, Scott2019, Seong2019a, Shin2020, Singh2016, Sridharan2015, Wang2017f, Williams2012, Xie2022, Yang2017b, Yang2022b, Zhang2017c, Zhang2019b, Zhao2017a, Zhao2017b, Zhao2018, Zhou2010a, Zhou2010b, Zi2021, vanOmmen2015}\\ 
 \midrule
Electrocatalysis & 52 & \citenum{Chen2016, Cheng2014, Cheng2015a, Cheng2015b, Cheng2016, Gan2020, Godoy2021, Han2020, He2020a, Hsu2011, Hsu2012, Hsu2015, Jo2023, Kamitaka2021, Kim2015c, Kim2018b, Lee2019c, Lee2020a, Lee2020b, Lee2022c, Lee2022d, Li2022b, Liu2016a, Liu2023c, Lubers2016, Lubers2017, Luo2019, McNeary2018, McNeary2019, McNeary2020, Page2019, Palmer2018, Ray2012, Rikkinen2011, Saha2015, Saha2016, Sairanen2014, Shi2019, Song2017, Song2018, Song2020, Sun2013b, Sun2019, Tsai2020, Wang2016a, Wang2016f, Wang2019c, Xu2018a, Zaccarine2022, Zhang2015a, Zhang2019c, Zhang2019e}\\ 
 \midrule
Medical & 32 & \citenum{Barros2023, Duong2022, Garcea2020, Gupta2022, HaghshenasLari2018, Hautala2017, Hellrup2017, Hellrup2019, Hilton2017, Hirschberg2019, Kaariainen2017, LaZara2021a, LaZara2021b, Liang2009b, LopezdeDicastillo2019, Mestres2016, Mittal2022, Moret2020a, Moseson2022, Moseson2023a, Sengar2019, Seong2016, Seong2019b, Sosa2023, Swaminathan2023, Uudekull2017, Vasudevan2015, Witeof2022, Witeof2023, Zhang2017b, Zhang2019d, Zhang2023}\\ 
 \midrule
Phosphors & 20 & \citenum{Huang2022, Jeong2009, Karacaoglu2020, Karacaoglu2023, Kim2007a, Kim2007b, Kim2009b, Kim2011d, Rauwel2011, Rauwel2012a, Sosnov2012, Verstraete2019, Yoon2011, Zhang2018a, Zhao2019b, Zhao2020, Zhao2022, Zhao2023, Zhou2016, tenKate2019}\\ 
 \midrule
Thermoelectric materials & 10 & \citenum{He2022, Jung2022, Kim2020, Lee2022e, Lee2023, Lehmann2023, Li2018d, Li2020, Lim2020b, Zhang2019f}\\ 
 \midrule
Adsorbents \& separations & 7 & \citenum{Armutlulu2017, Dong2019, Kim2019a, Lai2021, Li2018c, Liang2008b, Wang2017d}\\ 
 \midrule
Colloidal interaction & 7 & \citenum{Eom2015, Ermakova2002, Hakim2007b, Kim2011b, Ok2017, Ugur2021, Xiang2022}\\ 
 \midrule
Nanofabrication & 6 & \citenum{BorbonNunez2017, Lin2013, Meng2011, MunozMunoz2015, Sree2017, Zhao2015}\\ 
 \midrule
Oxidation control & 6 & \citenum{Cremers2018, Hakim2007a, Hoskins2018, Hoskins2019a, Nam2012, Qin2019}\\ 
 \midrule
Capacitors & 5 & \citenum{Li2019a, Liu2020a, Sun2012, Sun2013a, Wang2017g}\\ 
 \midrule
Ceramic materials & 4 & \citenum{Hering2021, Lichty2013, OToole2019, OToole2020}\\ 
 \midrule
Thermites & 4 & \citenum{Ferguson2005, Qin2013, Qin2017, Yan2017a}\\ 
 \midrule
Photo-electrochemistry & 4 & \citenum{He2021a, Kim2019c, Lim2020a, Sharma2023}\\ 
 \midrule
Hydrogen storage & 3 & \citenum{Bull2020, Chen2017, Leick2021}\\ 
 \midrule
Nanofluids & 3 & \citenum{Bohus2022, GilFont2020, Varady2022}\\ 
 \midrule
Magnetic nanoparticles & 3 & \citenum{Choi2011, Duan2016b, Smirnov2008}\\ 
 \midrule
Ceramic polymer composite materials & 3 & \citenum{Liang2007a, Liang2007b, Nevalainen2012}\\ 
 \midrule
Bandgap engineering & 3 & \citenum{King2009b, King2009c, King2009d}\\ 
 \midrule
Soft magnetic composities & 3 & \citenum{Wang2022b, Wang2022c, Wang2023b}\\ 
 \midrule
Nuclear power & 2 & \citenum{Bhattacharya2019, Zhang2018c}\\ 
 \midrule
Corrosion protection & 2 & \citenum{Cremers2019, Gao2020a}\\ 
 \midrule
Fillers for additive manufacturing & 1 & \citenum{Miller2020}\\ 
 \midrule
Lasers & 1 & \citenum{Chazot2022}\\ 
 \midrule
Heat sinks & 1 & \citenum{OToole2021a}\\ 
 \midrule
Combustion prevention & 1 & \citenum{Moghtaderi2006}\\ 
 \midrule
Schottky contacts & 1 & \citenum{Kim2022a}\\ 
 \midrule
Toxicology & 1 & \citenum{Devine2011}\\ 
 \midrule
Nuclear materials & 1 & \citenum{Housauer2020}\\ 
 \midrule
Lubrication & 1 & \citenum{Kilbury2012}\\ 
 \midrule
Multiferroic ceramics & 1 & \citenum{Duran2016}\\ 
 \midrule
Nuclear propulsion via H2 heating & 1 & \citenum{Bull2021}\\ 
 \midrule
Oxygen scavengers & 1 & \citenum{He2021b}\\ 
 \midrule
Phase change materials & 1 & \citenum{FornerEscrig2021}\\ 
 \midrule
Photovoltaics & 1 & \citenum{Park2012}\\ 
 \midrule
Polymer composite material & 1 & \citenum{Nevalainen2009}\\ 
 \midrule
Powder coating (varnish) & 1 & \citenum{Valdesueiro2017}\\ 
 \midrule
Radioactive particle tracking & 1 & \citenum{Valdesueiro2016}\\ 
 \midrule
Raman spectroscopy & 1 & \citenum{Zhang2015c}\\ 
 \midrule
Rocket fuels & 1 & \citenum{Qin2018}\\ 
 \midrule
Semiconductors & 1 & \citenum{Weimer2008}\\ 
 \midrule
Sensors & 1 & \citenum{Qi2022}\\ 
 \midrule
Thermal energy storage & 1 & \citenum{Navarrete2020}\\ 
 \midrule
Thermochemical characterization & 1 & \citenum{Shvareva2008}\\ 
 \midrule
QD as phosphor alternatives & 1 & \citenum{Fang2020}\\ 
\bottomrule
\end{longtable}
\end{footnotesize}

\begin{footnotesize}
\begin{longtable}{c c >{\raggedright\arraybackslash}p{10cm}}
\caption{List of Top 9 countries based on the number of articles published throughout this review.}\label{tab:allcountry}\\
\toprule
\textbf{Country} & \textbf{Number of publications} & \textbf{References} \\
\midrule
\endfirsthead

\toprule
\textbf{Country} & \textbf{Number of publications} & \textbf{References} \\
\midrule

\endhead

\midrule
\endfoot

\bottomrule
\endlastfoot

    America & 291 & \citenum{CamachoBunquin2018a,King2008a, King2008b, King2008c, King2008d, Liang2008a, Liang2008b, Fang2018b, Wank2004a, Wank2004b, Ferguson2005, Hakim2005a, Hakim2005b, Hakim2006, Zhou2010a, Zhou2010b, Jung2010a, Jung2010b, Gertsch2023, Moseson2023a, Swaminathan2023, Canlas2022, Jin2022, Kang2022, Lai2021, Lee2022b, Hakim2007a, Hakim2007b, Hakim2007c, OToole2021a, OToole2021b, Onn2017a, Onn2017b, Onn2017c, Wang2017d, Wang2017f, Wang2017b, Wang2017c, Wang2017e, Zhang2014b, Zhang2014a, Godoy2021, McNeary2021, Zaccarine2022, Bull2021, Xu2018a, Chen2016, Liu2017b, Yang2017a, Yang2017c, Leick2021, Myers2021, Shah2021, Bull2020, Wang2016a, Wang2016g, Greenberg2020, Coile2020, He2020a, Krishna2020, Lu2020, McNeary2020, McNeary2019, McNeary2018, Lubers2015, Lubers2016, Lubers2017, Miller2020, OToole2020, Patel2016a, Patel2015a, Patel2016b, Zhao2017b, Pan2020, Bhattacharya2019, King2007, Liang2007a, McCormick2007a, McCormick2007b, Snyder2007, Weimer2008, Wilson2008, Zhan2008, Cavanagh2009, King2009a, King2009b, King2009c, King2009d, Liang2009a, Liang2009b, Liang2009c, Liang2009d, Scheffe2009, Jackson2016, Gould2015a, Gould2015b, Gould2013, Christensen2009, Christensen2010, Elam2007, Elam2010, Feng2010, Libera2008, Goldstein2008, Scheffe2013, Scheffe2011, Lichty2013, Lu2009, Shvareva2008, Palaparty2016, PiernaviejaHermida2016, Gao2019b, Li2010, Liang2010a, Liang2010b, Lu2010, Riley2011, Hoskins2019b, Hoskins2019a, Riley2010, Scheffe2010, Setthapun2010, Cronauer2011, Deng2011, Enterkin2011b, Feng2011a, Feng2011b, Hsu2011, Liang2011a, Liang2011b, PaganTorres2011, Arifin2012, Cronauer2012, Hsu2012, Kilbury2012, Lei2012, Liang2007b, Jackson2019, Liang2012a, Liang2012b, Liang2012c, Lichty2012, Lobo2012, Lu2012a, Lu2012b, Luan2012, Ray2012, Jin2019, Sun2012, Wiedmann2012, Zhao2012, Zhou2012, Guan2013, Jackson2013, Lei2013a, Lei2013b, Liang2013b, Liang2013a, Lu2013b, Muntean2013, Shang2013, Sun2013a, Xie2013, Zhang2013, Zhao2013a, Zhao2013b, Mohanty2016, Lin2019b, AlbaRubio2014, Aziz2014, Kennedy2018, Fu2014, Gould2014, He2014, Jackson2014, Xie2015a, vanNorman2015, Littlewood2019, Lu2019, Mao2019a, Mao2019b, Page2019, Settle2019, Hoskins2018, Vinado2018, Kim2018a, Ke2023, Qin2023, Weng2019, Weng2018a, Zhang2019a, Zhang2019b, Peters2015, CamachoBunquin2015, Kim2015c, Onn2015, Woo2015, Clancey2015, OToole2019, Feng2018, Gao2018, Manandhar2017, CamachoBunquin2017, Sarkar2017b, Patel2015b, Sarkar2017a, PlateroPrats2017, Patel2017, Hedlund2016, Manandhar2016, Onn2016, Saha2016, CamachoBunquin2018b, Lin2018, Liu2018a, Lu2018, Onn2018, Palmer2018, Wang2018, Zhang2018c, Zhao2018, Weng2018b, Gao2017b, Li2017b, Laskar2017, Park2017, Shang2017, Hilton2017, Ro2016, Mao2020a, Jackson2015, Meng2015, Saha2015, Hsu2015, Wise2015, Tan2017, Wu2017, Li2016b, Lu2016a, Moseson2022, Duong2022, Kwon2021, Huang2021, Wang2016e, Feng2015, Luo2015, Zhao2015, Zhang2015b, Kim2014b, Masango2014, Liang2014, Wang2014, Sun2014, ONeill2014, Lee2014, Lei2014, Clary2020, Gao2020a, Wang2020, Garcea2020, Devine2011, Hackler2020, Witeof2022, Witeof2023, Sosa2023, Yu2022b, Yu2021, Gao2020b, Li2018e, Han2019, Scott2019, Gao2020c, Halder2020, He2020b, Jang2020, Lin2020a, Lin2020b, Mao2020b, Tang2021, Chazot2022, Lu2022a, Yu2022a, Fan2023, Hood2023, Shen2023a, Shen2023b, Lee2022h, Adebayo2022, Asundi2021, Jin2023}\\ 
    \midrule
    China & 153 & \citenum{Cao2018, Zang2006, Cai2023, Lin2023, Liu2023a, Sun2023, Tian2023, Zhang2023, Han2022, Li2022a, Qi2022, Tian2022, Wang2022a, Liu2020b, Zhang2018a, Zuo2022, Zhang2017a, Gong2015, Liu2017a, Liang2017, Li2021, Liu2021a, Bao2020, Gan2020, Song2020, Fang2020, Feng2020, Liu2020a, Duan2016b, Duan2016a, Zhao2017a, Zhao2017c, Zhao2020, Lu2007, Duan2015, Cao2019, Chen2019, Cheng2019, Chu2019, Ding2016, Duan2019, Sun2019, Gong2019, Huang2019, Kong2016, Fu2013, Qin2013, Kong2019, Li2019a, Liu2016a, Wang2016c, Jiang2014, Xie2015b, Luo2019, Qin2019, Zhang2015a, Li2018b, Wang2019a, Wang2019b, Wang2019c, Yan2019, Yang2019b, Zhao2019a, Zhao2019b, Zhao2019c, Gong2018, Yan2017a, Yan2017b, Li2018a, Li2018c, He2018, Qin2016, Qin2018, Xu2018b, Yan2018a, Yan2018b, Zhang2018b, urrehman2018, Cao2017, Chen2017, Du2017, Gao2017a, Huang2017, Li2017a, Liu2016b, Gong2016, Zhou2016, Lv2016, Yao2016, Dai2016, Wang2015, Yan2015, Lu2017, Lv2017, Qin2017, Wang2017a, Wang2017g, Yi2017, Yang2017b, Zhang2017c, Bai2016, Wang2023a, Wang2016f, Wang2016b, Wang2016d, Li2015, Yi2015, Zhang2015c, Lu2014, Duan2017, Yang2022b, Li2018d, Zhang2019f, Li2020, Zhao2023, Hu2019, Mao2020d, Zhu2019, Zhou2020, Ahmad2020, Cao2020, Yang2020, Liu2018b, Cao2021, He2021b, Cai2021, Wang2021, Zhang2021, Zi2021, Li2022c, Liu2022, Wang2022b, Wang2022c, Yang2022c, Zhao2022, Shi2023, Wang2023b, Xing2023, Zhang2023b, Xiang2022, Wang2022d, Li2022d, Li2023, Liu2023b, Liu2023c, Liu2021b, Meng2023, Shen2023c, Zhong2023, Xie2022, Deng2022, Guo2022, Zhang2020}\\ 
     \midrule
    Finland & 74 & \citenum{Backman1998, Lakomaa1992, Backman2000, Haukka1993a, Haukka1993b, Puurunen2000a, Haukka1994a, Haukka1994b, Backman2001, Hirva1994, Kanervo2001, Lindblad1994, Haukka1995a, Kytokivi1996a, Iiskola1997a, Lindblad1997, Rautiainen2002, Ek2003a, Ek2003b, Lashdaf2003a, Puurunen2003, Lashdaf2004a, Lindfors1994, Lakomaa1996, Haukka1997a, Iiskola1997b, Suvanto1997, Suvanto1998, Nieminen1999, Hakuli2000, Puurunen2000b, Uusitalo2000a, KrogerLaukkanen2001, Lashdaf2003b, Ek2004, Haukka1996, Smeds1996, Juvaste2000a, Juvaste2000b, Juvaste2000c, Uusitalo2000b, Haukka1994c, Arandia2023, Yim2023, Kytokivi1996b, Kytokivi1997a, Kytokivi1997b, Puurunen2001, Puurunen2002a, Puurunen2002b, MakiArvela2003, Juvaste1999a, Juvaste1999b, Timonen1999, Heikkinen2021, Korhonen2007, Silvennoinen2007a, Silvennoinen2007b, Backman2009, Nevalainen2009, Vuori2009, Karinen2011, Rikkinen2011, Vuori2011, Nevalainen2012, Sairanen2012, Voigt2019, Hautala2017, Kaariainen2017, Sairanen2014, Lindblad1993, Lindblad1998, Lashdaf2004b, Heikkinen2022}\\ 
     \midrule
    Korea & 71 & \citenum{Lee2023, Yoon2023a, Jung2022, Kim2022a, Lee2022a, Lee2022c, Yang2022a, Yoon2022b, Lee2020a, Lee2020b, Lee2019c, Jeong2016a, Jeong2016b, Jeong2016c, Han2016a, Han2016b, Han2016c, Lim2020a, Han2020, Shin2020, Kim2007b, Jeong2009, Kim2015a, Kim2007a, Kim2009b, Kim2011d, Kim2011b, Kim2011a, Kim2011c, Chae2018, Choi2011, Kang2011, Yoon2011, Jang2019a, Jang2019b, Nam2012, Park2012, Choi2013, Kim2013, Seo2013, Lee2019a, Kim2019c, Seong2016, Jeong2014a, Jeong2014b, Kim2014a, Seong2019a, Shi2019, Kim2018b, Sridharan2015, Kim2015b, Kim2015d, Nguyen2019, Yang2018, Kim2016, Jang2016, Ok2017, Jo2023, Lee2022d, Lee2016, Park2014, Yoon2023b, Kim2020, Lee2022e, Lee2022f, Lee2022g, Lim2020b, Seong2019b, Kim2021, Lee2021, Kim2023}\\ 
     \midrule
    The Netherlands & 34 & \citenum{Jacobs1994a, Jacobs1994b, Moulijn2023, Benz2022, Li2022b, LaZara2021a, LaZara2021b, Benz2020b, Benz2020a, Benz2021, Guo2020, Hashemi2020, LaZara2020, Moret2020a, Beetstra2009, Hashemi2019, Goulas2013, Valdesueiro2016, Valdesueiro2015, Goulas2014, Didden2014, vanOmmen2015, tenKate2019, Vasudevan2015, Grillo2018a, Grillo2018b, Guo2018, VanBui2017, Didden2016, Valdesueiro2017, Grillo2017, Mahtabani2023, Mahtabani2021, Moret2020b}\\ 
     \midrule
    Canada & 25 & \citenum{Deng2019, Sun2013b, Gao2019a, Zhang2019e, Meng2010, Hall2019, Meng2011, Li2012, Liu2013, Li2019b, Cheng2014, Liu2015b, Liu2015a, Kaliyappan2015, Song2018, Zhang2019c, Cheng2016, Xiao2015, Cheng2015a, Cheng2015b, Song2017, Xiao2017, Pallister2014, Liang2020, Nizami2023}\\ 
     \midrule
    Germany & 22 & \citenum{Dannehl2018, Lehmann2023, Basak2022, Baumgarten2022, He2022, He2021a, Knemeyer2021a, Knemeyer2020, Ingale2020, Strempel2016, Strempel2018, Ruff2019, Strempel2017, Ingale2021, Shen2020, Knemeyer2021b, Hussain2022, Moryson2021, Hering2021, Lange2021, Wolff2023, Ikuno2017}\\ 
     \midrule
    Russia & 17 & \citenum{Tsyganenko2000, Malygin1997, Ermakova2002, Malygin2002, Bodalyov2019, Smirnov2008, Malkov2010, Sosnov2010, Sosnov2011, Malygin2012, Sosnov2012, Nasriddinov2019, Zhilyaeva2018, Sosnov2017, Koshtyal2014, Podurets2020, Zemtsova2020}\\ 
     \midrule
    Belgium & 16 & \citenum{Cremers2018, Baltes2001, Cremers2019, Ramachandran2016, DePrins2019, Verheyen2014, Longrie2012, Sree2012, Meledina2016, Verstraete2019, Sree2016, Sree2017, Longrie2014a, Rampelberg2014, Vandegehuchte2014, Srinath2021}\\ 
    \bottomrule
\end{longtable}
\end{footnotesize}

\begin{table}
    \centering
    \begin{tabular}{cc}
           \makecell{Precursor/compound/ligand\\abbreviation} & Name\\
           \hline
           DEZ & Diethylzinc \\
           TMA & Trimethylaluminum \\
           LiOtBu & Lithium \textit{tert}-butoxide\\
           TTIP & Titanium(IV) isopropoxide \\
           NbOEt & Niobium(V) ethoxide\\
           APTS& gamma-(aminopropyl)triethoxysilane\\
           -acac& -acetylacetonate\\
           -hfac& -hexafluoroacetylacetonate\\
           -tmhd& -(2,2,6,6-tetramethyl-3,5-heptanedionate)\\
           TDMA-& Tetrakis(dimethylamino)-\\
           -Cp& Cyclopentadienyl-\\
           MeCpPtMe$_3$ & Trimethyl(methylcyclopentadienyl)platinum(IV) \\
           Mg(EtCp)$_2$ & Bis(ethylcyclopentadienyl)magnesium \\
           Ce(iPrCp)$_3$& Tris(isopropylcyclopentadienyl)cerium(III)
           
    \end{tabular}
    \caption{Abbreviations used in discussion of precursors in the main manuscript.}
    \label{tab:precursor_chem}
\end{table}

\begin{figure}
    \centering
    \includegraphics[width=0.75\linewidth]{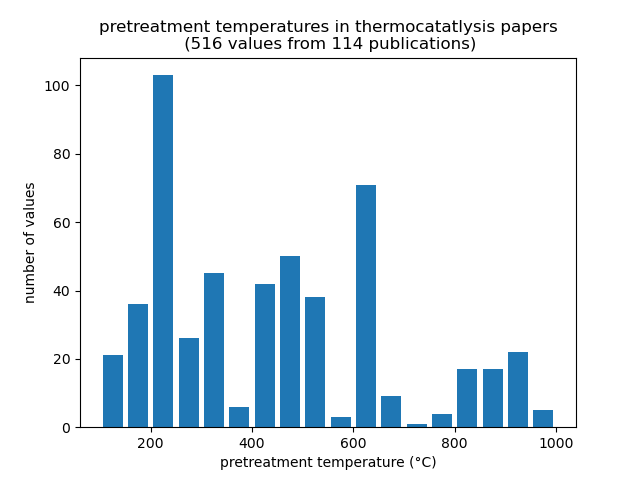}
    \caption{Pretreatment temperature of support materials in publications from application category thermocatalysis.}
    \label{SI:pretreatT_thermo}
\end{figure}

\begin{figure}
  \includegraphics[width=0.75\textwidth]{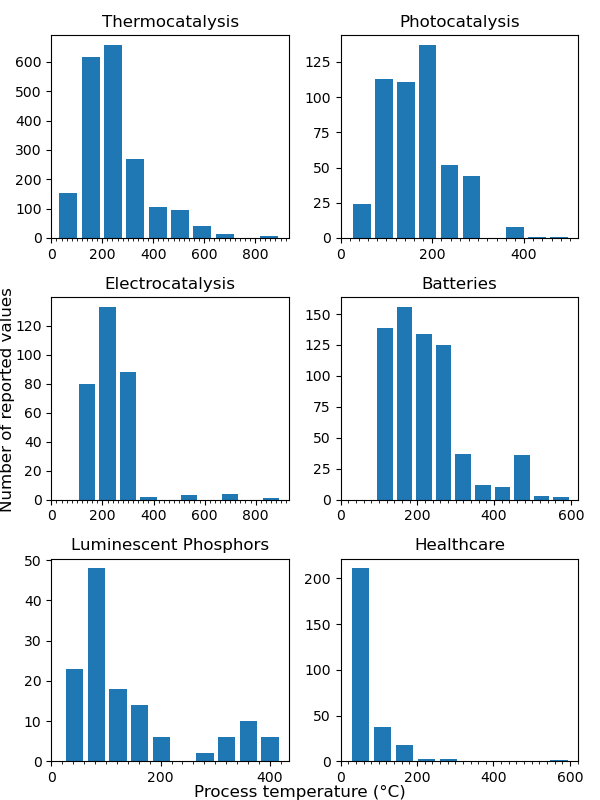}
  \caption{Reported process temperature according to application. Note that in some cases, the process temperature may change from one reactant to the other.}
  \label{SI:histotempapp}
\end{figure}

\begin{figure}
  \includegraphics[width=0.75\textwidth]{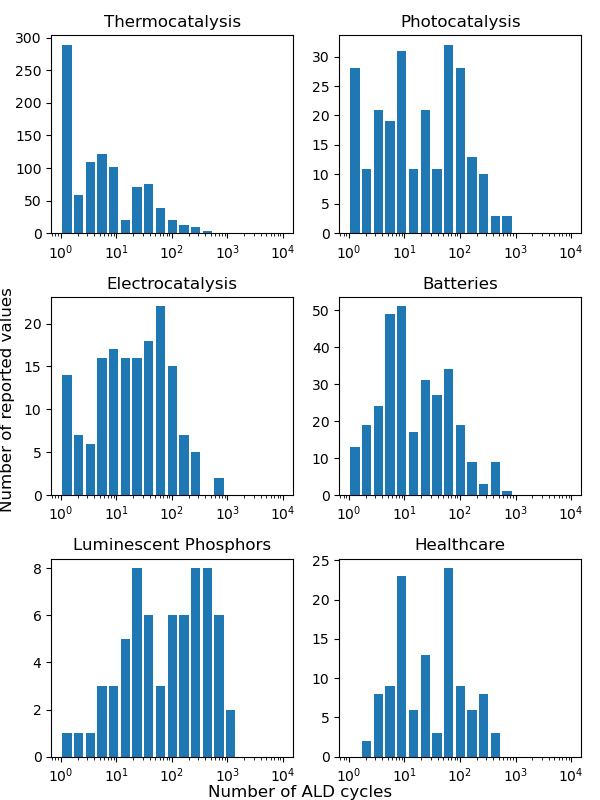}
  \caption{Reported number of ALD cycles for applications discussed in the main manuscript.}
  \label{SI:histcyclesreactor}
\end{figure}

\begin{figure}
  \includegraphics[width=0.75\textwidth]{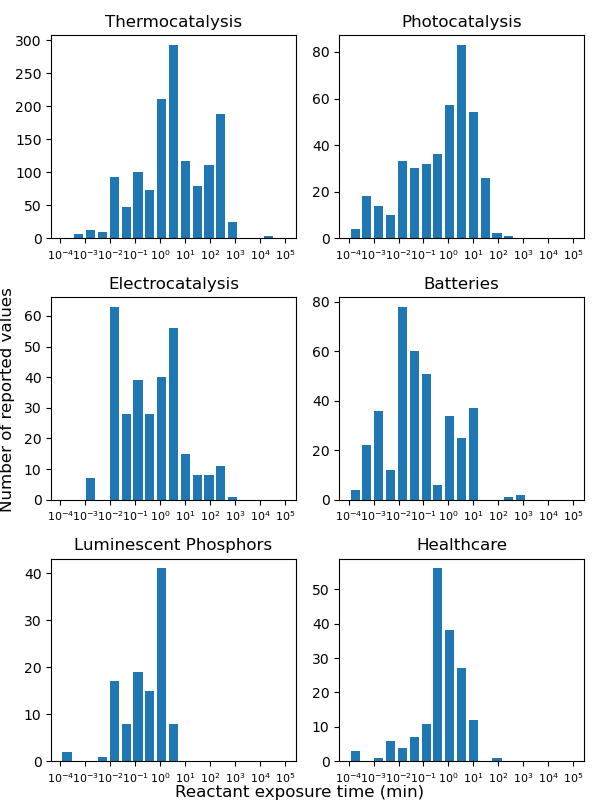}
  \caption{Reported exposure time (irrespective of reactant) according to application.}
  \label{SI:histoexposuretime}
\end{figure}

\begin{figure}
    \centering
    \includegraphics[width=0.75\linewidth]{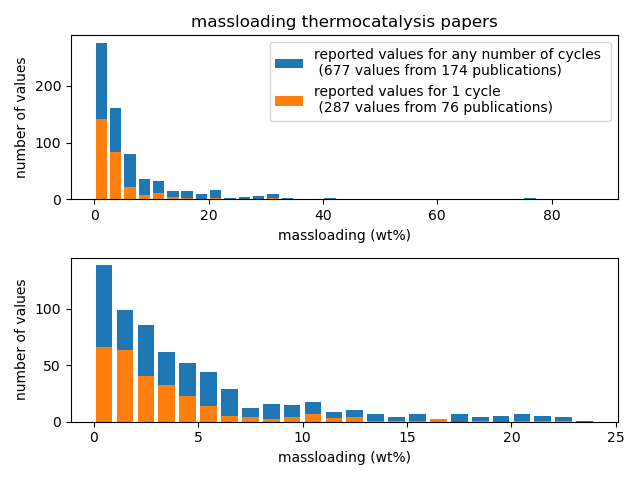}
    \caption{Reported mass loadings for one ALD cycle and any number of ALD cycles, respectively, in publications from application category thermocatalysis.}
    \label{SI:massloading_thermo}
\end{figure}

\clearpage
\bibliography{ALDpmBibliography.bib}